%% file: regret2clmn.tex
\newtheorem{theorem}{\itshape Theorem}

\newtheorem{lemma}{\itshape Lemma}
\newtheorem{proposition}{\itshape Proposition}
\newtheorem{definition}{\itshape Definition}
\newtheorem{assumption}{\itshape Assumption}
\newtheorem{example}{\itshape Example}
\newcommand{\nan}{N \hspace{-4pt} I}
\newcommand{\al}{{\cal X}}
\newcommand{\deff}{\stackrel{\mbox{\scriptsize \rm def}}{=}}
\newcommand{\ubar}[1]{\mbox {\b {$#1$}}}

\newcommand{\thetas}{{K_0}}
\newcommand{\Thetac}{K}
\newcommand{\ep}{\eta}
\newcommand{\phii}{\phi}

\newcommand{\trc}{{\rm TR}}
\newcommand{\var}{{\rm Var}}
\newcommand{\cov}{{\rm Cov}}
\newcommand{\ratio}{\rho}

\newcounter{forassumption}
\newcounter{forlemma}
\newcounter{forlemmaasymptnormality}
\newcounter{forlemmaratio}
\newcounter{forlemmaideal}
\newcounter{forlemmafornormalization}
\newcounter{forlemmamonotone}

\documentclass[journal]{IEEEtran}

\usepackage{amssymb}
\usepackage{amsmath}
\usepackage{mathrsfs}
\usepackage{euscript}
\usepackage[dvipdfmx]{graphicx,color}
\usepackage{url}

\ifCLASSINFOpdf
\else
\fi
\begin{document}

\newcommand{\Limsup}{{\overline{\lim}}}

%
\title{Asymptotically Minimax Regret by Bayes Mixtures}
%
%
%

\author{Jun'ichi~Takeuchi,~\IEEEmembership{Member,~IEEE,}
        and~Andrew~R.~Barron,~\IEEEmembership{Fellow,~IEEE}
\thanks{J. Takeuchi is with Faculty of Information Science
  and Electrical Engineering, Kyushu University,
Fukuoka 819-0395, Japan. E-mail: (see 
http://www.me.inf.kyushu-u.ac.jp/\~{}tak/homepage.en.html).}
\thanks{A. R. Barron is with Department of Statistics and Data Science, Yale University, CT 06511, USA.}
\thanks{This research was supported in part
by JSPS KAKENHI, Grant Numbers 
JP19300051,
JP24500018, JP18H03291, and JP23H05492.}}

%
%

\markboth{IEEE Transactions on Information Theory,~Vol.~X,
  No.~X, XXXX 201X}%
{Shell \MakeLowercase{\textit{et al.}}: Bare Demo of IEEEtran.cls for Journals}
%




\maketitle

\begin{abstract}
We study the problems of data compression, gambling
and prediction of a sequence $x^n=x_1x_2...x_n$
from an alphabet ${\cal X}$,
in terms of regret and expected regret (redundancy) 
with respect to various smooth families of probability distributions.
We evaluate
the regret of  Bayes mixture distributions compared to maximum likelihood,
under the condition that
the maximum likelihood estimate is in the interior of the parameter space.
For general exponential families (including the non-i.i.d.\ case) the 
asymptotically mimimax value is achieved 
when variants of the prior of Jeffreys are used.
Interestingly, we also obtain 
a modification of Jeffreys prior which has measure
outside the given family of densities,
to achieve minimax regret with respect to non-exponential type
families. This modification enlarges the family
using local exponential tilting (a fiber bundle). Our conditions are confirmed for certain non-exponential families, including curved families and mixture families (where either the mixture components or their weights of combination are parameterized) as well as contamination models.
Furthermore for mixture families we show how to deal with the full
simplex of parameters.
These results also
provide characterization of Rissanen's stochastic complexity.
\end{abstract}

\begin{IEEEkeywords}
universal coding, universal prediction, regret, redundancy, 
exponential family, Bayes mixture, Jeffreys prior
\end{IEEEkeywords}

%
\IEEEpeerreviewmaketitle

\section{Introduction}

We study the problem of data compression, gambling
and prediction of a string $x^n=x_1,x_2,...,x_n$
from a given alphabet ${\cal X}$,
in terms of regret and expected regret (redundancy) with respect to 
various families of probability distributions.
We evaluate
the regret and expected regret of Bayes mixture distributions
and
show that it asymptotically achieves 
the minimax value when variants of Jeffreys prior are used.
Our results contain
generalization of the results in \cite{xb96a,xb96b,cb92}.
Our results
provide evaluation of stochastic complexity 
defined by Rissanen \cite{riss96}.

This paper's main concern is the regret
of a coding or prediction strategy.
This regret is defined as
the difference of the loss incurred and the loss
of an ideal coding or prediction strategy for each sequence.
A coding scheme for strings of length $n$
is equivalent to
a probability mass function $q(x^n)$ on ${\cal X}^n$.
We can also use $q$ for prediction and gambling, that is,
its conditionals $q(x_{i+1}|x^i)$ 
provide a distribution for the coding or prediction
of the next symbol given the past.
The minimax regret with respect to
a target family of probability mass functions
$S=\{p(\cdot|\theta):\theta \in \Theta \}$ 
and a given set of parameters $K \subset \Theta$ is defined as
\[
\min_{q}\max_{x^n \in {\cal K}}
\max_{\theta \in \Theta} 
\Bigl(
\log \frac{1}{q(x^n)}
-
\log \frac{1}{p(x^n|\theta)}\Bigr),
\]
where 
${\cal K}$ denotes the set
$\{x^n: \hat{\theta}(x^n) \in K \}$,
which
is the set of strings for which
the maximum likelihood estimate $\hat{\theta}$ is in 
$K$.
Typically $K$ is either all of $\Theta$ or a subset that 
excludes points near the boundary.
{
The maximum is taken for all strings $x^n$ in $\cal K$.}
The regret
$\log (1/q(x^n))-\log (1/p(x^n|\hat{\theta}))$
in the data compression context is
also called the (pointwise) redundancy:
the difference between
the code length based on $q$ and the minimum
of the codelengths $\log (1/p(x^n|\theta))$
achieved by distributions in the family.
Also, 
$\log (1/q(x^n))-\log (1/p(x^n|\theta))$
is the sum of the incremental regrets of prediction
$\log (1/q(x_{i+1}|x^i))-\log (1/p(x_{i+1}|x^i,\theta))$,
with what is sometimes called the log-loss.
This regret is also called the pointwise regret, to
emphasize the distinction from the expected regret also
discussed below.

The heart of our analysis is the consideration 
of Bayes mixtures and the use of the Laplace method to
approximate them.
A Bayes mixture takes 
the form
$q(x^n) = \int p(x^n|\theta)w(\theta)d\theta$,
also called the 
Bayes factor or the marginal density of $x^n$
obtained by integrating out $ \theta$ from the joint
distribution.

Though certain coding and prediction settings have a 
discrete alphabet $\al$, we are interested in minimax regret
problems also for continuous spaces $\al$.
Then the $p(x^n|\theta)$ as well as $q(x^n)$
are understood to be probability density functions
with respect to a given reference measure.

For the case that $S$ is the class of all discrete memoryless sources,
it was proved \cite{xb96b} that the minimax regret asymptotically equals
$(d/2) \log (n / 2\pi)+\log C_{J}(K) + o(1)$,
where $d$ equals the size of alphabet minus $1$
and 
$C_{J}(K)$ is the integral of the square root of the 
determinant of the Fisher information matrix
over $K$.
An important point in the above is that
$K$ is taken there to be $\Theta$ itself, i.e.\
we do not have to have any restriction for the sequence
$x^n$.
To obtain this asymptotically minimax regret,
they use sequences of Bayes mixtures 
with certain prior distributions that weakly converge to
the Jeffreys prior.
The reason why one needs such variants of the Jeffreys prior
is as follows:
If we use the Jeffreys prior,
the regret is asymptotically higher than the minimax value,
for $x^n$ such that $\hat{\theta}$ is near the boundary of $\Theta$.
Priors which have higher density
near the boundaries than the Jeffreys prior
give more prior attention to these boundary regions
and thereby pull the regret down to not more than the asymptotically minimax level.

In this paper, we provide
such regret results for more general parametric families of
densities.
For exponential families of arbitrary dimension, the Jeffreys
mixture is shown to be asymptotically minimax, if $K$ is a
compact subset included in the interior of $\Theta$.
A boundary modification is shown to produce a variant of
Jeffreys mixture asymptotically minimax for $K =\Theta$
in the one-dimensional exponential family case. For general
smooth families that are not of exponential type, we find that any Bayes mixture
that uses a prior restricted to the family is not asymptotically minimax, but a slight
modification (enlargement) of the family allows for priors for which the regret of Bayes mixtures does achieve the asymptotically minimax value.

The related
problem of minimax expected regret (redundancy),
which is defined as
\[
\min_q \max_{\theta \in K} 
E_\theta\Bigl(
\log \frac{1}{q(x^n)}
-
\log \frac{1}{p(x^n|\theta)}
\Bigr)
\]
was studied by Clarke and Barron \cite{cb92}.
They considered fairly general classes of i.i.d.\
processes and
showed that the minimax expected regret
asymptotically equals
$
(d/2)\log(n/2\pi e)
+\log C_{J}(K) +o(1)
$,
when $K$ is a compact subset of the
interior of $\Theta$.
Moreover the expected regret is related to the value of
\[
\min_q \max_{\theta \in K}
E_\theta
\Bigl( 
\log\frac{1}{q(x^n)}
-
\log\frac{1}{p(x^n|\hat{\theta})}
\Bigr)
\]
using the target code length with $1/p(x^n|\hat{\theta})$ instead of
$1/p(x^n|\theta)$.
It has a corresponding minimax value
$
(d/2)\log(n/2\pi )
+\log C_{J}(K) +o(1)
$.
Preceding work \cite{xb96b},
\cite{xb96a} evaluated
the minimax expected regret for
the class of discrete memoryless sources
and
showed that sequences of slightly modified
Jeffreys mixtures achieve the minimax value asymptotically
for the whole probability simplex $\Theta$.
As we shall see, the answer for the minimax regret and
the minimax expected regret are similar.

To obtain the minimax regret results,
we employ the Laplace integration method,
which was used by \cite{cb90,cb92}
to evaluate the expected regret of the Bayes procedures.
Especially in \cite{cb92},
they succeeded to uniformly evaluate the expected regret
by the Laplace integration for compact subsets $K$ of
$\Theta^\circ$, the interior of $\Theta$.
To handle all of $\Theta$, 
careful modification of the Laplace method is required to handle 
behaviour near the boundary.

Since the Jeffreys mixture
achieves asymptotically the minimax expected regret,
one might expect that
it also achieves asymptotically the minimax pointwise regret.
However, it does not in general.
Namely, we can see that
for processes which are not exponential type,
the worst case regret of the Jeffreys mixture is
asymptotically higher than the minimax value,
even if strings are restricted such that
the maximum likelihood estimate (MLE)
is in the interior of the parameter space.

The reason follows from examination
of the Laplace approximation to the Bayes mixture density.
This leads to an approximation to 
the regret of the Jeffreys mixture which converges to the minimax value,
if and only if 
the difference
between 
the determinant of the empirical Fisher information matrix
and
that of the Fisher information matrix at 
the MLE converges to $0$.
For exponential families,
the 
empirical and expected
Fisher informations are the same, hence the Jeffreys mixture
is asymptotically minimax.
The situation is different
for families which are not exponential type, for then
there exist sequence $x^n$
for which the determinants of these matrices are 
asymptotically different.

Even though the Jeffreys mixture is not asymptotically minimax,
we can obtain an asymptotic minimax regret, 
by a modified Jeffreys mixture 
obtained by adding a small contribution from
a mixture of an enlarged set of densities,
whose dimension is higher than the original set.
The added components deal with the strings for which
the empirical Fisher information differs from
the Fisher information.
When the original set is a curved exponential family embedded in
an exponential family,
we have the option to use that family as the enlarged model.
Furthermore, this method can be applied to non-i.i.d.\ families under certain assumptions.

Our result about minimax regret provides 
an alternative way to evaluate, more generally,
the stochastic complexity in Rissanen \cite{riss96},
where he used the normalized maximum likelihood.
Our results show that the codelength of
the minimax strategy retains asymptotically the mixture codelength
interpretation of earlier incarnations of his criterion. The work \cite{BRW14} shows 
that in some cases the exact minimax strategy has a signed mixture interpretation 
without resorting to asymptotics.

The codelength of a code based on a mixture
$q(x^n) = \int_K p(x^n|\theta)w(\theta)d\theta$ has Laplace
approximation (for $\hat{\theta}$ away from the boundary of $K$) given by
\[
\log \frac{1}{q(x^n)}
\sim
\log \frac{1}{p(x^n|\hat{\theta})}+
\frac{d}{2}\log\frac{n}{2\pi}
+\log \frac{|\hat{J}(\hat{\theta},x^n)|^{1/2}}{w(\hat{\theta})\,\;},
\]
where $\hat{J}(\hat{\theta},x^n)$ is the empirical Fisher information,
defined as $(1/n)$ times the second derivative matrix
of the minus log-likelihood, evaluated at the MLE
$\hat{\theta}$.
With Jeffreys prior
in which $w(\theta)$ is proportional to 
$|J(\theta)|^{1/2}$,
the Laplace approximation for $\log(1/q(x^n))$ is
\[
\log \frac{1}{p(x^n|\hat{\theta})}
+
\frac{d}{2}\log\frac{n}{2\pi}
+\log C_{J}(K)
+\frac{1}{2}\log \frac{{|\hat{J}(\hat{\theta})|}}
{{|J(\hat{\theta})|}},
\]
where
\[
C_{J}(K) = \int_K {|J(\theta)|^{1/2}}d\theta.
\]
Accordingly, the regret of the code based on
Jeffreys prior takes the approximation form
\[
\frac{d}{2}\log\frac{n}{2\pi}
+\log C_{J}(K)
+\frac{1}{2}\log \frac{{|\hat{J}(\hat{\theta})|}}
{{|J(\hat{\theta})|}}.
\]
Our modification of $q(x^n)$ by enlargement 
of the family with an associated modification to the prior
shows that
\begin{eqnarray}\label{star}
\frac{d}{2}\log\frac{n}{2\pi}
+\log C_{J}(K)
\end{eqnarray}
remains the 
asymptotically minimax regret
even though in the non-exponential family cases 
there are sequences $x^n$ for which 
$\log \bigl({{|\hat{J}(\hat{\theta},x^n)|}}/
{{|J(\hat{\theta})|}}\bigr)$
does not converge to $0$.

This presents a challenge for the formulation and analysis
of our asymptotic minimax procedures with the maximum of the
regret taken over all $x^n$.
In contrast the minimax expected regret is easier to achieve
because
the expectation washes out the effect of $\hat{J}$ different
from $J$.
In either formulation, with $\log(1/p(x^n|\hat{\theta}))$ as
the target,
the minimax value takes the asymptotic form (\ref{star}).

{
In Section~II, we
formally introduce the notion of minimax and maximin regret.
Section III gives keys to the bounds.
In Section~IV,
we give the lower bound on the maximin regret
for general smooth families in the i.i.d. setting (Theorems 1 and 2) and for families of
non-i.i.d.\ densities (Theorems 3 and 4).  These hold for
any subset $K$ of $\Theta$ with finite Jeffreys integral.
Likewise, in Section V, we give corresponding upper bound on the minimax regret
for general smooth families
with a compact subset $K$ in the interior of $\Theta$ 
(Theorem 5), which incorporate the discussed innovations.
In Section~VI, we provide various types of families as concrete examples, including certain curved exponential families and certain mixture families that are not representable as exponential families.

}


\section{Preliminaries}

Let $( {\cal X},{\cal B},\nu )$ be a measurable space 
with a reference measure $\nu$, assumed to be sigma finite 
(such as counting measure or Lebesgue measure).
Let $S=\{p(\cdot|\theta):\theta \in \Theta \}$
denote a parametric family of 
probability densities over ${\cal X}$ with respect to $\nu$.
Assume that $\Theta \subseteq \Re^d$.
We let $p(x^n|\theta)$ denote
$
\prod_{i=1}^n p(x_i|\theta)
$.
Also, we let $\nu(dx^n)$ denote
$
\prod_{i=1}^n \nu(dx_i)
$.
Here, we are treating models for
independently identically distributed (i.i.d.)\ random variables.
(Section~IV treats certain non i.i.d.\ cases.)
We let $P_\theta$ denote the 
distribution function with density $p(\cdot|\theta)$
and $E_\theta$ denote expectation with respect to $P_\theta$.


Define the Fisher information matrix by
\[
J_{ij}(\theta)=E_\theta\Bigl(-
\frac{\partial^2 \log p(x|\theta)}
{\partial \theta^i\partial \theta^j}\Bigr).
\]
(We let $\log$ denote the natural logarithm.) 
We assume that $J(\theta)$ exists 
and is strictly positive definite in 
the interior of $\Theta$.
Let $K$ denote a subset of $\Theta$.
We let
$
C_J(K)\deff \int_K|J(\theta)|^{1/2}d\theta
$, which we call the Jeffreys integral on $K$.
Our interest is in cases in which $C_J(K)$ is finite.
The Jeffreys prior over $K$ 
is a prior distribution with density function
\[
w_{K}(\theta)=\frac{{|J(\theta)|^{1/2}}}{C_J(K)}
1_{K}(\theta).
\]
Define the Jeffreys mixture for $K$ as
\[
m_K(x^n)
=
\int_K p(x^n|\theta)w_{K}(\theta)d\theta.
\]
We also introduce the empirical Fisher information as a function of $x^n$:
\[
\hat{J}_{ij}(\theta) = \hat{J}_{ij,n}(\theta)
=
\hat{J}_{ij}(\theta,x^n)
=\frac{-1\:\:\; }{n}\frac{\partial^2 \log p(x^n|\theta)}
{\partial \theta^i\partial \theta^j}.
\]
Note that $J_{ij}(\theta)=E_\theta [\hat J_{i,j}(\theta)]$. Moreover, in the present i.i.d. setting, $\hat {J}_{ij,n}(\theta)$ is near $J_{ij}(\theta)$ with high probability for large $n$, by the law of large numbers, when $J_{ij}(\theta)$ is finite, if we were to have the $X_i$ distributed according to $P_\theta$.  We work with the arbitrary sequence perspective, so the empirical $\hat{J}_{ij,n}(\theta)$ need not be close to $J_{ij}(\theta)$.  Nevertheless we will find a role for the expected value formulation of Fisher information in characterization of the minimax regret among arbitrary sequences. 

Let $\hat{\theta}(x^n)$ be
the maximum likelihood estimate (MLE), 
which is
\[
\arg \max_{\theta \in \Theta}p(x^n|\theta).
\]
More strictly, we define $\hat{\theta}$ to be an element of the set $
\{ \theta : p(x^n|\theta) = \max_{\theta} p(x^n|\theta)  \}$.
For the case where this set has more than one element,
we presume there is an arbitrarily specified rule to choose one element.

For example, 
in the case of Bernoulli sources, 
we have
$\hat{\theta}=\sum_{i=1}^nx_i/n$,
where $\theta$ is the parameter denoting
the probability that `$1$' occurs.
The function $\hat{\theta}$ mapping 
each $x^n$ in ${\cal X}^n$ to the value $\hat{\theta}(x^n)$ 
is the maximum likelihood estimator.
In the above case of Bernoulli sources,
$\hat{\theta}(x^n)$
is defined on the whole ${\cal X}^n$.
However, in certain cases,
there may exist strings for which 
$\max p(x^n|\theta)$
does not exists.
In such cases, 
we restrict the domain of $\hat{\theta}$
to the set of $x^n$'s such that the MLE does exist. 

We introduce the notion of minimax regret and maximin regret.
Let ${\cal K}_n$ denote a fixed subset of ${\cal X}^n$.
Let ${\cal P}_{n}$ denote the set which consists 
of all probability densities over $\al^n$ 
with respect to the fixed sigma-finite reference measure.

For sample size $n$ and set ${\cal K}_n$, 
define the maximum regret of a $q$ in ${\cal P}_{n}$
(denoted by $\bar{r}_n(q,{\cal K}_n)$) as
\[
\bar{r}_n(q,{\cal K}_n)
=
\sup_{x^n \in {\cal K}_n}
\sup_{\theta \in \Theta}
\log \frac{p(x^n|\theta)}{q(x^n)}
=
\sup_{x^n \in {\cal K}_n}
\log \frac{p(x^n|\hat{\theta})}{q(x^n)}
\]
and the \emph{minimax} regret as 
\[
\bar{r}_n({\cal K}_n)
=\inf_{q \in {\cal P}_{n}}\bar{r}_n(q,{\cal K}_n).
\]
For each $q \in {\cal P}_{n}$,
the minimum average regret for sample size $n$
denoted 
\[
\ubar{\; r}_n(q)=
\inf_{\tilde{q} \in {\cal P}_{n}}
\int q(x^n)
\Bigl(\log \frac{p(x^n|\hat{\theta})}{\tilde{q}(x^n)} \Bigr) \nu(dx^n)
\]
is achieved by $\tilde{q}=q$. 
We define the \emph{maximin} regret for the set ${\cal K}_n$ as
\[
\ubar{\; r}_n({\cal K}_n)=\sup_{q \in {\cal P}_{{\cal
      K}_n}}\ubar{\; r}_n(q),
\]
where ${\cal P}_{{\cal K}_n}$ is the set of all probability
densities supported on ${\cal K}_n$.

By the above definitions,
$\bar{r}_n({\cal K}_n) \geq \ubar{\; r}_n({\cal K}_n)$ holds.
(In fact, 
$\bar{r}_n({\cal K}_n)=\ubar{\; r}_n({\cal K}_n)$ 
holds \cite{shtarkov,xb96b}.)
In this paper, we usually consider the
minimax regret problem for 
the following form of the set ${\cal K}_n$:
given $K \subseteq \Theta$,
\[
{\cal K}_n ={\cal X}^n(K) =\{x^n:\hat{\theta} \in K  \}.
\]

Next, we remind the reader of the notions of the minimax and maximin value of 
expected regret (redundancy).
Let a sample size $n$ and a parameter set $K$ be given.
For each $q \in {\cal P}_{n}$,
we define the maximum expected regret
(denoted by $\bar{R}_n(q,K)$) as
\[
\bar{R}_n(q,K)=
\sup_{\theta \in K}
E_\theta \Bigl(\log \frac{p(x^n|\theta)}{q(x^n)} \Bigr)
\]
and
the minimax expected regret as 
\[
\bar{R}_n(K) = \inf_{q \in {\cal P}_n}\bar{R}_n(q,K),
\]
where
$E_\theta$ denotes the expectation with respect to $p(x^n|\theta)$.
Let $\Omega(K)$ denote the set which consists
of all prior probability measures over $K$.
For each prior $w \in \Omega(\Theta)$,
the minimum Bayes expected regret
for sample size $n$ denoted
\[
\ubar{R}_n(w)=
\inf_{q \in {\cal P}_n}
\int E_\theta\Bigl(\log \frac{p(x^n|\theta)}{q(x^n)}\Bigr)w(d\theta)
\]
is achieved by $q(x^n) = m(x^n)= \int p(x^n|\theta)w(d\theta)$.
We define the maximin expected regret for the parameter set $K$
as
\[
\ubar{R}_n(K)=\sup_{w \in \Omega(K)}\ubar{R}_n(w).
\]
We also have $\bar{R}_n(K) \geq \ubar{R}_n(K)$.
(In fact, $\bar{R}_n(K) = \ubar{R}_n(K)$ holds
\cite{dl80,haussler97}.)

Finally, we note some terminology.
We let $1_A$ denote the characteristic function of a set $A$.
We write $o(1)$ for an expression upper bounded by a positive quantity tending to zero.
In the proofs,
we denote 
certain positive constants by $C_i$ ($i=1,2,...$),
where $i$ is `local' in each proof.

\section{The Key to Handling Worst Case Regret}

The notion
that the worst case regret is the same asymptotically 
as an average case
regret should come as a surprise.
Indeed, if one inspects the worst case regret of an average
case optimal procedure, e.g.\ with the Jeffreys mixture,
there will be strings for which the worst case performance
of the procedure is
quantitatively different in its asymptotics 
than the average case performance.

To succeed in bringing the worst case
regret value down to
the average case optimum requires a sequence of 
somewhat new specialized procedures.
These specializations are modifications of Jeffreys mixtures
that address two difficulties.
One is that
Laplace approximation is no longer valid for
strings for which the MLE 
$\hat{\theta}$ is at or very close to the boundary of the
parameter space.
That is a familiar issue, addressed by modification of the
prior to give small additional prior weights to parameter values
near the boundary.
Such boundary modification
was needed in the special
case of \cite{xb96a,xb96b}.
Here we provide an improved near-boundary fix by using a multiplicative correction factor
for the prior (see Definition~\ref{def_of_ideal_prior}).  
This improved correction factor comes from the probability of 
$K$ assigned by a normal distributions that arises in the Laplace approximation. This just-right probability correction factor from Laplace approximation is near $1$ away from the boundary (by an amount of larger order than $1/\sqrt n)$, and typically approaches $1/2$ as the parameter approaches the boundary (at a rate faster than order $1/\sqrt n$).

The more fundamental difficulty
is the discrepancy between the worst case and
average values of the regret of mixture procedures that can
exist even when $\hat{\theta}$ is in the interior,
which, as exhibited above, is due to individual strings
having 
$(1/2) \log \bigl(|\hat{J}(\hat{\theta})|/|J(\hat{\theta})|\bigr)$ 
not near zero when not in an exponential family.
We present and analyze ideas to overcome that difficulty
which
were initiated by the authors as discussed in \cite{tb98j,bt98,bry98}.

For general smooth families we form a direct
enlargement by a exponential tilting using linear
combinations of the entries of the differences
$\hat{J}(\theta)-J(\theta)$.
It is formed as
\begin{eqnarray}
  \label{eq:enlargeform}
p(x^n|\theta,\beta)
=
p(x^n|\theta) e^{n\beta 
\cdot (\hat{J}(\theta)-J(\theta))-\psi_n(\theta,\beta)} 
\end{eqnarray}
Where $\beta \cdot M$ for matrices $\beta$ and $M$
denotes the Frobenius inner product (the sum of products across all
$d^2$ entries).

The idea for this enlargement in addressing minimax regret
originate in preliminary form in
\cite{tb98j,bt98}
as informally discussed in \cite{bry98,patent}. 
Here
$\psi_n(\theta,\beta)$ is the log of the required
normalization
factor, so that
$p(x^n|\theta,\beta)$ sums (integrates with respect to
$\nu^n$)
to the value $1$ for every $\theta \in K$ and every $\beta$ in a 
neighborhood around $0$.
The prior assigns most of its weight to a
Jeffreys prior on $\theta$ (with $\beta$ set to $0$)
and small weight on a smooth prior on $(\theta,\beta)$
with $\beta$ in a neighborhood of $0$.
Now demonstration of the success of this modification is 
based on demonstration of increased likelihood
beyond that which is available at
$\beta= 0$ 
when $\hat{J}(\hat{\theta})-J(\hat{\theta})$ 
is not equal to $0$.
Indeed, 
as we show in Section~\ref{section:generalization:upperbounds},
the contribution to the mixture from $\beta$ 
with $\beta \cdot
(\hat{J}(\hat{\theta})-J(\hat{\theta}))$ 
positive
is 
sufficient
to increase the value of $m(x^n)$ to again overcome the
discrepancy in
$(1/2)\log (|\hat{J}(\hat{\theta})|/|J(\hat{\theta})|)$
from Laplace approximation.

We have a somewhat different modification that 
can be used in the case of curved exponential families embedded in a
full exponential family as developed in examples in  Section VI.
For both of the modifications we study, there is,
in the analysis, the consideration of values of $\beta$ in a neighborhood of a
small multiple of $\hat{J}(\hat{\theta})-J(\hat{\theta})$
which are sufficient to accomplish our objectives.
We see
the similarity of effect. In both cases there is
opportunity
to create a likelihood increase from a suitable linear
combination of these statistics.

These ideas emanate from an underlying
principle.
Of importance in smooth statistical families is
the parametric enlargement
\begin{align}
  \label{eq:tangent}
p(x^n|\theta,\beta)
=
p(x^n|\theta)e^{\beta\cdot \nabla \log p(x^n|\theta)-\psi_n(\theta,\beta)},
\end{align} 
where 
$\nabla \log p(x^n|\theta)$ is the score function at
$\theta$
and $\psi_n(\theta,\beta)$ is the log normalizing constant
near $(1/2)\beta^t J(\theta)\beta$ for small $\beta$.
Traditionally 
such a family 
 arises  in {local asymptotic expansion} of
likelihood ratios, 
evaluated at a perturbation $\theta + \beta$ of a
given $\theta$,
as used in demonstration of {\itshape local asymptotic normality}
\cite{IH81,lecam86,pollardonline}.
In Amari's information geometry it is a local exponential
 tangent to the family at $\theta$.
Again, in this setting, a prior can be put jointly on $\theta$ and $\beta$,
where most of the prior weight is concentrated at
$\beta = 0$ with Jeffreys prior on $\theta$.

The use of the first derivative of $\log p(x^n|\theta)$
in the exponent would seem to produce a different sort of enlargement
than occurs with equation (2) which uses the second derivative.
Nevertheless,
improvement in $m(x^n)$ still arises by suitable characterization
of the increased likelihood available with the enlarged
family.
For each small $\beta$ the maximum likelihood
value $\hat{\theta}_{\beta}$ achieving 
$\max_\theta p(x^n|\theta,\beta)$ in (\ref{eq:tangent}) satisfies the approximate
relationship
\begin{eqnarray}\label{star2}
p(x^n|\hat{\theta}_\beta,\beta)
=p(x^n|\hat{\theta}_0)
e^{(n/2)\beta^T (\hat{J}(\hat{\theta}_0)-J(\hat{\theta}_0))\beta}  
\end{eqnarray}
to within term of order $n|\beta|^3$ in the
exponent.
This is the desired effect, where 
$\hat{\theta}_0 = \hat{\theta}$
here
is the MLE in the original family
and
$\hat{\theta}_\beta$
is the MLE in the tilted family (\ref{eq:tangent}).
The likelihood is larger at some non-zero $\beta$ than at $\beta=0$
provided 
$\beta^T(\hat{J}(\hat{\theta})-J(\hat{\theta}))\beta$ can be
strongly positive, or equivalently provided
$\hat{J}(\hat{\theta})(J(\hat{\theta}))^{-1}$ has
some eigenvalue greater than $1$.
Consequently, when 
$(1/2)\log (|\hat{J}(\hat{\theta})|/|J(\hat{\theta})|)$
is positive (which is the only case of concern)
optimization in this tangent family is sufficient to
realize similar likelihood gain
to optimization in the family tilted by second derivatives 
of log-likelihood.

This provides an enlargement of parametric families by
exponential tilting using score functions suitable to endow
the family with the property we need, comparable to the
curved exponential family case.
Analogous conditions are in Amari and Nagaoka \cite{an2000},
who use tilting by first and second derivatives.
A distinction thought is that he is considering local families
with a fixed true $\theta$
whereas 
we are considering the joint family
parametrized by $\theta$ and $\beta$.

For general families in the present paper we use the
enlargement of the form
(\ref{eq:enlargeform})
in our analysis, now further 
motivated
by the connections to locally asymptotically normal familiers we have discussed here.

\section{Lower Bounds}
\label{section:generalization}

Our targets are
fairly general smooth parametric families under some conditions
and
some classes of non-i.i.d.\ stochastic processes.
First we describe our results about lower bounds.

\subsection{Lower Bounds for General i.i.d.\ Families}

Let $S \deff \{ p(\cdot|\theta) : \theta \in \Theta  \}$ be a family of 
probability densities
with respect to the reference measure $\nu$, with a $d$-dimensional parameter $\theta$.
Assume that 
$\Theta \subseteq \Re^d$
and let $\Theta^\circ$ denote its interior.
We use $K$ to denote subsets of $\Theta$, with particular interest in
sets for which $C_J(K)$ is finite and for which the boundary measure of $K$ is zero.
We handle both the case of bounded sets $K$ (Lemma~1)
and more general cases of possibly unbounded sets $K$ (Theorem~1).

For each $\delta > 0$,
define Mahalanobis neighborhoods of $\theta \in \Theta$ as
\[
B_\delta(\theta) 
=
\{\theta'\in \Theta : 
(\theta' - \theta)^t J_\theta(\theta' - \theta) \leq \delta^2 \}.
\]

We employ the assumptions described below.
\begin{assumption}\label{assume:gen:1}
The density $p(x|\theta)$ is twice continuously
differentiable for $\theta \in \Theta^\circ$ for every $x$.
Moreover, for every $\theta \in \Theta^\circ$ there is a $r=r(\theta)$
such that, for every $i$, $j$,
\begin{align}\label{eq:assume:gen:1}
E_{\theta} \Biggl[
\sup_{\theta' \in B_{r}(\theta) } 
|\hat{J}_{ij,1}(\theta')|\Biggr]
\end{align}
is finite. 
\end{assumption}
\begin{assumption}\label{assume:gen:2}
The Fisher information $J(\theta)$  is continuous
and positive definite 
in $\Theta^\circ$.
\end{assumption}

Consequences of continuity and positive definiteness are that the following two quantities tends to $0$ as $\delta \rightarrow 0$ for any $\theta$ in $\Theta^\circ$
\begin{align}
\label{eq:Jsecond}
\sup_{\theta' \in B_\delta(\theta)}
 \frac{|J(\theta')|^{1/2}}{|J(\theta)|^{1/2}}-1
\end{align}
and
\begin{align}\label{eq:Jfirst}
\sup_{\theta' \in B_\delta(\theta)}
 \max_{z \neq 0}
 \frac{z^tJ(\theta)z}
 {z^tJ(\theta')z}
 -1.
\end{align}

Since $J(\theta)$ is symmetric and positive definite it has a real eigendecomposition with positive eigenvalues. Taking the positive square roots of the eigenvalues provides representation of the principle square root matrix $J(\theta)^{1/2}$ and taking their reciprocal provides representation of $J(\theta)^{-1}$. With $J(\theta)$ continuous these inverses and square roots remain symmetric and continuous in $\theta$ (\cite{HornJohnson},p.411). Armed with these we can standardize the empirical Fisher information as $J(\theta)^{-1/2}\hat{J}(\theta')J(\theta)^{-1/2}$, which has expectation equal to the identity matrix $I$ at $\theta'=\theta$.

From the continuity from Assumption 2 and the domination (finiteness of expected supremum) from Assumption 1, it follows by the monotone convergence theorem that for every $i,j$ the following quantity also tends to $0$ as $\delta \rightarrow 0$, for each $\theta$ in $\Theta^{\circ}$,
\begin{align}
\label{eq:expectedsupremum}
E_\theta \sup_{\theta' \in B_{\delta}(\theta) } 
\pm ((J(\theta)^{-1/2}\hat{J}_{1}(\theta',x)J(\theta)^{-1/2})_{ij} -I_{ij}),
\end{align}
where 
the $\pm$ indicates that the statement is true with each choice of sign.

\begin{assumption}\label{assume:gen:3}
The maximum likelihood estimator $\hat{\theta}$
 is a consistent estimator of $\theta$,
for each $\theta$ in $\Theta^\circ$, so that 
\[
P_\theta \{ ||\hat{\theta}(x^n)-\theta||_{J(\theta)} > \delta  \} 
=o(1),
\]
for each $\delta > 0$.
\end{assumption}

{\em Remark:}
We give a demonstration of consistency under suitable conditions 
in Lemma~\ref{lemma:uce} in the
Appendix. 
Assumptions 1, 2, and 3 are related to the conditions 
pioneered by Cram\'er (see p.\ 501 of \cite{cramer1946})
for local asymptotic properties of maximum likelihood estimators
building on earlier assumptions by Wald \cite{wald1949} for
consistency of the maximum likelihood estimator.

\noindent\begin{assumption}\label{assume:gen:K}
The set $K \subseteq \Theta$ has positive and finite 
\[
C_J(K)=\int_K|J(\theta)|^{1/2}d\theta
\]
and the measure of $K \setminus K^\circ$ is zero. 
\end{assumption}

We remark that the zero measure condition is to make $C_J(K^\circ)=C_J(K)$.
It is clearly satisfied if $K$ is open or if $K$ is a closed set
with boundary measure zero.

From the theory of finite measures, convergent functions are uniformly convergent except in sets of arbitrarily small measure. Consequently, as elaborated in the appendix, if Assumptions~\ref{assume:gen:1}, 
\ref{assume:gen:2}, and \ref{assume:gen:K} hold, then, for any $\epsilon >0$, there is a $\delta >0$ and a good set $G$ of parameters, such that for all $\theta$ in $G$ the quantities in (\ref{eq:Jsecond}) and (\ref{eq:Jfirst}) are less than $\epsilon$, the expected suprema in (\ref{eq:expectedsupremum}) are less than $\epsilon/(2d)$ and the Jeffreys measure of the complement of $G$, which is $\int_{K \smallsetminus G}  |J(\theta)|^{1/2} d\theta$, is less than $\epsilon\, C_J (K)$.  

Moreover, using a law of large numbers, 
it is shown in the appendix, as a consequence of the expected suprema being less than $\epsilon/(2d)$, that for $\theta$ in $G$,
\begin{align}\label{seriesC}
P_\theta\Bigl(
\inf_{z \neq 0}
\inf_{\theta' \in B_{\delta}(\theta) } 
\frac{z^t\hat{J}(\theta',x^n)z}{z^tJ(\theta)z}
< 1-\epsilon
\Bigr)
=
o(1)
\end{align}
as $n$ goes to infinity.

To see the connection with the standardized empirical information, note that the infimum here among $z$ is the same as $$\inf_{\zeta: ||\zeta||=1}
\inf_{\theta' \in B_{\delta}(\theta) } 
{\zeta^t  J(\theta)^{-1/2}\hat{J}(\theta')J(\theta)^{-1/2} \zeta}$$ as can be seen by setting $\zeta$ to correspond to $J(\theta)^{1/2}z$ divided by its norm.


Two lower bounds related to maximin regret will be given. The first, using Assumption 1, is a general lower bound showing the asymptotically validity of the $(d/2) \log (n/2\pi) + \log C_J(K)$ expression, but it somewhat less constructive. It uses the fact mentioned above that convergent functions are nearly uniformly convergent, so there exists a good set $G$ with the indicated properties, while having a small Jeffreys measure for its complement.
Also, as we said, it uses a law of large numbers for a sample average of such suprema, but, assuming only the finite expectation, there is, in general, no assurance of rates of approach in such a law of large numbers, and this translates to a lack of explicit rate of approach in the general lower bound. Moreover, for the purpose of obtaining the general lower bound on the minimax value it suffices to appeal to an approximation argument with Jeffreys mixtures that live on $G$.

The other lower bound we give is specific to the use of the Jeffreys prior on $K$ without modification. In that case a stronger condition, refining Assumption 1, will be used with finite expected square (to permit application of Chebyshev's inequality for control of rate of convergence in the law of large numbers), as well as a continuity assumption of the expected suprema, to get uniform convergence (uniform closeness to zero of the expected suprema) within any compact subset of $K$.  Armed with this more refined Assumption $1^\prime$ (to be specified later below) we have more explicit control of rates of approach to the maximin value.

Now let $A \subseteq K$ be a parameter set to which Lemma~1
will appeal.  This set $A$ can be the good set $G$ as discussed above, whose existence is a consequence of Assumptions 1, 2, and
\ref{assume:gen:K}. Or if $K$ is compact and additional assumptions are satisfied we may have $A=K$. 

Define
\begin{align}
\eta_{1,\delta}(A)
= &
\sup_{\theta \in A}
\sup_{\theta' \in B_\delta(\theta)}
 \max_{z \neq 0}
 \frac{z^tJ(\theta)z}
 {z^tJ(\theta')z}
 -1,
\end{align}
\begin{align}
\eta_{2,\delta}(A)
= &
\sup_{\theta \in A}
\sup_{\theta' \in B_\delta(\theta)}
 \frac{|J(\theta')|^{1/2}}{|J(\theta)|^{1/2}}-1,
\end{align}
and, similarly, let $\eta_{0,\delta}(A)$ be $2d$ times the following supremum 
\begin{align}
\sup_{\theta \in A} E_\theta \sup_{\theta' \in B_{\delta}(\theta) } 
\pm (J(\theta)^{-1/2}\hat{J}_{1}(\theta',x)J(\theta)^{-1/2})_{ij} -I_{ij}).
\end{align}
As previously mentioned the $2d$ factor is so that when this $\eta_{0,\delta}(A) \le \epsilon$ the conclusion of (\ref{seriesC}) holds.

We extract nice sets within $\Theta \times \al^n$,
which we will use in obtaining lower bounds on the maximin value.
Define
\begin{align}
\mathscr{B} & = \nonumber
\{ (\theta,x^n) :
 \theta,\hat{\theta}(x^n) \in K, 
||\theta - \hat{\theta}(x^n)||_{J(\theta)} \leq \delta/4  \}  \\
\mathscr{N} & =
\Bigl\{
(\theta,x^n) : \theta \in A,
\inf_{\theta' \in B_{\delta}(\theta) } 
\inf_{z \neq 0} 
\frac{z^t\hat{J}(\theta',x^n)z}{z^tJ(\theta)z}
\geq
1-\epsilon
\Bigr\}.
\end{align}
Here, for any subset $\cal G$ of $\Theta \times \al^n$,
we let ${\cal G}_{x^n}$ and ${\cal G}_{\theta}$ denote the section of $\cal G$ given $x^n$ and
$\theta$, respectively,
that is,
\begin{align*}
  {\cal G}_{x^n} & = \{ \theta : (\theta,x^n) \in \cal G \}, \\
  {\cal G}_{\theta} & = \{ x^n : (\theta,x^n) \in \cal G \}. 
\end{align*}




Now fix a pair $(\theta,x^n) \in \mathscr{N}$.
We show that the section $\mathscr{B}_{x^n}$
is included in $B_\delta(\theta)$.
Assume $\theta' \in \mathscr{B}_{x^n}$.
Then, 
\[
||\theta' -\hat{\theta} ||_{J(\theta')} \leq \delta/4
\]
holds. Likewise, since $(\theta,x^n) \in \mathscr{B}$,
\[
||\theta -\hat{\theta} ||_{J(\theta)} \leq \delta/4.
\]
Since $\theta,\theta' \in \mathscr{B}_{x^n}$, and $\theta \in A$ we have,
by the definition of $\eta_{1,\delta}=\eta_{1,\delta}(A)$,
\[
||\theta' -\hat{\theta} ||_{J(\theta)} \leq 
(1+\eta_{1,\delta})^{1/2}
||\theta' -\hat{\theta} ||_{J(\theta')} \leq \delta/2,
\]
assuming $\eta_{1,\delta} \le 3$. Whence
\[
||\theta' -\theta ||_{J(\theta)} < \delta,
\]
which means $\theta' \in B_{\delta}(\theta)$
and $\mathscr{B}_{x^n}  \subset B_{\delta}(\theta)$.
By this, if $x^n \in \mathscr{N}_\theta$
and $(\theta, x^n) \in \mathscr{B}$,
the following holds,
\begin{align}\label{iidfisherlower}
\min_{\theta' \in \mathscr{B}_{x^n}} 
\min_{z \neq 0} 
\frac{z^t\hat{J}(\theta',x^n)z}{z^tJ(\theta)z}
\geq
1-\epsilon.
\end{align}




We can prove the following Lemma,
which is the heart of our main results for the lower bounds.
\begin{lemma}\label{gen:lemma:lower}
Let $S=\{ p(\cdot|\theta) : \theta \in \Theta\}$
be a $d$-dimensional family of probability densities.
We suppose that 
Assumptions~\ref{assume:gen:1}-\ref{assume:gen:K} hold for $S$. Let $0<\epsilon<1$ be given and assume
for a specified $A \subset K$ that there is a $\delta >0$ such that $\eta_{0,\delta}(A)\le \epsilon$. Assume also that $\eta_{1,\delta}(A)\le 3$ and let $\eta_{2,\delta}=\eta_{2,\delta}(A)$.
Define a probability density $m_{\delta}$ supported on the set
\[
\mathcal{A}_\delta=
\{ x^n : (\mathscr{N} \cap \mathscr{B})_{x^n} \neq \emptyset \}
\subset \mathcal{K},
\]
by
\begin{align}
 \bar{m}_{\delta}(x^n)
&= 
\int_{(\mathscr{N}\cap\mathscr{B})_{x^n}}p(x^n|\theta)w_{A}(\theta)d\theta,\\
\label{eq:codedistribution}
m_{\delta}(x^n) 
&=
\frac{\bar{m}_{\delta}(x^n) }{\bar{C}_{n}},
\end{align}
where
{
\begin{align}\label{def_of_cbar_n}
\bar{C}_{n}
= 
\int_{\mathcal{K}}
\bar{m}_{\delta}(x^n)\nu(dx^n)
\end{align}}
Then
\[
E_{m_{\delta}}
\log\frac{p(x^n|\hat{\theta})}{m_{\delta}(x^n)}
\geq
\log  
\frac{(1-\epsilon)^{d/2}C_J(A)n^{d/2}}{(1+\eta_{2,\delta})(2\pi)^{d/2}}
-o(1)
\]
holds as $n$ goes to infinity.
\end{lemma}

{\em Remark:} The support of the density $m_\delta$ is
the set of $x^n$ such that
the section $(\mathscr{N}\cap\mathscr{B})_{x^n}$ is not empty.

{\em Proof:} 
We have
\begin{align}
E_{m_{\delta}}
\log\frac{p(x^n|\hat{\theta})}{m_{\delta}(x^n)}
\label{decomp_of_maximin}
=
E_{m_{\delta}}
\log\frac{p(x^n|\hat{\theta})}{\bar{m}_{\delta}(x^n)}
+\log \bar{C}_{n}.
\end{align}
Note that
the log likelihood ratio in the first term of the right hand side is positive,
while $\log 
(p(x^n|\hat{\theta})/m_{\delta}(x^n))
$ can be negative.

For the first term.
we have
\begin{align*}
& 
E_{m_{\delta}}\log\frac{p(x^n|\hat{\theta})}{\bar{m}_{\delta}(x^n)}  
\\
= &
\int_{{\mathcal{K}}}
\frac{\bar{m}_{\delta}(x^n)}{\bar{C}_{n}}
\log\frac{p(x^n|\hat{\theta})}{\bar{m}_{\delta}(x^n)}  
\nu(dx^n) \\
= &
\frac{1}{\bar{C}_{n}}
\int_{{\mathcal{K}}}
\int_{(\mathscr{N}\cap \mathscr{B})_{x^n}}
\! \! \!
p(x^n|\theta)w_{{A}}(\theta)
d\theta
\log\frac{p(x^n|\hat{\theta})}{\bar{m}_{\delta}(x^n)}  
\nu(dx^n) \\
= &
\frac{1}{\bar{C}_{n}}
\int_{{K}} \! \!
w_{{A}}(\theta)
\Bigl[
\int_{(\mathscr{N}\cap \mathscr{B})_{\theta}}
\!\!
p(x^n|\theta)
\log \frac{p(x^n|\hat{\theta})}{\bar{m}_{\delta}(x^n)}\nu(dx^n)\Bigr]
d\theta .
\end{align*}
Here, using the sections, reversed order integration is possible
by the Fubini-Tonelli theorem because of positivity of the integrand.

Let $i(\theta)$ denote the quantity in the brackets $[$ $]$.
We will show a lower bound on $i(\theta)$.

Note that 
$x^n \in (\mathscr{N} \cap \mathscr{B})_{\theta}$
is equivalent to
$(\theta,x^n) \in \mathscr{N} \cap \mathscr{B}$
and
to 
$\theta \in (\mathscr{N} \cap \mathscr{B})_{x^n}$.
By Taylor expansion 
of the log likelihood ratio around $\hat{\theta}$
under the condition that $x^n \in (\mathscr{N} \cap \mathscr{B})_{\theta}$,
we have 
\begin{align} \nonumber
 &\frac{\bar{m}_{\delta}(x^n)}{p(x^n|\hat{\theta})}\\ \nonumber
=&
\int_{(\mathscr{N} \cap \mathscr{B})_{x^n}}
\frac{p(x^n|\theta')w_{A}(\theta')}{p(x^n|\hat{\theta})}d\theta'\\ \nonumber
= &
\int_{(\mathscr{N} \cap \mathscr{B})_{x^n}}
e^{-n(\theta'-\hat{\theta})^t\hat{J}(\theta'',x^n)(\theta'-\hat{\theta})/2}
w_{A}(\theta')d\theta',
\end{align}
where $\theta''$ is a point between $\hat{\theta}$ and $\theta'$.
Since
$x^n \in (\mathscr{N} \cap \mathscr{B})_{\theta}$,
we have $\theta \in (\mathscr{N} \cap \mathscr{B})_{x^n}$.
Now
\begin{align*}
\max_
{\theta' \in B_\delta(\theta)}
\frac{w_{A}(\theta')}{w_{A}(\theta)}
&\leq (1+\eta_{2,\delta})
\end{align*}
and
\begin{align*}
\min_
{\theta' \in B_\delta(\theta)}
\min_{z \neq 0}
\frac{z^t\hat{J}(\theta',x^n)z}{z^tJ(\theta)z}
&\geq
1-\epsilon
\end{align*}
hold for $\theta \in A$, where $\eta_{2,\delta}=\eta_{2,\delta}(A)$.
Hence 
we have for all $x^n \in (\mathscr{N} \cap \mathscr{B})_{\theta}$,
\begin{align*}
& \frac{\bar{m}_\delta(x^n)}{p(x^n|\hat{\theta})}\\
\leq &
 (1+\eta_{J,\delta})w_{A}(\theta)
\int_{\mathscr{N} _{x^n}}
e^{-n(1-\epsilon)(\theta'-\hat{\theta})^tJ(\theta)(\theta'-\hat{\theta})/2}
d\theta'\\
\leq &
(1+\eta_{2,\delta})w_{A}(\theta)
\int
e^{-n(1-\epsilon)(\theta'-\hat{\theta})^tJ(\theta)(\theta'-\hat{\theta})/2}
d\theta'\\
= &
\frac{(1+\eta_{2,\delta})w_{A}(\theta)}{(1-\epsilon)^{d/2}}
\frac{(2\pi)^{d/2}}{n^{d/2}|J(\theta)|^{1/2}}\\
= &
\frac{(1+\eta_{2,\delta})}{(1-\epsilon)^{d/2}}
\frac{(2\pi)^{d/2}}{n^{d/2}C_J(A)}.
\end{align*}
Hence, for all $\theta \in A$, we have
\begin{align*}
i(\theta)
\!
\geq
\!
\log \! 
\frac{(1-\epsilon)^{d/2}C_J(A)n^{d/2}}{(1+\eta_{2,\delta})(2\pi)^{d/2}}
\!\!
\int_{(\mathscr{N} \cap \mathscr{B})_{\theta}} \!\!\!\!
p(x^n|\theta)
\nu(dx^n),
\end{align*}
which implies
\begin{align*}
&\int_{{K}}w_{A}(\theta)i(\theta)d\theta \\
\geq &
\Bigl(
\log \frac{(1-\epsilon)^{d/2}C_J(A)n^{d/2}}{(1+\eta_{2,\delta})(2\pi)^{d/2}}
\Bigr)
\int_{{\mathcal{K}}}\bar{m}(x^n)\nu(dx^n)\\
= &
\Bigl(
\log \frac{(1-\epsilon)^{d/2}C_J(A)n^{d/2}}{(1+\eta_{2,\delta})(2\pi)^{d/2}}
\Bigr)
\bar{C}_{n}.
\end{align*}
Dividing both sides by $\bar{C}_{n}$, we have
\begin{align}\label{lower_barm}
E_{m_\delta}\log \frac{p(x^n|\hat{\theta})}{\bar{m}_\delta(x^n)}
\geq 
\log  
\frac{(1-\epsilon)^{d/2}C_J(A)n^{d/2}}{(1+\eta_{2,\delta})(2\pi)^{d/2}}.
\end{align}

Now we will proceed to a lower bound on $\bar{C}_{n}$.
We have
\begin{align}\nonumber
\bar{C}_{n}&= 
%
%
\int_{\mathcal{K}}
\int_{(\mathscr{N} \cap \mathscr{B})_{x^n}}
p(x^n|\theta)w_{A}(\theta)
d\theta
\nu(dx^n)\\ \nonumber
&= 
\int_{{K}}
\int_{(\mathscr{N} \cap \mathscr{B})_{\theta}}
p(x^n|\theta)
\nu(dx^n)
w_{A}(\theta)
d\theta\\ \nonumber
& =
\int_{{K}}
P_{\theta}( (\mathscr{N} 
\cap 
\mathscr{B})_\theta )w_{A}(\theta)d\theta\\ \label{barCbound_pre}
& \geq
\int_{{K}}(
P_{\theta}( \mathscr{N}_\theta )
+
P_{\theta}( \mathscr{B}_\theta )
-1)
w_{A}(\theta)d\theta\\
&=
\int_{{K^\circ}}(
P_{\theta}( \mathscr{N}_\theta )
+
P_{\theta}( \mathscr{B}_\theta )
-1)
w_{A}(\theta)d\theta,
\end{align}
where the last equality holds because the boundary of $K$
is assumed to have zero measure.
Here, because of (\ref{seriesC}),
\[
P_{\theta}( \mathscr{N}_\theta )
= 1-o(1)
\]
holds 
as $n$ goes to infinity,
for each $\theta \in {A}$.
Now
\[
\mathscr{B}_\theta
= \{ x^n  : \hat{\theta} \in B_{\delta/4}(\theta) \cap K   \}.
\]
For $\theta$ on the boundary, the probability of this event generally converges
to number less than one. But for $\theta$  in the interior, the
consistency implies that $\hat{\theta}$ is in $K$ with probability
tending to $1$.
Hence for each $\theta \in {K^\circ}$,
we have
\[
P_\theta(\mathscr{B}_\theta) 
= 1- o(1)
\]
as $n$ goes to infinity,
because of Assumption~\ref{assume:gen:3}. Obviously, $\bar{C}_n \le 1$.
Together with (\ref{barCbound_pre}) and the pointwise convergence of
$
P_{\theta}( \mathscr{N}_\theta )
$
and
$
P_{\theta}( \mathscr{B}_\theta )
$ to $1$, by Fatou's Lemma (or the dominated convergence theorem)
we have that  $\liminf_{n} \bar{C}_n \ge 1$. Thus
\begin{align}\label{barCbound}
\bar{C}_{n} = 1-o(1)
\end{align}
as $n$ goes to infinity.
Hence, with (\ref{decomp_of_maximin}) and (\ref{lower_barm}), we have
\[
E_{m_\delta}
\log\frac{p(x^n|\hat{\theta})}{m_\delta(x^n)}
\geq
\log  
\frac{(1-\epsilon)^{d/2}C_J(A)n^{d/2}}{(1+\eta_{2,\delta})(2\pi)^{d/2}}
-o(1),
\]
as $n$ goes to infinity.
{\em The proof is completed.}

Now we state the theorem about the lower bound on
the maximin regret for general i.i.d.\ families.
\begin{theorem}\label{gen:thm:lower}
Let $S=\{ p(\cdot|\theta) : \theta \in \Theta\}$
be a $d$-dimensional family of probability densities.
Suppose that 
Assumptions~\ref{assume:gen:1}-\ref{assume:gen:3} hold for $S$.
Let
$K$ be a subset of $\Theta$ (possibly $K =\Theta$)
which satisfies Assumption~\ref{assume:gen:K}.
Then the following holds
\begin{eqnarray}\label{lower_limit}
\liminf_{n \rightarrow \infty}
\Bigl(\ubar{r}_n(\mathcal{K})
-\frac{d}{2}\log\frac{n}{2\pi}
\Bigr)
\geq \log C_J(K).
\end{eqnarray}

\end{theorem}

{\it Remark:}
This theorem allows for $\Theta$ and $K$ to be unbounded
as long as $C_J(K)$ is finite.


{\it Proof of Theorem~1:}
We are given $K \subset \Theta$ satisfying the finite $C_J(K)$ assumption. Let $\epsilon >0$ be arbitrarily small. We have the existence of the set $A=G$ for which $\int_{K-G}w_K(\theta)d\theta \le \epsilon$ and within which the properties of convergence to zero of expressions (\ref{eq:Jsecond}), (\ref{eq:Jfirst}) and (\ref{eq:expectedsupremum}) hold uniformly within $G$. This is an appeal to Egorov's theorem for finite measures that pointwise convergence implies uniform convergence except in a set of negligible measure.  Accordingly, we can arrange positive $\delta$ sufficiently small that $\eta_{1,\delta}(A)\le \epsilon \le 3$, $\eta_{2,\delta}(A)\le \epsilon$, and $\eta_{0,\delta} \le \epsilon$.  As detailed in the appendix, the last inequality implies the in-probability behavior (\ref{seriesC}).  By Lemma 1 we have a procedure providing a lower bound on the maximin regret
$$
\underline {r}_n(K)\ge
\log  
\frac{(1-\epsilon)^{d/2}C_J(A)n^{d/2}}{(1+\eta_{2,\delta})(2\pi)^{d/2}}
-o(1)
$$
Accordingly, 
$$\liminf \left[ \underline {r}_n(K) - (d/2)\log \frac{n}{2\pi} \right]$$
is at least $$\log C_J(K) +(1+d/2) \log (1-\epsilon) - \log(1+\epsilon).$$
Now since $\epsilon$ is arbitrarily small it implies that 
$$\liminf \left[ \underline {r}_n(K) - (d/2)\log \frac{n}{2\pi} \right] \ge \log C_J(K).$$
{\it This completes the proof of Theorem 1.}

Under the following stronger assumptions
than those for
Theorem~\ref{gen:thm:lower}, 
we can obtain a lower bound on 
the average regret for mixtures using Jeffreys prior on all of $K$, when $K$ is compact.
%
\setcounter{forassumption}{\theassumption}
\renewcommand{\theassumption}{\arabic{assumption}$^{\prime}$}
\setcounter{assumption}{0}
\begin{assumption}\label{assume:gen:dash:1}
The density $p(x|\theta)$ is twice continuously
differentiable in $\theta$ for all $x$,
and 
there is a $\bar{r} = \bar{r}(K) > 0$ 
so that for each $i$, $j$ and every $\theta \in K$,
\begin{align}\label{eq:strongassumption1}
E_{\theta} \biggl[
\sup_{\theta' : |\theta'-\theta| \leq
  {r}} 
|\hat{J}_{ij,1}(\theta')|^2
\biggr]
\end{align}
is finite and continuous as a function of $\theta$ for $r \le \bar {r}$.
\end{assumption}

\begin{assumption}\label{assume:gen:dash:2}
$J(\theta)$ 
is continuous and positive definite in $\Theta$.
Furthermore, 
the gradient $\nabla \log p(x|\theta)$
has expectation $E[\nabla \log p(X|\theta)] = \int \nabla p(x|\theta)dx$
equal to zero
with covariance matrix 
\[
I(\theta) = E_\theta[ \nabla \log p(X|\theta)\nabla^t \log p(X|\theta)]
\]
also finite and continuous in $\Theta$.
\end{assumption}

Here $I(\theta)$ and $J(\theta)$ are the two forms of Fisher Information. Usually $I(\theta)$ equals $J(\theta)$,
though we do not need that here.

\begin{assumption}\label{assume:gen:dash:3}
The maximum likelihood estimator $\hat{\theta}$
is a consistent estimator of $\theta$,
indeed consistent uniformly in $K$,
with tail probability tending to zero faster than $1/\log n$,
that is,
\[
\sup_{\theta \in K}
P_\theta \{ |\hat{\theta}(x^n)-\theta| > \delta  \} 
=o(1/\log n),
\]
for each $\delta > 0$.
\end{assumption}

Demonstration of consistency, indeed with tail probability tending to zero at rate $O(1/n)$ is given in the Appendix under sensible conditions.

Assumption~\ref{assume:gen:dash:1}
differs from Assumption~1 because of the presence
of the square of the empirical Fisher information in (\ref{eq:strongassumption1}).
With Assumption~1$^\prime$ using Chebyshev's inequality 
we can obtain the following property which strengthens (\ref{seriesC}) 
\begin{align}\label{seriesBdash}
& \max_{\theta \in K}
P_\theta \Bigl(
\inf_{z \neq 0}
\inf_{\theta' : |\theta'-\theta| \leq \delta  }
\frac{z^t\hat{J}(\theta',x^n)z}{z^tJ(\theta)z}
\leq  1-\epsilon
 \Bigr) \\ \nonumber
&  =  \: O(1/n) .
\end{align}



\setcounter{forlemmaasymptnormality}{\thelemma}
\begin{lemma}\label{lemma:Lower:Jeffreys}
\input{lemma_asympt_normality}
\end{lemma}

This Lemma~\ref{lemma:Lower:Jeffreys} is proved in the Appendix.

Let $K_\delta= \{ \theta \in \Theta : B(\theta,\delta)\subset K \}$ be
the $\delta$ interior of $K$,
where we are using the ordinary Euclidean ball
$B(\theta,r)=\{\theta' \in R^d : ||\theta' -\theta|| \le r \}$
and let $K\setminus K_\delta$ be the corresponding $\delta$ shell of
$K$.
(Here we could have used the ellipses based on $J(\theta)$ in place of
Euclidean balls in defining an analogous shell.)

\begin{assumption}\label{assume:gen:K;prime}
The set $K \subseteq \Theta$ has positive and finite 
\[
C_J(K)=\int_K|J(\theta)|^{1/2}d\theta
\]
and, with $\delta_n$ of order $\sqrt{\log n}/\sqrt{n}$, the contribution 
of the shell $\int_{K\setminus K_{\delta_n}}|J(\theta)|^{1/2}d\theta$ 
is of order $o(1/\log n)$.
\end{assumption}

\renewcommand{\theassumption}{\arabic{assumption}}
\setcounter{assumption}{\theforassumption}

Note that, if 
the boundary of $K$ has finite surface measure, then the
measure
of the $\delta$ shell is of order $\delta$, which is of order
$\sqrt{\log n}/\sqrt{n}$
with the chosen $\delta_n$, so the shell measure vanishes at rate much
faster than the required $o(1/\log n)$.

The asymptotic lower  bound of Theorem~1 was established by using
a code distribution (\ref{eq:codedistribution})
analogous to the Jeffreys mixture code,
but with modification by restriction for each $x^n$
to $\theta$ in $(\mathscr{B} \cap \mathscr{N})_{x^n}$.
It is of interest to know whether a Bayes mixture using Jeffreys prior
on all of $K$ has the same asymptotic lower bound.
The next result (Theorem~2) shows this to be true for compact $K$
with slightly more stringent assumptions. Indeed with these 
assumptions 
the Jeffreys mixture provides a lower bound
which is as good as the asymptotically
maximin value.
\begin{theorem}\label{gen:thm:lower:new}
Let $S=\{ p(\cdot|\theta) : \theta \in \Theta\}$
be a $d$-dimensional family of probability densities.
Suppose that 
$K$ is a compact subset of $\Theta$ for which 
Assumptions~1$^\prime$, 2$^\prime$, 3$^\prime$, and 4$^\prime$ hold. 
Define a density $m_{{K}}$ over $\mathcal{K} = \al^n(K)$ as
\begin{align*}
m_{{K}}(x^n) &= \frac{\bar{m}_{{K}}(x^n)}{C_{n}}  \\
\bar{m}_{{K}}(x^n) &= 1_{\mathcal{K}}(x^n)\int_{{K}}
p(x^n|\theta)
w_{{K}}(\theta)d\theta,
\end{align*}
where $C_{n}$ is the normalization constant
and $w_{{K}}(\theta)$ is the Jeffreys prior on ${K}$.
Then the following holds
\begin{eqnarray*}
\liminf_{n \rightarrow \infty}
\Bigl(
\ubar{r}_n(m_{{K}})
-\frac{d}{2}\log\frac{n}{2\pi}
\Bigr)
\geq \log C_J({K}).
\end{eqnarray*}
\end{theorem}


{\it Remark:} The code distribution $m_K(x^n)$ here differs from
code distribution $m_\delta$ used in Theorem~1 because here
we do not make the restriction to $\mathscr{B} \cap \mathscr{N}$.

{\em Proof:} 
We examine the value $\ubar{r}_n(m_{{K}})$.
We have
\begin{align*}
  & E_{m_K} \log \frac{p(x^n|\hat{\theta})}{m_K(x^n)} \\
= &
 \int_{\mathcal{K}} m_K(x^n) \log \frac{p(x^n|\hat{\theta})}{m_K(x^n)}\nu(dx^n)\\
= &
 \int_{\mathcal{K}} m_K(x^n) \log \frac{p(x^n|\hat{\theta})}{\bar{m}_K(x^n)}\nu(dx^n)
+ \log C_n \\
= &
\frac{1}{C_n} \int_{\mathcal{K} } \int_K p(x^n|\theta)w_K(\theta)d\theta 
\log \frac{p(x^n|\hat{\theta})}{\bar{m}_K(x^n)}\nu(dx^n)\\
& + \log C_n .\\
\end{align*}
We proceed to lower bound it as follows, where $m_\delta$ is the same as in Lemma~1 and
$1_{\mathscr{G}\cap\mathscr{N}}=1_{\mathscr{G}\cap\mathscr{N}}(\theta,x^n)$, 
\begin{align*}
\ge &
\frac{1}{C_n}
 \int_{\mathcal{K}} \int_K
1_{\mathscr{G}\cap\mathscr{N}}
p(x^n|\theta)w_K(\theta)d\theta
\log \frac{p(x^n|\hat{\theta})}{\bar{m}_K(x^n)} \nu(dx^n)\\
& + \log C_n \\
= & \frac{\bar{C}_n}{C_n}
 \int_{\mathcal{K}} 
m_\delta(x^n)
\log \frac{p(x^n|\hat{\theta})}{\bar{m}_K(x^n)} \nu(dx^n)
+ \log C_n \\
= & \frac{\bar{C}_n}{C_n}
 \int_{\mathcal{K}} 
m_\delta(x^n)
\log \frac{p(x^n|\hat{\theta})}{m_K(x^n)} \nu(dx^n)
+ (1-\frac{\bar{C}_n}{C_n})\log C_n \\
\ge &  
\frac{\bar{C}_n}{C_n}
 E_{m_\delta}
\log \frac{p(x^n|\hat{\theta})}{m_\delta(x^n)}
+ (1-\frac{\bar{C}_n}{{C}_n})\log C_n \\
\ge &  
\bar{C}_n
 E_{m_\delta}
\log \frac{p(x^n|\hat{\theta})}{m_\delta(x^n)}
+ \log C_n,
\end{align*}
where the first inequality in the above manipulation follows
from
${p(x^n|\hat{\theta})/\bar{m}_K(x^n)} \geq 1$,
the second inequality follows from the positivity 
of Kullback divergence of $m_\delta$ from $m_K$,
and the last inequality follows from
$C_n \leq 1$,
which implies $\log C_n \le 0$.

We appeal to Lemma~1 using $A=K$.  By Assumption 1$^\prime$, the expected supremum in (\ref{eq:expectedsupremum}) is continuous in $\theta$ and by the monotone convergence theorem it converges monotonically to $0$ as $\delta \rightarrow 0$, so by Dini's Theorem (see e.g. (\cite{Rudin}), p.150) the convergence is uniform on the compact $K$. 
Accordingly, $\eta_{0,\delta}(K)$ tends to zero as $\delta\rightarrow 0$.  Likewise, by compactness, $\eta_{1,\delta}(K)$ and $\eta_{2,\delta}=\eta_{2,\delta}(K)$ tend to zero as $\delta \rightarrow 0$. Given $\epsilon>0$, pick $\delta$ such that $\eta_{0,\delta}(K) \le \epsilon$ and $\eta_{1,\delta}(K)\le 3.$

By Lemma~1, the expectation in the last line above
is lower bounded as
\[
E_{m_{\delta}}
\log\frac{p(x^n|\hat{\theta})}{m_{\delta}(x^n)}
\geq
\log  
\frac{(1-\epsilon)^{d/2}C_J({K})n^{d/2}}{(1+\eta_{2,\delta})(2\pi)^{d/2}}
-o(1).
\]
Picking $\delta > 0$ and $\epsilon > 0$
sufficiently small, the above lower bound is sufficient for our purpose, 
provided $\bar{C}_n$ and $C_n$ converges to $1$
at a rate sufficiently fast.
In fact, noting that the lower bound contains the term $(d/2)\log n$,
we need to prove
$\bar{C}_n=1-o(1/\log n)$
and
$C_n=1-o(1)$,
which imply the claim of Theorem~2.

First examine $\bar{C}_n$.
Here, we already have
$\bar{C}_n=1-o(1)$ in the proof of Lemma~1,
but it is not sufficient for our purpose.
However, using Assumptions 1$^\prime$-3$^\prime$, we can show
$\bar{C}_n=1-o(1/\log n)$.
Indeed, by (\ref{barCbound_pre}), we have
\begin{align}\label{eq:cbarnlower}
\bar{C}_{n} \ge
\int_{{K^\circ}}(
P_{\theta}( \mathscr{N}_\theta )
+
P_{\theta}( \mathscr{B}_\theta )
-1)
w_{{K}}(\theta)d\theta.
\end{align}
Here,
by (\ref{seriesBdash}), we have
\[
\min_{\theta \in K}P_{\theta}( \mathscr{N}_\theta )
\ge 1 - o(1/\log n).
\]
Hence
\begin{align}\label{eq:Nlower}
\int_K P_\theta(\mathscr{N}_\theta )w_{K}(\theta)d\theta
\ge 1 - o(1/\log n).
\end{align}
For evaluation of $P_\theta(\mathscr{B}_\theta)$, we employ
Lemma~2.
Let $\delta_n = \sqrt{b\log n}/\sqrt{n}$.
Then by Lemma~2,
\[
\min_{\theta \in K}
P_\theta( \hat{\theta} \in B(\theta,\delta_n/2) )
\ge 1- \frac{4c}{b^2\log n} -o\Bigl(\frac{1}{\log n}  \Bigr).
\]
Provided $\theta \in K_{\delta_n}$,
$\hat{\theta} \in B(\theta,\delta_n/2)$
implies 
$\hat{\theta} \in K$ and
$\hat{\theta} \in B_\delta(\theta)$ for sufficiently large $n$,
that is, $x^n \in \mathscr{B}_\theta$.
Then, we have
\[
\min_{\theta \in K_{\delta_n}}P_{\theta}( \mathscr{B}_\theta )
\ge 1- \frac{4c}{b^2\log n} -o\Bigl(\frac{1}{\log n}  \Bigr).
\]
Hence we have
\begin{align*}
& \int_K P_\theta(\mathscr{B}_\theta )w_K(\theta)d\theta \\
& \quad \ge 
\int_{K_{\delta_n}} P_\theta(\mathscr{B}_\theta )w_K(\theta)d\theta\\
& \quad \ge 
\frac{C_J(K_{\delta_n})}{C_J(K)}
\Bigl(1- \frac{4c}{b^2\log n} -o\Bigl(\frac{1}{\log n}  \Bigr)\Bigr).
\end{align*}
Here, by Assumption~4$^\prime$,
${C_J(K_{\delta_n})/C_J(K)} = 1- o(1/\log n)$ holds.
Hence, we have
\begin{align*}
\int_K P_\theta(\mathscr{B}_\theta )w_K(\theta)d\theta
\ge 
1- \frac{4c}{b^2\log n} -o\Bigl(\frac{1}{\log n}  \Bigr).
\end{align*}
Together with (\ref{eq:cbarnlower}) and (\ref{eq:Nlower}), we have
\[
\bar{C}_n
\ge
1- \frac{4c}{b^2\log n} -o\Bigl(\frac{1}{\log n}  \Bigr).
\]
Noting that $b$ can be arbitrarily large, we have
$\bar{C}_n \ge 1-o(1/\log n)$.

Next, we will examine $C_n$, for which we 
have
\begin{align}
  C_n  = & \int_{\mathcal{K}}\int_{K} 1_{\mathcal{K}}(x^n)
  p(x^n|\theta)w_K(\theta)d\theta \\
 = & \int_{K} P_\theta(\hat{\theta} \in K )w_K(\theta)d\theta.
\end{align}
By the same way as for $\bar{C}_n$,
we can prove this is $1-o(1/\log n )$.
{\em The proof is completed.}

\subsection{Lower Bounds for Stochastic Processes} 

Here, we consider the case in which
the model consists of
non i.i.d.\ stochastic processes.

Let $S=\{p(\cdot|\theta): \theta \in \Theta\}$ be
a parameterized family of
joint densities of a
stochastic process, i.e.\
$p(x^n|\theta)$
satisfies
\[
p(x^n|\theta)=\int_\al p(x^{n+1}|\theta)\nu(dx_{n+1})
\]
and
\[
\int_\al p(x_1|\theta)\nu(dx_1)=1.
\]
We introduce the following notation
\[
J_{n,\theta}
=
J_n(\theta) \deff E_\theta \hat{J}(\theta,x^n).
\]
In this subsection, 
define the Jeffreys prior for each $n$ over $K$ as
\begin{align*}
w^{(n)}_{K}(\theta) & = \frac{|J_{n,\theta}|^{1/2}}{C_{J,n}(K)}
\end{align*}
where $C_{J,n}(K)$ is the Jeffreys integral
\begin{align*} 
C_{J,n}(K) & = \int_K |J_{n,\theta}|^{1/2}d\theta
\end{align*}
based on the Fisher information indexed by $n$.

With the use of $J_{n,\theta}$ we modify the definition of $B_\delta(\theta) $ 
as
\[
B_{\delta}(\theta) = B_{n,\delta}(\theta)
=
\{\theta'\in \Theta : 
(\theta' - \theta)^t J_{n,\theta}(\theta' - \theta) \leq \delta^2 \}.
\]

We use the following assumptions.

%
\setcounter{forassumption}{\theassumption}
\renewcommand{\theassumption}{\arabic{assumption}$^{\prime\prime}$}
\setcounter{assumption}{0}

\begin{assumption}\label{assume:st:-1}
For each $i$, $j$, for each $\theta \in K$, and for each $\tilde \epsilon > 0$, there is a $\delta = \delta(\theta,\tilde \epsilon) > 0$, such that
\begin{align}\label{eq:assumption_for_sp}
\sup_{\theta' \in B_{\delta}(\theta) } 
\bigl( \pm (
J_{n,ij}(\theta)
-
\hat{J}_{ij}(\theta',x^n)
) \bigr)
\leq \tilde \epsilon \:\:\: 
\end{align}
holds, with $P_\theta$ probability converging to $1$ as $n\rightarrow \infty$.
\end{assumption}

To define $B_\delta(\theta)$
properly, the smallest eigenvalue of $J_n(\theta)$
must be lower bounded by a positive constant.
Hence we use the following assumption,
which is used in the upper bound results, too.

\begin{assumption}\label{assume:st:-0}
The matrix
$
J_n(\theta)
$
is positive definite,
and
the collection of functions $J_n(\theta)$, $n \in \mathbb{N}$
is equicontinuous for $\theta \in \Theta$, with a continuous $\lambda(\theta)>0$ serving as a lower bound on its smallest eigenvalue, uniformly for $n \in \mathbb{N}$.
\end{assumption}

\renewcommand{\theassumption}{\arabic{assumption}}
\setcounter{assumption}{\theforassumption}

It follows from these two assumption that the assertion of Assumption $1^{\prime\prime}$ can be refined.  Indeed, it follows that for for each $\theta$ in $K$ and for any $\epsilon>0$, there is an $\tilde \epsilon = \tilde \epsilon(\epsilon,\theta)>0$ such that at $\delta=\delta(\tilde \epsilon,\theta)$,
$$\sup_{\theta' \in B_{\delta}(\theta) } 
\Bigl( \pm \bigl(J_n(\theta)^{-1/2}\bigl(
J_{n}(\theta)
-
\hat{J}(\theta',x^n)\bigr) J_n(\theta)^{-1/2}
\bigr)_{ij} \Bigr)
\leq \frac{\epsilon}{d} \:\:\: 
$$
with $P_\theta$ probability converging to $1$ as $n\rightarrow \infty$, and consequently,
\begin{align}\label{seriesC_st}
P_\theta\Bigl(
\inf_{z \neq 0}
\inf_{\theta' \in B_{\delta}(\theta) } 
\frac{z^t\hat{J}(\theta',x^n)z}{z^tJ_n(\theta)z}
< 1-\epsilon
\Bigr)
=
o(1)
\end{align}
as $n$ goes to infinity.
This corresponds to (\ref{seriesC})
for the i.i.d.\ case.

From Assumption $2^{\prime\prime}$, we also have
\begin{align}\label{eq:lowerofprocessFisher}
    \min_{n \in \mathbb{N}} \min_{\theta \in K}\min_{z:|z|=1}z^tJ_n(\theta)z > 0.
\end{align}
for compact subsets $K$ of $\Theta$. This is because $\lambda(\theta)$, being positive and continuous, 
is therefore strictly positive on any compact set.

Sufficient condition for Assumptions~$1^{\prime\prime}$ and $2^{\prime\prime}$ 
are given in the appendix for Markov processes.

Define two nice sets as
\begin{align}
\mathscr{B} & =
\{ (\theta,x^n) :
\theta, \hat{\theta}(x^n) \in {K}, 
||\theta - \hat{\theta}(x^n)||_{J_{n,\theta}} \leq \delta/4  \}  \\
\mathscr{N} & =
\Bigl\{
(\theta,x^n) :
\inf_{\theta' \in B_{\delta}(\theta) } 
\inf_{z \neq 0} 
\frac{z^t\hat{J}(\theta',x^n)z}{z^tJ_n(\theta)z}
\geq
1-\epsilon
\Bigr\}.
\end{align}
Further define
\begin{align}\label{Jfirst_st_1}
\eta_{1,J,\delta}
=
\sup_{\theta \in K}
\sup_{\theta' \in B_\delta(\theta)}
\max_{z \neq 0}
\frac{z^tJ_n(\theta)z}
{z^tJ_n(\theta')z}
-1,
\end{align}
\begin{align}\label{Jfirst_st_2}
\eta_{2,J,\delta}
=
\sup_{\theta \in K}
\sup_{\theta' \in B_\delta(\theta)}
\max_{z \neq 0}
\frac{z^tJ_n(\theta')z}
{z^tJ_n(\theta)z}
-1,
\end{align}
and $\eta_{J,\delta}= \max\{\eta_{1,J,\delta},\eta_{2,J,\delta} \}$.
By equicontinuity of $J_n(\theta)$
and compactness of ${K}_\delta$,
this $\eta_{J,\delta}$ converges to $0$
as $\delta$ tends to zero, uniformly for $n \in \cal{N}$.

For $\epsilon > 0$,
let $\delta=\delta_\epsilon > 0$, with $(1+\eta_{J,\delta})^{1/2} < 2$, be sufficiently small that (\ref{seriesC_st})
holds for each $\theta \in {K}$, and
$\delta_\epsilon \rightarrow 0$ as $\epsilon \rightarrow 0$.

As before, when (\ref{eq:assumption_for_sp}) holds we have
\begin{align}
\min_{\theta' \in \mathscr{B}_{x^n}} 
\min_{z \neq 0} 
\frac{z^t\hat{J}(\theta',x^n)z}{z^tJ_n(\theta)z}
\geq
1-\epsilon.
\end{align}

%
\setcounter{forassumption}{\theassumption}
\renewcommand{\theassumption}{\arabic{assumption}$^{\prime\prime}$}
\setcounter{assumption}{2}

\begin{assumption}
For each $\theta$ in $\Theta^\circ$, the maximum likelihood estimator $\hat \theta(x^n)$ is consistent.  Namely, $\lim_n |\hat \theta (x^n) - \theta| = 0$ in $P_\theta$ probability, where $P_\theta$ is the distribution determined by the $x^n$ with densities $p(x^n)$ for $n \ge 1$.
\end{assumption}

\begin{assumption}
The Jeffreys integral $C_{J,n}(K)$ is finite.
\end{assumption}

\renewcommand{\theassumption}{\arabic{assumption}}
\setcounter{assumption}{\theforassumption}

We can prove the following.
\begin{lemma}\label{st:lemma:lower}
Let $S=\{ p(\cdot|\theta) : \theta \in \Theta\}$
be a $d$-dimensional family of stochastic processes.
We suppose that 
Assumptions~$1^{\prime\prime}$, $2^{\prime\prime}$, $3^{\prime\prime}$, and $4^{\prime\prime}$
hold for $S$.
Define a density $m_{\delta}$ supported on the set
\[
\{ x^n : (\mathscr{N} \cap \mathscr{B})_{x^n} \neq \emptyset \}
\subset {\mathcal{K}},
\]
by
\begin{align*}
 \bar{m}_{\delta}(x^n)
&= 
\int_{(\mathscr{N}\cap\mathscr{B})_{x^n}}p(x^n|\theta)
w^{(n)}_{{K}}(\theta)d\theta,\\
m_{\delta}(x^n) 
&=
\frac{\bar{m}_{\delta}(x^n) }{\bar{C}_{n}},
\end{align*}
where
{
\begin{align}\label{def_of_cbar_n:2}
\bar{C}_{n}
= 
\int_{{\mathcal{K}}}
\bar{m}_{\delta}(x^n)\nu(dx^n).
\end{align}}
Then
\[
E_{m_{\delta}}
\log\frac{p(x^n|\hat{\theta})}{m_{\delta}(x^n)}
\geq
\log  
\frac{(1-\epsilon)^{d/2}C_{J,n}({K})n^{d/2}}{(1+\eta_{J,\delta})(2\pi)^{d/2}}
-o(1)
\]
holds as $n$ goes to infinity.
\end{lemma}

{\em Proof:} 
Similar to the proof of
Lemma~\ref{gen:lemma:lower}.

The following result, under the stated assumptions in the non i.i.d. setting, states that $\log C_{J,n}(K)$ provides control on the minimax regret from below. It corresponds to Theorem 1 in the i.i.d. setting.

\begin{theorem}\label{st:thm:lower}
Let $S=\{ p(\cdot|\theta) : \theta \in \Theta\}$
be a $d$-dimensional family of stochastic processes.
Suppose that 
Assumptions~$1^{\prime\prime}$, $2^{\prime\prime}$, $3^{\prime\prime}$, and $4^{\prime\prime}$
hold for $S$.
Let $K$ be an arbitrary measurable subset of $\Theta$.
Then the following holds
\begin{eqnarray*}
\liminf_{n \rightarrow \infty}
\Bigl(\ubar{r}_n(\mathcal{K})
-\frac{d}{2}\log\frac{n}{2\pi}
-\log C_{J,n}(K)
\Bigr)
\geq 0.
\end{eqnarray*}
\end{theorem}

{\em Proof:} 
The conclusion is a consequence of 
Lemma~\ref{st:lemma:lower} and is shown in the same fashion as Theorem 1.

\section{Upper Bounds}

\label{section:generalization:upperbounds}

For the upper bound 
we handle 
both i.i.d.\ and stochastic process cases
at once.
Similarly as for the lower bound results,
we use a compact subset $K$ of $\Theta^\circ$.
In this section,
$K$ denotes a convex set which is the closure of an open set.

To make the new modification of the Jeffreys mixture
for the boundary issue (Definition~\ref{def_of_ideal_prior})
to work well, 
we need to strengthen Assumption~\ref{assume:st:-0}, 
which is a certain equi-continuity of Fisher information.
In fact, we employ the following stronger assumption than
Assumption~\ref{assume:st:-0}.
%
\setcounter{forassumption}{\theassumption}
\renewcommand{\theassumption}{\arabic{assumption}$^{\prime\prime\prime}$}
\setcounter{assumption}{1}
\begin{assumption}\label{assume:st:-upper2}
The matrix
$
J_n(\theta)
$
is positive definite,
and
the collection of functions $J_{n,ij}(\theta)$ 
is Lipschitz continuous in $K$,
uniformly for all $n \in \mathbb{N}$
and for all $i,j$,
\[
|J_{n,ij}(\theta')-J_{n,ij}(\theta)| \le M|\theta' - \theta|
\]
with a continuous $\lambda(\theta)>0$ serving as a lower bound on its smallest eigenvalue, uniformly for $n \in \mathbb{N}$.
\end{assumption}
\renewcommand{\theassumption}{\arabic{assumption}}
\setcounter{assumption}{\theforassumption}

The main theorem gives
the upper bound which matches the lower bound we
have obtained.

To resolve the issue due to the discrepancy between the Fisher information
and the empirical Fisher information, 
we will construct an enlarged model
of probability densities from the model $S$.

Define a matrix-valued random variable $V \in \mathbb{R}^{d\times d}$ as
\[
V_{n}(\theta)
=
V(x^n|\theta)
= 
J_{n,\theta}^{-1/2}\hat{J}(\theta,x^n)J_{n,\theta}^{-1/2} -I.
\]
We also use the notation $V_{ij,n}(\theta)$ to denote the $(i,j)$-entry of $V_n(\theta)$.
Let $\mathcal{B}=(-b/2,b/2)^{d \times d}$ for some small $b > 0 $.
We consider the case where the following assumption holds.
\begin{assumption}\label{assume:process:upper:0}
For every compact set $K$ included in $\Theta^\circ$,
there exists a $b > 0$ and a $C_1 = C_{1,K} > 0$, such that the following holds.
\begin{align}\nonumber
& \forall n \in  \mathbb{N},  \:\:
\forall \theta \in K,\:\:
\forall \beta \in \mathcal{B},   \\ \label{sup11}
&
\Bigl(
\int
p(x^n|\theta)\exp(n V_{n}(\theta)\cdot \beta  ) \nu_n(dx^n)
\Bigr)^{1/n} < C_1,
\end{align}
where
$V(x^n|\theta)\cdot \beta$ denotes the matrix inner product
of Frobenius
\[
\mathrm{trace}(
V_n(\theta)\beta^t)=\sum_{ij} V_{ij,n}(\theta)\beta_{ij}.
\]
\end{assumption}

Define a function $\psi_n(\theta,\beta)$ as
\[
\psi_n(\theta,\beta)
=
\frac{1}{n}\log \!
\int
p(x^n|\theta)\exp(n V_n(\theta) \cdot \beta  ) \nu_n(dx^n).
\]
Note that Assumption~\ref{assume:process:upper:0} is the boundedness of $\psi_n(\theta,\beta)$.

We define the enlarged family $\bar{S}$ by
\begin{align*}
\bar{S} 
&  =  \{ p_e(x^n|\theta,\beta)
:
\theta \in \Theta, \beta \in \mathcal{B} \},\\
p_e(x^n|\theta,\beta)
&  = 
p(x^n|\theta)
\exp\Bigl( n (V_n(\theta) \cdot \beta  -\psi_n(\theta,\beta))   \Bigr).
\end{align*}
Let $u=(\theta,\beta)$
and ${\cal U}=\{u=(\theta,\beta): \theta 
\in \Theta, \beta \in \mathcal{B}  \} $.
We have
\begin{eqnarray*}
\frac{\partial \psi_n(\theta,\beta)}{ \partial \beta_{ij}}
=
E_{\theta,\beta} V_{ij,n}(\theta)
\end{eqnarray*}
and
\begin{align*}
&\frac{\partial^2 \psi_n(\theta,\beta) }{ \partial
  \beta_{kl} \partial\beta_{ij}} \\
 = &
n\Bigl(
E_{\theta,\beta} V_{kl,n}(\theta)V_{ij,n}(\theta)
-
E_{\theta,\beta}V_{kl,n}(\theta)
E_{\theta,\beta}V_{ij,n}(\theta)\Bigr),
\end{align*}
where $E_{\theta,\beta}$ denotes expectation with respect to
$p_e (x^n|\theta,\beta)$.
The latter is the covariance of
$\sqrt{n}V_{ij,n}(\theta)$ and $\sqrt{n}V_{kl,n}(\theta)$ at $p_e(x^n|\theta,\beta)$.
Correspondingly, let $\cov_n(\theta,\beta)$ denote the matrix whose
$(ij,kl)$-entry is
\[
[\cov_n(\theta,\beta)]_{ij,kl}
=
\frac
{\partial^2 \psi_n(\theta,\beta)}{\partial \beta_{ij} \beta_{kl}}.
\]
We let $\lambda_n^*$ denote the maximum
of the largest eigenvalue of
$\cov_n(\theta,\beta)$ among $(\theta,\beta) \in \Theta \times \mathcal{B}$.

When the model is i.i.d.,
the enlarged family can be single-letterized as
\begin{align*}
\bar{S} 
&  =  \{ p_e(x|\theta,\beta)
:
\theta \in \Theta, \beta \in \mathcal{B} \},\\
p_e(x|\theta,\beta)
&  = 
p(x|\theta)
\exp\Bigl( V_1(\theta) \cdot \beta  -\psi(\theta,\beta)   \Bigr),
\end{align*}
where
(\ref{sup11}) is reduced to
\begin{align}\nonumber
\int
p(x|\theta)\exp( V_1(\theta)\cdot \beta  ) \nu(dx)
 < C_1,
\end{align}
and
\[
\psi(\theta,\beta)
=
\log \!
\int
p(x|\theta)\exp( V_1(\theta) \cdot \beta  ) \nu(dx).
\]
Note that the enlarged family $p_e(x^n|\theta,\beta)$ is still i.i.d.,
that is,
$p_e(x^n|\theta,\beta)=\prod_{t=1}^n p_e(x_t|\theta,\beta)$.
Further we have
\begin{align*}
p_e(x^n|\theta,\beta)
=
p(x^n|\theta)
\exp\Bigl(n ( V_n(\theta) \cdot \beta  -\psi(\theta,\beta) )  \Bigr).
\end{align*}
where the following holds.
\[
V(x^n|\theta)=\frac{1}{n}\sum_{t=1}^{n} V(x_t|\theta).
\]
Then,
$\partial^2 \psi(\theta,\beta) / \partial \beta_{kl} \partial \beta_{ij}$
is the covariance of
$V_{ij}(x|\theta)$ 
and
$V_{kl}(x|\theta)$ 
at $p_e(x|\theta,\beta)$,
and we have
\[
[\cov_n(\theta,\beta)]_{ij,kl}
=
\frac{\partial^2 \psi(\theta,\beta) }{ \partial \beta_{kl} \partial \beta_{ij}}.
\]
In the i.i.d.\ case,
$[\cov_n(\theta,\beta)]_{ij,kl} = \partial^2 \psi(\theta,\beta) / \partial \beta_{kl} \partial \beta_{ij}$
dose not depend on $n$, $\lambda^*_n$ also does not depend on $n$.
Hence, we use the symbols $\cov(\theta,\beta)$ and $\lambda^*$ for i.i.d.\ cases.

For a given compact  $K$ included in $\Theta^\circ$,
define subsets $G_{n,\delta}$ 
and $G_{n,\delta}^c$ of $\mathcal{K}$ as follows.
\begin{align*}
G_{n,\delta} &=  \Bigl\{ x^n : 
\hat{\theta} \in K , \:
||V(x^n|\hat{\theta})||_s \leq \delta
\Bigr\},\\
G_{n,\delta}^c &= \Bigl\{ x^n : 
\hat{\theta} \in K, \:
||V(x^n|\hat{\theta})||_s > \delta
\Bigr\},
\end{align*}
where $||A||_s$ for a symmetric matrix $A \in \Re^{d \times d}$ denotes the 
spectral norm of $A$ defined as 
\[
||A||_s = \max_{z: |z|=1} ||Az|| = \max_{z: |z|=1} |z^t Az|.
\]
Note the following relation between the spectral norm $||A||_s$ and the Frobenius norm $||A||$ (which is the Euclidean norm of the associated vector with $d^2$ elements).
\[
\frac{||A||}{\sqrt{d}} \leq ||A||_s \leq  ||A||.
\]
We call elements of $G_{n,\delta}$ and $G_{n,\delta}^c$ 
good strings and not good strings, respectively.


We also define 
the two kinds of neighborhood of $\theta$ 
as
\begin{align}
B_{\epsilon}(\theta)
&=
\{\theta' : (\theta'-\theta)^t J_{n,\theta}(\theta'-\theta) \leq \epsilon^2 \},\\
\hat{B}_{\epsilon}(\theta)
&=
\{\theta' : (\theta'-\theta)^t \hat{J}(\hat{\theta})(\theta'-\theta) \leq \epsilon^2 \},
\end{align}
where the latter depends on $x^n$. 

We further put the following assumptions.
\begin{assumption}\label{assume:process:upper:prior}
Let $w_n(\theta)$ be the prior density function we use
for $p(x^n|\theta)$.
Assume the following equi-semicontinuity, that is,
\[
w_n(\theta') \geq (1-\eta_{r})w_n(\theta)
\]
holds 
for all large $n$, 
for all small $r > 0$,
for all $\theta \in K$, and for all $\theta'$
in $B_{r}(\theta)$, 
where
$\eta_{r}=\eta(r)$ 
is a positive valued function defined on $(0,\infty)$ with
$\lim_{r \rightarrow 0}\eta_{r}=0$.
\end{assumption}



{

\begin{assumption}\label{assume:process:upper:2:for_all}
There exist a positive number $\bar{h}_K \ge 1$ 
and a small positive number $\epsilon = \epsilon_K$
such that
the following inequality holds
for all large $n$, for all $x^n \in {\mathcal K}$,
for all $\tilde{\theta} \in B_\epsilon(\hat{\theta}) \cap K$,
and for all $z$.
\begin{align}\label{assume:process:upper:2:ineq1:for_all}
z^t\hat{J}(\tilde{\theta},x^n)z
\le
\bar{h}_K
z^t\hat{J}(\hat{\theta},x^n)z.
\end{align}
\end{assumption}

This assumption does hold for the mixture family with compact $K$ 
$\subset \Theta^\circ$.
However, there are other families for which we cannot show it holds.
To treat those cases too, we use the following weakened assumption.
In fact, this assumption is sufficient for our tasks.

\renewcommand{\theassumption}{\arabic{assumption}$^\prime$}
\setcounter{assumption}{6}

\begin{assumption}\label{assume:process:upper:2:for_all:dash}
There exist a positive number $\bar{h}_K \ge 1$ 
and a small positive number $\epsilon = \epsilon_K$
such that
the following inequality holds
for all large $n$, for all $x^n \in {\mathcal K}$,
for all $\tilde{\theta} \in B_\epsilon(\hat{\theta}) \cap K$,
and for all $z$.
\begin{align}\label{assume:process:upper:2:ineq1:for_all_2}
z^t\hat{J}(\tilde{\theta},x^n)z
\le
\bar{h}_K
\max \{
z^t\hat{J}(\hat{\theta},x^n)z,
z^tJ_{n,\hat{\theta}}z
 \}.
\end{align}
\end{assumption}
\renewcommand{\theassumption}{\arabic{assumption}}

\begin{assumption}\label{assume:process:upper:2:for_good}
There exist a number $\kappa_J$ $=\kappa_J(K) > 0$ 
and a small number $\delta_0 = \delta_0(K) > 0$ 
such that
the following inequality holds
for all large $n$, for all $x^n \in G_{n,\delta_0}$,
for all small $\epsilon > 0$,
for all $\tilde{\theta} \in B_\epsilon(\hat{\theta}) \cap K$,
and for all $z$.
\begin{align}\label{assume:process:upper:2:ineq1:for_good}
z^t\hat{J}(\tilde{\theta},x^n)z
\le
(1+\kappa_J \epsilon)
z^t\hat{J}(\hat{\theta},x^n)z.
\end{align}
\end{assumption}


\begin{assumption}\label{assume:process:upper:3}
There exists an $\epsilon > 0$ and a $\zeta \in (0,1)$, such that,
\[
{\forall n \in \mathbb{N}},\;
{\forall x^n \in \mathcal{K}}, \;
\inf_{\tilde{\theta} \in \hat{B}_\epsilon(\hat{\theta})\cap K}
||V(x^n|\tilde{\theta})||_s
\geq
\zeta
||V(x^n|\hat{\theta})||_s.
\]
\end{assumption}

Assumptions~\ref{assume:process:upper:2:for_all:dash}, \ref{assume:process:upper:2:for_all},
and \ref{assume:process:upper:2:for_good}
are used to control the Laplace approximation.
In particular, Assumptions~\ref{assume:process:upper:2:for_all:dash} and \ref{assume:process:upper:2:for_all}
guarantee that the regret of the Jeffreys mixtures is finite for all the strings,
while Assumption~\ref{assume:process:upper:2:for_good} guarantees that
the regret for the good strings almost equals the minimax value. 
As for Assumption~\ref{assume:process:upper:2:for_good},
it is used to control the regret of the mixture of the extended model
(the fibre bundle of local exponential families) on the not good strings.

If the collection of empirical Fisher information matrices
$\hat{J}(\theta)=\hat{J}(\theta,x^n)$ for all $x^n$ and all $n$,
is equicontinuous as functions of $\theta$ in $K$, 
then Assumptions~\ref{assume:process:upper:2:for_all:dash}, 
\ref{assume:process:upper:2:for_good}, 
and
\ref{assume:process:upper:3}
hold.
For Assumption~\ref{assume:process:upper:2:for_all}
, if also the minimum of the smallest eigenvalue of $\hat{J}_n(\hat{\theta})$ 
for all $n$ and for all $x^n \in \mathcal{K}$
is lower bounded 
by a positive constant, 
then it holds.

The following is used to control the behavior of the ideal prior
for stochastic processes. Note that it holds automatically for i.i.d.\ cases.
\begin{assumption}\label{assumption:upper:process:Fisher}
For each $\theta \in K$, 
$
\sup_{n \in \mathbb{N}}||J_n(\theta)||_s
$ is finite.
\end{assumption}

Under this assumption 
and Assumption~\ref{assume:st:-upper2}, we
can show that $
\sup_{n \in \mathbb{N}}\sup_{\theta \in K}||J_n(\theta)||_s
$ is finite, similarly as (\ref{eq:lowerofprocessFisher}).

Hereafter we use the symbol $\Phi$ to denote the probability measure
of the $d$-dimensional standard normal distribution.

\begin{lemma}\label{newlemma1}
Let $m_{w_n}$ be the mixture of $S$ with respect to a prior density $w_n(\theta)$
over a convex and compact set $K$ included in $\Theta^\circ$. 
Suppose that Assumptions~
\ref{assume:st:-upper2}, \ref{assume:process:upper:prior}, 
and \ref{assume:process:upper:2:for_good} hold.
Then, for an arbitrary $\epsilon > 0$,
and for all $x^n \in G_{n,\delta}$ with $\delta \le \delta_0$, 
the following holds
\begin{align*}
\log \frac{p(x^n|\hat{\theta})}{m_{w_n}(x^n)}
&\leq
\frac{d}{2}\log\frac{n}{2\pi} 
+\log
\frac{|J_n(\hat{\theta})|^{1/2}}{w_n(\hat{\theta})
\Phi(U_{K,\epsilon,n}(\hat{\theta}))}\\
&+\log\frac{(1+\kappa_J \epsilon)^{d/2}(1+\delta)^{d/2}}{1-\eta_{\epsilon}},
\end{align*}
where 
\[
U_{K,\epsilon,n}(\theta)= 
N_{\sqrt{n}\epsilon}(0) \cap (n J_n(\theta))^{1/2}(K-\theta),
\]
and
\[
N_r(0)=\{z : |z| \leq r \}.
\]
\end{lemma}

  

Ignoring the restriction to $N_{\sqrt{n}\epsilon}(0)$ 
the 
$\Phi(U_{K,\epsilon,n}(\hat{\theta}))$ 
may be interpreted as the probability that a Gaussian variable
of mean $\hat{\theta}$ and covariance
$(nJ_n(\hat{\theta}))^{-1}$
belongs to the parameter set $K$. When $\hat{\theta}$
is sufficiently in the interior, 
this $\Phi(U_{K,\epsilon,n}(\hat{\theta}))$ 
is near $1$ for large $n$.
Whereas, for $\hat{\theta}$ at the boundary
or within order $n^{-1/2}$ of the boundary, it provides the
modification for the classical Laplace approximation to account for
the fact that the near boundary case makes the normal integral
approximation be an incomplete normal integral.

{\em Proof of Lemma~\ref{newlemma1}:}
By a Laplace integration for the mixture $m_w$
restricted in $B_\epsilon(\hat{\theta})$, we have
\begin{align*}
  \frac{m_{w_n}(x^n)}{p(x^n|\hat{\theta})}
&\geq
\int_{B_\epsilon(\hat{\theta})\cap K}
e^{-n(\theta-\hat{\theta})^t\hat{J}(\tilde{\theta})
(\theta-\hat{\theta})/2}w_n(\theta)d\theta \\
&\geq
(1-\eta_{\epsilon})w_n(\hat{\theta}) \\
& \cdot \int_{B_\epsilon(\hat{\theta})\cap K}
e^{-n(1+\kappa_J \epsilon)(\theta-\hat{\theta})^t\hat{J}(\hat{\theta})
(\theta-\hat{\theta}))/2} d\theta,
\end{align*}
by Assumption~\ref{assume:process:upper:prior}.
Since $x^n \in G_{n,\delta}$, by Assumption~\ref{assume:process:upper:2:for_good},
\begin{align*}
\frac{m_{w_n}(x^n)}{p(x^n|\hat{\theta})}
&\geq
(1-\eta_{\epsilon})w_n(\hat{\theta}) \\
& \cdot \int_{B_\epsilon(\hat{\theta})\cap K}
e^{-n(1+\kappa_J \epsilon)(1+\delta)(\theta-\hat{\theta})^tJ_n(\hat{\theta})
(\theta-\hat{\theta}))/2} d\theta .
\end{align*}
Noting
\[
(n J_n(\hat{\theta}))^{1/2}
(B_\epsilon(\hat{\theta})-\hat{\theta}) = N_{\sqrt{n}\epsilon}(0),
\]
by change of variables 
\[
z=\bigl((1+\kappa_J\epsilon)(1+\delta)n
J_n(\hat{\theta})\bigr)^{1/2}(\theta-\hat{\theta}),
\]
the integration in the last line is transformed as 
\begin{align*}
&\int_{B_\epsilon(\hat{\theta})\cap K}
e^{-n(1+\kappa_J \epsilon)(1+\delta)(\theta-\hat{\theta})^tJ_n(\hat{\theta})
(\theta-\hat{\theta}))/2} d\theta\\
=&  
\frac{\int_{(1+\kappa_J \epsilon)^{1/2}(1+\delta)^{1/2}U_{K,\epsilon,n}(\theta)}e^{-|z|^2/2} dz}
{(1+\kappa_J \epsilon)^{d/2}(1+\delta)^{d/2}n ^{d/2}|J_n(\hat{\theta})|^{1/2}},
\end{align*}
which is larger than
\begin{align*}
\frac{(2\pi)^{d/2}
\Phi(U_{K,\epsilon,n}(\theta))}{
(1+\kappa_J \epsilon)^{d/2}(1+\delta)^{d/2}
n^{d/2}
|J_n(\hat{\theta})|^{1/2}},
\end{align*}
since
$U_{K,\epsilon,n}(\theta)$ is included in
\begin{align*}
(1+\kappa_J \epsilon)^{1/2}(1+\delta)^{1/2}
\bigl(N_{\sqrt{n}\epsilon}(0) \cap (n J_n(\theta))^{1/2}(K-\theta) \bigr)
\end{align*}
because $K$ is a convex set. 
Hence we have
\[
\frac{m_{w_n}(x^n)}{p(x^n|\hat{\theta})}
\geq
\frac{(1-\eta_{\epsilon})w_n(\hat{\theta}) (2\pi)^{d/2}
\Phi(U_{K,\epsilon,n}(\theta))}
{
(1+\kappa_J \epsilon)^{d/2}(1+\delta)^{d/2}n^{d/2}|J_n(\hat{\theta})|^{1/2}
},
\]
which yields  the claim of the lemma.
{\em This completes the proof of
Lemma~\ref{newlemma1}.}

Lemma~\ref{newlemma1} suggests the following choice of a prior density:
\[
w_{K,\epsilon,n}(\theta)
=
\frac{|J_n(\theta)|^{1/2}}{C_{K,\epsilon,n} \Phi(U_{K,\epsilon,n}(\theta))},
\]
where
$
C_{K,\epsilon,n}
$ is the normalization constant.
This idea is that the prior proportional to 
$|J_n(\theta)|^{1/2}/\Phi(U_{K,\epsilon,n}(\theta))$
provides the approximate modification to Jeffreys prior to equalize
the regret in accordance with Laplace approximation, not only in the
interior, but also for $\hat{\theta}$ near the boundary of $K$.

However, unfortunately this prior $w_{K,\epsilon,n}$
does not satisfy Assumption~\ref{assume:process:upper:prior}, as seen below.
Recall that the assumption requires that the prior density $w_n(\theta')$ keeps its value 
over the range $B_\epsilon(\theta)$.
We can make a counter example as follows.
Assume that $\theta$ is at the boundary of $K$
and that $\Phi(N_{\sqrt{n}\epsilon}(0))$ almost equals $1$.
Then, $\Phi(U_{K,\epsilon,n}(\theta))$ is smaller than a half,
which means that $w_{K,\epsilon,n}(\theta)$ is larger than twice the value of Jeffreys prior.
Note that there exists $\theta'$ in $B_\epsilon(\theta)$
such that $B_\epsilon(\theta') \subset K$, provided $\epsilon$ is sufficiently small.
For that $\theta'$,  
$\Phi(U_{K,\epsilon,n}(\theta'))$ almost equals $1$.
Hence, $w_{K,\epsilon,n}(\theta') \lessapprox w_{K,\epsilon,n}(\theta)/2$ holds.

To overcome this problem, we employ the slightly modified
 prior than $w_{K,\epsilon,n}$, which we call the ideal priors
for the good strings.
To this end, we define a modified region of $U_{K,\epsilon,n}(\theta)$ as below.
\begin{align*}
U^{(\alpha)}_{K, \epsilon,n}(\theta)
=    N_{\sqrt{n}\epsilon}(0) \cap ((n/\alpha^2) J_n(\theta))^{1/2}(K-\theta)
\end{align*}
where $\alpha = \alpha_n$ is a positive real diverging to infinity.

{\it Remark:} 
Ignoring the restriction to $N_{\sqrt{n}\epsilon}(0)$ 
the 
$\Phi(U_{K,\epsilon,n}(\hat{\theta}))$ 
may be interpreted as the probability that a Gaussian variable
of mean $\hat{\theta}$ and covariance
$\alpha^2 (nJ_n(\hat{\theta}))^{-1}$
belongs to the parameter set $K$.

Note that
\[
U^{(\alpha)}_{K,\epsilon,n}(\theta)
\subset U_{K,\epsilon,n}(\theta)
\]
holds for $\alpha \ge 1$, since $K$ is a convex set,
which means
\begin{align}\label{eq:alphamodification}
   \Phi( U^{(\alpha)}_{K,\epsilon,n}(\theta))
\le 
\Phi(
U_{K,\epsilon,n}(\theta)).
\end{align}

\begin{definition}[The ideal prior for the good strings]
\label{def_of_ideal_prior}
Define the prior density $w^{(\alpha)}_{K,\epsilon,n}$ over $K$ as 
\[
w^{(\alpha)}_{K,\epsilon,n}(\theta)
=
\frac{|J_n(\theta)|^{1/2}}{C_{K,\epsilon,n}^{(\alpha)} \Phi(U^{(\alpha)}_{K, \epsilon,n}(\theta))},
\]
where
\[
C^{(\alpha)}_{K,\epsilon,n}
=
\int_K
\frac{|J_n(\theta)|^{1/2}}{\Phi(U^{(\alpha)}_{K,\epsilon,n}(\theta))}d\theta,
\]
where we assume 
$n \epsilon^2 \rightarrow \infty$,
$\alpha/\sqrt{n}\epsilon \rightarrow \infty$ 
and $\epsilon \alpha \rightarrow 0$ as $n$ goes to infinity.
\end{definition}

{\it Remark:}
As for setting of $\epsilon$ and $\alpha$,
we may let $\epsilon=\sqrt{\log n/n}$ and $\alpha = \log n$ for example.


For this prior we can show that 
Assumption~\ref{assume:process:upper:prior} holds
under Assumption~\ref{assume:st:-upper2}
and that
$C_{J,n}(K)/C_{K,\epsilon,n}^{(\alpha)} \rightarrow 1$ as $n$ goes to infinity.
First, we will show the latter.

Note that the following holds.
\begin{proposition}\label{factor}
\input{proposition1}
\end{proposition}

See Appendix~\ref{proofoffactorporoposition} for the proof.

Using this proposition, we have the following lemma, where
\begin{align*}
\ratio^{(\alpha)}_n(\epsilon,\theta)
&=
\Phi(U^{(\alpha)}_{K,\epsilon,n}(\theta)) 
\end{align*}
and
\begin{align*}
\ratio^{(\alpha)}_n(\epsilon)
=
\inf_{\theta \in K}\rho^{(\alpha)}_n(\epsilon,\theta).
\end{align*}

\setcounter{forlemmafornormalization}{\thelemma}
\begin{lemma}\label{newlemma12-0}
\input{lemma_for_normalization}
\end{lemma}

The proof is in Appendix~\ref{appendix:normalizaionconstant}.

We have a lower bound 
on $\ratio^{(\alpha)}_n(\epsilon)$ as follows,
where ${\rm vol}(A)$ denotes the volume of $A \in \mathbb{R}^d$,
${\rm diam}(A)$ the diameter of $A \in \mathbb{R}^d$,
and
$V_k$ the volume of the $k$-dimensional sphere of radius $1$,
that is, 
\[
V_k =
\frac{\pi^{k/2}}{\Gamma(k/2+1)}.
\]
\setcounter{forlemmaratio}{\thelemma}
\begin{lemma}\label{lemma:for_ratio}
\input{lemma_for_ratio}
\end{lemma}

Since Assumptions~\ref{assume:st:-upper2}
and \ref{assumption:upper:process:Fisher},
the minimum eigenvalue of $J_{n,\theta}$ has a positive
lower bound and the maximum eigenvalue has an upper bound,
which are uniform for $\theta \in K$.
Hence, by Lemma~\ref{lemma:for_ratio},
$\ratio^{(\alpha)}_n(\epsilon)$ is lower bounded 
by a positive constant when $\sqrt{n}\epsilon/\alpha$ is not small.

Noting this fact, we can show the following lemma.
\begin{lemma}\label{newlemma12-1}
Suppose that Assumptions~\ref{assume:st:-upper2}
and \ref{assumption:upper:process:Fisher} hold.
If $\epsilon/\alpha \rightarrow 0$
and $n\epsilon^2/\alpha^2 \rightarrow \infty$
as $n$ goes to infinity, then
\[
\lim_{n \rightarrow \infty}
\Bigl(
\log {C_{K,\epsilon,n}^{(\alpha)}} -\log {C_{J,n}(K)}\Bigr)
=
0.
\]
\end{lemma}

{\it Proof:}
Since $
\sup_{\theta \in K}
\sup_{n}|J_{n}(\theta)|^{1/2}$ is finite
under
Assumption~\ref{assumption:upper:process:Fisher}, 
$C_{J,n}(K\setminus K_{\epsilon'})$ converges to zero 
uniformly for all $n$,
as
$\epsilon'$ tends to zero.
Noting that $\ratio_n^{(\alpha)}$ is lower bounded by a certain positive constant,
since Lemma~\ref{lemma:for_ratio}.
Hence by Lemma~\ref{newlemma12-0}, which is shown under Assumption~\ref{assume:st:-upper2},
we have the claim of the lemma.
{\it This completes the proof.}

To show that Assumption~\ref{assume:process:upper:prior}
holds for the ideal priors, we will show the following lemma.
\setcounter{forlemmaideal}{\thelemma}
\begin{lemma}\label{lemma:idealpriorfactor}
\input{lemma_for_ideal_prior}
\end{lemma}

The proof is in Appendix~\ref{appendix_continuity_of_factor}.

Note that $\ratio_n^{(\alpha)}(\epsilon,\theta)$ 
is lower bounded by a positive constant,
since $\sqrt{n}\epsilon \rightarrow \infty$ when
$n$ goes to infinity
because of Lemma~\ref{lemma:for_ratio}.
Since $g_{\theta,r}$ is of order $r$ uniformly over $K$ by
Assumption~\ref{assume:st:-upper2}, Lemma~\ref{lemma:idealpriorfactor} implies
\begin{align}
\inf_{\theta \in K}
\inf_{\theta' \in K \cap B_r(\theta)}   
\frac
{\Phi(U^{(\alpha)}_{K,\epsilon,n}(\theta)}
{\Phi(U^{(\alpha)}_{K,\epsilon,n}(\theta'))}
\ge 
1 - O\Bigl( \frac{\sqrt{n}\epsilon}{\alpha} \Bigr),
\end{align}
when $\sqrt{n}\epsilon/\alpha$ tends to $0$.
From this, we have
\[
\inf_{\theta \in K}
\inf_{\theta' \in K \cap  B_r(\theta)} 
\frac{w_{K,\epsilon,n}^{(\alpha)}(\theta')}{w_{K,\epsilon,n}^{(\alpha)}(\theta)}
\ge 1 - O\Bigl( \frac{\sqrt{n}\epsilon}{\alpha} \Bigr),
\]
Since we assume $\sqrt{n}\epsilon/\alpha$ tends to $0$ as $n$ goes to infinity,
this means that $w_{K,\epsilon,n}^{(\alpha)}$ satisfies
Assumption~\ref{assume:process:upper:prior}.
Hence, 
we can immediately show the following lemma from
Lemma~\ref{newlemma1} and (\ref{eq:alphamodification}).
In fact, for the ideal priors, Assumption~\ref{assume:process:upper:prior} 
holds under Assumption~\ref{assume:st:-upper2} on Fisher information.
Therefore, we have the following lemma.
\begin{lemma}\label{newlemma2}
Let $m_{K,\epsilon,n}^{(\alpha)}$ denote the mixture of $S$ with respect to the
prior density $w_{K,\epsilon,n}^{(\alpha)}$.
Suppose that Assumptions~\ref{assume:st:-upper2}, 
\ref{assume:process:upper:prior}, 
\ref{assume:process:upper:2:for_good}, 
and \ref{assumption:upper:process:Fisher} 
hold.
Then, for all $x^n \in G_{n,\delta}$ and for all $\epsilon > 0$,
we have
\begin{eqnarray*}
\log \frac{p(x^n|\hat{\theta})}{m_{K,\epsilon,n}^{(\alpha)}(x^n)}
&\leq&
\frac{d}{2}\log\frac{n}{2\pi} +\log C_{K,\epsilon,n}^{(\alpha)}\\
&+&\log\frac{(1+\kappa_J \epsilon)^{d/2}(1+\delta)^{d/2}}{1-\eta_{\epsilon}}.
\end{eqnarray*}
\end{lemma}



Next, we will consider the strings $x^n \in G_{n,\delta}^c$. 
We use the mixture of the extended model $\bar{S}$.
Define a prior density $\bar{w}(\theta,\beta)$ over $\bar{S}$ as
\begin{eqnarray*}
  \bar{w}(\theta,\beta) = w_{K}^{(n)}(\theta)b^{-d^2}.
\end{eqnarray*}
Note that this is the direct product of 
the Jeffreys prior $w_{K}^{(n)}(\theta)$
and the uniform probability density over $\mathcal{B}$.

Recall that
$\lambda_n^*$ denotes the maximum
of the largest eigenvalue of
$\cov_n(\theta,\beta)$ among 
$(\theta,\beta) \in \Theta \times \mathcal{B}$.
We have the following lemma.
\begin{lemma}\label{newlemma3}
Let 
$\bar{m}$ 
denote the mixture of $\bar{S}$
with respect to the prior $\bar{w}$. Suppose 
Assumptions~\ref{assume:process:upper:prior},
\ref{assume:process:upper:2:for_all:dash}, 
and \ref{assume:process:upper:3} hold.
Set $\tilde{a} = \min \{1/2\lambda_n^*, 1  \}$
and $\tilde{\delta} = \zeta \delta/\sqrt{d}$.
Then for all $x^n \in G_{n,\delta}^c$, 
we have
\begin{align} \nonumber
\frac{\bar{m}(x^n)}{p(x^n|\hat{\theta})}
& \geq
\frac{(1-\eta_{\epsilon}) \pi^{d/2}\Phi(U_{K,\epsilon,n})
}
{\bigl(
\bar{h}_K
(1+||V(x^n|\hat{\theta})||_s)n \bigr)^{d/2} C_{J,n}(K) } \\ \label{eq:newlemma3}
&\cdot 
\frac{V_{d^2} (\tilde{a}\tilde{\delta})^{d^2}}{2^{d^2}}
\exp \Bigl( \frac{ \zeta n ||V(x^n|\hat{\theta})||_s \tilde{a}\tilde{\delta}}{16} \Bigr).
\end{align}
\end{lemma}

{\it Remark:}
The key factor in this expression is
the exponential in 
$n ||V(x^n|\hat{\theta})||\tilde{\delta}$,
which makes the coding distribution $\bar{m}(x^n)$
large when the empirical Fisher information deviation
$V(x^n|\hat{\theta})$
is not small.

{\it Proof:}
Note that
\begin{align*}
& \frac{\bar{m}(x^n)}{p(x^n|\hat{\theta})}\\
= &
\int \frac{p_e(x^n|\theta,\beta) \bar{w}(\theta,\beta)}{p(x^n|\hat{\theta})}d\theta d\beta \\
= &
b^{-d^2}\! \! \!
\int \int
e^{n(V_n(\theta) \cdot \beta -\psi_n(\theta,\beta))}d\beta
\frac{p(x^n|\theta)}{p(x^n|\hat{\theta})}
w_{K}^{(n)}(\theta)d\theta.
\end{align*}
First, we will evaluate the integral with respect to $\beta$
assuming $\theta \in B_\epsilon(\hat{\theta})$,
which implies $||V_n(\theta)||_s \ge 
\zeta ||V_n(\hat{\theta})||_s > \zeta \delta$,
by Assumption~\ref{assume:process:upper:3} and $x^n \in G_{n,\delta}^c$.
Hence, $||V_n(\theta)|| \geq \zeta \delta/\sqrt{d} = \tilde{\delta}$ holds.
Below, We will find a certain region where the integrand
$e^{n(V_n(\theta) \cdot \beta -\psi_n(\theta,\beta))}$
is exponentially large in the polynomial of $n$.

By Taylor expansion of $\psi_n(\theta,\beta)$ 
with respect to $\beta$ at $(\theta,0)$,
we have
\begin{align}\nonumber
& V_n(\theta) \cdot \beta -\psi_n(\theta,\beta)\\ \nonumber
= &
V_n(\theta) \cdot \beta
- \beta^t  n \cov_n(\theta,\beta')\beta/2 \\ \label{quadform}
\ge &
V_n(\theta) \cdot \beta - \lambda^*_n ||\beta||^2/2,
\end{align}
where $\beta'=t \beta$ with a certain $t \in [0,1]$, and
\[
\beta^t n \cov_n(\theta,\beta')\beta
=
\sum_{ij,kl}\beta_{ij} [n\cov_n(\theta,\beta')]_{ij,kl}\beta_{kl}.
\]
Let $f(\beta)$ denote the last side of (\ref{quadform}).
Introduce new parameters $(v,\beta_{\perp}) \in \Re \times \mathcal{B}$ by
\[
\beta = v \frac{V_n(\theta)}{||V_n(\theta)||} + \beta_{\perp},
\]
where $\beta_{\perp} \cdot V_n(\theta) = 0$ is assumed. 
Then, we have
\[
f(\beta) = ||V_n(\theta)|| v - \frac{\lambda_n^*}{2}(v^2+ ||\beta_{\perp}||^2).
\]
Now consider the region $N_{\tilde{a}\tilde{\delta}/2}(\tilde{\beta})$ 
with $\tilde{\beta}$ $= (\tilde{v},\tilde{\beta}_{\perp})$ $= (\tilde{a} \tilde{\delta},0)$.
Note that $\tilde{a} \lambda_n^* \le 1/2$ holds since $\tilde{a} = \min\{ 1/2\lambda_n^*, 1 \}$.
For all $\beta$ in $N_{\tilde{a}\tilde{\delta}/2}(\tilde{\beta})$, we have
\begin{align*}
    & f(\beta) \\
    = & ||V_n(\theta)|| v
    - \frac{\lambda_n^*}{2}\bigl((\tilde{a}\tilde{\delta} + v-\tilde{a}\tilde{\delta})^2+ ||\beta_{\perp}||^2\bigr)\\
= &||V_n(\theta)|| v \\
-& \frac{\lambda_n^*}{2}\bigl(\tilde{a}^22\tilde{\delta}^2 +2\tilde{a}\tilde{\delta}(v-\tilde{a}\tilde{\delta})
+ (v-\tilde{a}\tilde{\delta})^2+ ||\beta_{\perp}||^2 \bigr)\\
\ge &||V_n(\theta)|| v - \frac{\lambda_n^*}{2}\Bigl( \tilde{a}^22\tilde{\delta}^2+ 2\tilde{a}\tilde{\delta}(v-\tilde{a}\tilde{\delta})
+ \frac{\tilde{a}^22\tilde{\delta}^2}{4}\Bigr)\\
    = & \bigl(||V_n(\theta)||-\lambda_n^* a \tilde{\delta}\bigr) v -\frac{3\lambda_n^* \tilde{a}^22 \tilde{\delta}^2}{8}\\
    \ge & \frac{||V_n(\theta)||}{2} v -\frac{3\lambda_n^* \tilde{a}^22 \tilde{\delta}^2}{8}\\
    \ge & \frac{||V_n(\theta)||\tilde{a}\tilde{\delta}}{4} -\frac{3\lambda_n^* \tilde{a}^22 \tilde{\delta}^2}{8}\\
    = & \frac{\tilde{a}\tilde{\delta}}{4}\Bigl( ||V_n(\theta)||-\frac{3\lambda_n^*\tilde{a}\tilde{\delta}}{2} \Bigr)\\
    \ge & \frac{\tilde{a}\tilde{\delta}}{4}\Bigl( ||V_n(\theta)||-\frac{3\tilde{\delta}}{4} \Bigr)\\
    \ge & \frac{\tilde{a}\tilde{\delta}||V_n(\theta)||}{16}.
\end{align*}
To derive the first inequality above,
we have used $(v-\tilde{a}\tilde{\delta})^2+ ||\beta_{\perp}||^2 \le (\tilde{a}\tilde{\delta}/2)^2$ for 
$\beta \in N_{\tilde{a}\tilde{\delta/2}}(\tilde{\beta})$.
The second follows from $\lambda_n^* \tilde{a} \le 1/2$ and $\tilde{\delta} \le ||V_n(\theta)||$.
The others also follow from them and $v \ge \tilde{a}\tilde{\delta}/2$.
In this evaluation, we used the condition $a \lambda_n^* \le 1/2$, but
we actually defined
$ 
\tilde{a} = \min \{ 1/2\lambda_n^*, 1 \},
$ 
since too large $\tilde{a}$ may cause the problem that $N_{\tilde{a}\tilde{\delta}/2}(\tilde{\beta})$ is beyond $\mathcal{B}$. 

From this, the following lower bound 
provided $\theta \in B_\epsilon(\hat{\theta})$
is obtained.
\begin{align}\nonumber
&\int 
e^{n(V(x^n|\theta) \cdot \beta -\psi_n(\theta,\beta))}d\beta\\ \nonumber
&\ge
\int_{N_{\tilde{a}\tilde{\delta}/2}(\tilde{\beta})} 
e^{n(V(x^n|\theta) \cdot \beta -\psi_n(\theta,\beta))}d\beta\\ \nonumber
&\ge
\frac{V_{d^2} (\tilde{a}\tilde{\delta})^{d^2}}{2^{d^2}}
\exp \Bigl( \frac{ n ||V_n(\theta)||\tilde{a}\tilde{\delta}}{16} \Bigr)\\ \label{betaint}
&\ge
\frac{V_{d^2} (\tilde{a}\tilde{\delta})^{d^2}}{2^{d^2}}
\exp \Bigl( \frac{ n \zeta ||V_n(\hat{\theta})||_s \tilde{a}\tilde{\delta}}{16} \Bigr),
\end{align}
where 
we have used Assumption~\ref{assume:process:upper:3} and the inequality $||A||_s \le ||A||$.

Next we will evaluate the integral with respect to $\theta$. 
Since the bound for the integral with $\beta$
is uniform for $\theta \in B_\epsilon(\hat{\theta})$, 
it suffices to evaluate 
\[
\int_{B_\epsilon(\hat{\theta})}
\frac{p(x^n|\theta)w_{K}^{(n)}(\theta)}{p(x^n|\hat{\theta})}d\theta.
\]
For that purpose, we make use of
(\ref{assume:process:upper:2:ineq1:for_all}) in
Assumption~\ref{assume:process:upper:2:for_all:dash}.
We need a case argument depending on which one is the maximum
in the right side of (\ref{assume:process:upper:2:ineq1:for_all}).
For $\theta \in B_{\epsilon}(\hat{\theta})$,
first assume 
$(\theta - \hat{\theta})^t J_{n,\hat{\theta}}(\theta - \hat{\theta})
\le
(\theta - \hat{\theta})^t \hat{J}(\hat{\theta})
(\theta - \hat{\theta})$, then we have
\begin{align*}
\frac{p(x^n|\theta)}{p(x^n|\hat{\theta})}
&\ge
\exp \Bigl(  
\frac{-n (\theta-\hat{\theta})^t\hat{J}(\theta')(\theta-\hat{\theta})}{2}
\Bigr)\\
&\ge
\exp \Bigl(  
\frac{-n \bar{h}_K(\theta-\hat{\theta})^t\hat{J}(\hat{\theta})(\theta-\hat{\theta})}{2}
\Bigr),
\end{align*}
where 
$\theta' = t \theta + (1-t)\hat{\theta}$ with a certain $t \in [0,1]$.
The last inequality follows from Assumption~\ref{assume:process:upper:2:for_all:dash}.
Since
\[
||(J_{n,\hat{\theta}})^{-1/2}
\hat{J}(\hat{\theta})(J_{n,\hat{\theta}})^{-1/2}||_s
\leq
||V(x^n|\hat{\theta})||_s +1
\]
holds, we have 
\[
\frac{(\theta-\hat{\theta})^t\hat{J}(\hat{\theta})(\theta-\hat{\theta})}
{
(\theta-\hat{\theta})^tJ_{n,\hat{\theta}}(\theta-\hat{\theta})}
\leq
||\hat{V}||_s +1,
\]
where $\hat{V}$ denotes $V(x^n|\hat{\theta})$.
Hence for any $\theta \in B_\epsilon(\hat{\theta})$
with 
$(\theta - \hat{\theta})^t J_{n,\hat{\theta}}(\theta - \hat{\theta})
\le
(\theta - \hat{\theta})^t \hat{J}(\hat{\theta})
(\theta - \hat{\theta})$,
\begin{align}\label{for_integral}
\frac{p(x^n|\theta)}{p(x^n|\hat{\theta})}
\geq
e^{
- n \bar{h}_K(1+||\hat{V}||_s)(\theta-\hat{\theta})^tJ_{n,\hat{\theta}}(\theta-\hat{\theta})/2
}.
\end{align}
Next assume 
$(\theta - \hat{\theta})^t J_{n,\hat{\theta}}(\theta - \hat{\theta})
>
(\theta - \hat{\theta})^t \hat{J}(\hat{\theta})
(\theta - \hat{\theta})$ for $\theta \in B_\epsilon(\hat{\theta})$.
Then,
directly from Assumption~\ref{assume:process:upper:2:for_all:dash},
we have
\[
\frac{p(x^n|\theta)}{p(x^n|\hat{\theta})}
\geq
e^{
- n \bar{h}_K(\theta-\hat{\theta})^tJ_{n,\hat{\theta}}(\theta-\hat{\theta})/2
}.
\]
Hence, (\ref{for_integral}) holds for any $\theta \in B_\epsilon(\hat{\theta})$.

Then,
in a manner similar to the proof of Lemma~\ref{newlemma1}, 
by Laplace approximation over $B_\epsilon(\hat{\theta})$,
we have
\begin{align*}
&\int_{B_\epsilon(\hat{\theta})}
\frac{p(x^n|\theta)w_{K,n}(\theta)}{p(x^n|\hat{\theta})}d\theta \\
&\geq
\frac{(1-\eta_{\epsilon}) (2\pi)^{d/2}
\Phi(U_{K,\epsilon,n})w_K^{(n)}(\hat{\theta})
}{
\bar{h}_{K}^{d/2}n^{d/2}(1+||\hat{V}||_s)^{d/2}|J_{n,\hat{\theta}}|^{1/2}}\\
&=
\frac{(1-\eta_{\epsilon}) (2\pi)^{d/2}
\Phi(U_{K,\epsilon,n})
}{
\bar{h}_K^{d/2}n^{d/2}(1+||\hat{V}||_s)^{d/2}C_{J,n}(K)}
\end{align*}
Together with (\ref{betaint}), we have
(\ref{eq:newlemma3}).
{\em This completes the proof of Lemma~\ref{newlemma3}.}

Now we can state the main theorem for 
upper bounds on regret.
\begin{theorem}\label{thm:main:process}
Let $S=\{ p(\cdot|\theta) : \theta \in  \Theta \}$
be a $d$-dimensional family 
of stochastic processes.
Suppose that 
Assumptions~\ref{assume:process:upper:prior},
\ref{assume:process:upper:2:for_all:dash}, 
\ref{assume:process:upper:2:for_good},
\ref{assume:process:upper:3},
and \ref{assumption:upper:process:Fisher}.
hold.
Fix a sequence $\{ \epsilon_n \}$
which converges to $0$ slower than $1/\sqrt{n}$.
Define a mixture $m_n$ over $\bar{S}$ as
\[
m_n(x^n) =
(1-n^{-r})
m_{K,\epsilon_n,n}(x^n)
+n^{-r} \bar{m}(x^n).
\]
Then, the following holds.
\begin{align}\nonumber
\varlimsup_{n \rightarrow \infty} 
\Bigl(\sup_{x^n \in \mathcal{K}}&
\log \frac{p(x^n|\hat{\theta})}{m_{n}(x^n)}
-\frac{d}{2}\log\frac{n}{2\pi}
-\log C_{J,n}(K)
\Bigr)
  \\  \label{eqn:general:main}
&\leq 0.
\end{align}
\end{theorem}

{\em Proof:}
In this proof, we use Lemmas~5 and 8, 
plugging in
$\delta_n=n^{-1/2 +\gamma}$ in place of $\delta$,
which means that
$\tilde{\delta}$ is replaced by
\[
\tilde{\delta}_n = \frac{\zeta \delta_n}{\sqrt{d}}.
\]
When $x^n \in G_{n,\delta_n}$, by Lemma~\ref{newlemma2}, 
we have
\begin{align*}
\log \frac{p(x^n|\hat{\theta})}{m_{n}(x^n)}
&\leq
\frac{d}{2}\log\frac{n}{2\pi} +\log C_{K,\epsilon_n,n}\\
+\log & \frac{(1+\kappa_J \epsilon_n)^{d/2}(1+\delta_n)^{d/2}}{1-\eta(\epsilon_n)}
-\log(1-n^{-r}).
\end{align*}
Here, $\delta_n$, $n^{-r}$ and
$|\log C_{K,\epsilon_n,n}-\log C_{J,n}(K)|$ converge to $0$
as $n$ goes to infinity (the last one follows from Lemma~7),
hence we have
\begin{align*}
\varlimsup_{n \rightarrow \infty}
&\sup_{x^n \in G_{n,\delta_n}}
\Bigl(\log \frac{p(x^n|\hat{\theta})}{m_{n}(x^n)}
-
\frac{d}{2}\log\frac{n}{2\pi} -\log C_{J,n}(K) \Bigr)\\
&\leq 0.
\end{align*}

Next, we move to the case that  $x^n \in G_{n,\delta_n}^c$. 
Since
\[
\frac{m_n(x^n)}{p(x^n|\hat{\theta})}
\ge
\frac{n^{-r}\bar{m}(x^n)}{p(x^n|\hat{\theta})},
\]
we evaluate the lower bound of 
$\bar{m}(x^n)/p(x^n|\hat{\theta})$
using Lemma~\ref{newlemma3}.
When $n$ is large and $\epsilon$ is small,
the right side of 
(\ref{eq:newlemma3}) 
times $C_{J,n}(K)$
is not less than $g(||\hat{V}||_s)$ with
\[
g(\xi)
=
\frac{A(\tilde{a}\tilde{\delta}_n)^{d^2}}{(1+\xi )^{d/2}n^{d/2}}
\exp\Bigl( 
\frac{\zeta n \xi \tilde{a}\tilde{\delta}_n}{16}
\Bigr),
\]
where $A$ is a constant 
which does not depend on $||\hat{V}||_s$, $n$, and $\tilde{\delta}_n$.
The first derivative of $\log g(\xi)$ is
\[
-\frac{d}{2(1+\xi)} + \frac{\zeta n \tilde{a} \tilde{\delta}_n}{16}
\ge
-\frac{d}{2} + \frac{\zeta n \tilde{a} \tilde{\delta}_n}{16},
\]
which is positive when $n$ is large,
since $n \tilde{\delta}_n$ diverges to infinity as $n$ goes to infinity.
Hence, $g(||\hat{V}||_s)$ is not less than
\begin{align*}
g(\tilde{\delta}_n)
&=
\frac{A(\tilde{a}\tilde{\delta}_n)^{d^2}}{(1+\delta_n)^{d/2}n^{d/2}}
\exp\Bigl( 
\frac{\zeta n \delta_n \tilde{a} \tilde{\delta}_n}{16}
\Bigr) \\
&=
\frac{A(\tilde{a} \delta_n)^{d^2}}{(1+\delta_n)^{d/2}n^{d/2}d^{d^2/2}}
\exp\Bigl( 
\frac{\zeta^2 n \tilde{a} \delta_n^2 }{16\sqrt{d}}
\Bigr)
\end{align*}
for large $n$.
Therefore, we have
\begin{align}\label{eq:upper_for_not_good}
&\frac{C_{J,n}(K)m_n(x^n)}{p(x^n|\hat{\theta})}\\ \nonumber
&\ge
\frac{n^{-r}A\tilde{a}^{d^2} n^{d^2 (-1/2+\gamma)}}{(1+\delta_n)^{d/2}n^{d/2}d^{d^2/2}}
\exp\Bigl( 
\frac{\zeta^2 \tilde{a} n^{2\gamma} }{16\sqrt{d}}
\Bigr),
\end{align}
which diverges to infinity as $n$ goes to infinity.
It means that the regret for not bad sequences is negative
and can be ignored.
{\it This completes the proof.}



}

\section{Regret Bounds for Special Cases}

In this section, 
employing the main theorems for general cases,
we will show
asymptotic minimax bounds on regret
of some special models:
exponential families,
curved exponential families,
models with hidden variables
including mixture families (with fixed components),
and contaminated Gaussian location families.
Concerning the mixture families, we show
a stronger upper bound than that obtained by Theorem~5, that is,
we show that the same form of the asymptotic upper bound is valid without restricting the strings
to $\mathcal{K}$.

In the first subsection below, we
describe the formal definition of exponential families
and review their basic properties.
In the remaining subsections, we discuss regret bounds for each example.

\subsection{Definition of  Exponential Families}


Define the exponential family, following \cite{br86,amari90}.
\begin{definition}[Exponential Family]
\label{def:exponential}
Given Borel measurable functions 
$T : {\cal X} \rightarrow \Re^d$
and 
$U : {\cal X} \rightarrow \Re$,
define the natural parameter space
\[
\Theta  = \Bigl\{  \theta \in \Re^d :
\int_{{\cal X}}
\exp\bigl(\theta \cdot T(x)+U(x))\nu(dx) < \infty \Bigr\}
\] 
a subset of $\Re^d$,
assumed to have a non-empty interior $\Theta^\circ$.
Define 
$\psi(\theta)= \log \int_{\cal X}
\exp(\theta \cdot T(x)+U(x))\nu(dx)$
and 
a probability density on ${\cal X}$
with respect to $\nu$ by
\[
p(x|\theta) 
= \exp\bigl(\theta \cdot T(x)-\psi(\theta)+U(x) \bigr).
\]
Let $K$ denote a subset of $\Theta$ such that
$\bar{K}=\bar{K^\circ}$.
We refer to the set 
$S(K) = \{p(x|\theta): \theta \in K \}$
as an {\itshape exponential family} of densities on ${\cal X}$.
\end{definition}

The class of exponential families includes many common
statistical models such as Gaussian,
Poisson, 
Bernoulli sources and so on.
The following familiar case expresses how
the Gaussian family is an exponential family.
\begin{example}[Gaussian distributions]
Let ${\cal X }=\Re$
and $\nu(dx)$ be the Lebesgue measure $dx$.
The density of $N(\mu,\sigma^2)$ is
\begin{align*}
&\frac{1}{\sqrt{2\pi\sigma^2}}\exp\Bigl(-\frac{(x-\mu)^2}{2\sigma^2}\Bigr)\\
&=
\exp\bigl(
\frac{\mu x}{\sigma^2}
-\frac{x^2}{2\sigma^2}
-\frac{\mu^2}{ 2\sigma^2}
-\frac{\log(2\pi\sigma^2)}{2} \Bigr).
\end{align*}
Let $\theta=(\mu/\sigma^2 ,-1/(2\sigma^2))$,
$T(x)=(x,x^2)$, $U(x)=0$ and
$\psi(\theta)= \mu^2/(2\sigma^2) +(1/2)\log(2\pi\sigma^2)$.
Then, we see that the Gaussian is an exponential family,
where $\Theta=\Re \times (-\infty,0)$.
\end{example}

Hereafter,
we absorb the factor $\exp(U(x))$
into the reference measure $\nu(dx)$ 
(that is, we change $\exp(U(x))\nu(dx)$ to $\nu(dx)$).

It is known that $\Theta$ is a convex set.
Let  $\cal T$ denote 
the range of $T(x)$ for $x$ in the support of $\nu$ and let
{$\overline{\mbox{conv}}({\cal T}$)} 
be the closure of its convex hull.
We can assume 
that 
the model is arranged to be full rank such that
$\dim (\overline{\mbox{hull}}({\cal T})) = \dim \Theta=d$ holds
without loss of generality \cite{br86}.
An exponential family which satisfies this condition
is said to be minimal.
We assume $S(\Theta)$ is minimal in this paper.

It is known that $\psi(\theta)$ is of class $C^\infty$
and strictly convex on $\Theta^\circ$.
We refer to $\theta$ as the natural parameter 
(or $\theta$-coordinates).
We define the expectation parameter (or $\eta$-coordinates) as
$
\eta_i = E_\theta(T_i).
$
It is known that
the function on $\Theta^\circ$ mapping $\theta \mapsto \eta$
is an injection and of class $C^\infty$.
Let ${\cal H}=\{\eta(\theta):\theta \in \Theta^\circ \}$ be 
the range of this map.
We also assume that $S(\Theta)$ is steep:
that is
$E_{\theta}(|T(x)|)=\infty$ holds for 
any $\theta \in \Theta-\Theta^\circ$.
It is known that
${\cal H}$ is the interior of $\mbox{conv}({\cal T})$
for steep exponential families.
Further, if $\Theta$ is open, then $S(\Theta)$
is said to be regular.
Note that any regular family is steep automatically (vacuously) since $\Theta-\Theta^\circ$ is empty.

Note that
$\partial \psi /\partial \theta^i=E_\theta(T_i(x)) =\eta_i$
and
$\partial^2 \psi /\partial \theta^j \partial \theta^i 
= E_\theta((T_i(x)-\eta_i)(T_j(x)-\eta_j))$
hold in $\Theta^\circ$. 
Here, this $E_\theta((T_i(x)-\eta_i)(T_j(x)-\eta_j))$ 
is the Fisher information matrix with respect to $\theta$
denoted $J(\theta)$. 
Let $I(\eta)$ denote the Fisher information matrix of $\eta$.
This $I(\eta)$ is known to be the inverse matrix of $J(\theta)$.
It is known that $J(\theta)$ is strictly positive definite
in $\Theta^\circ$.

Given the string $x^n$, we have
$
p(x^n|\theta)=
\prod_{\tau=1}^n 
p(x_\tau|\theta)
=
\prod_{\tau=1}^n 
\exp(\theta \cdot T(x_\tau) -\psi(\theta))
=
\exp(n(\theta \cdot \bar{T} -\psi(\theta)))$,
where
\[
\bar{T}=\bar{T}(x^n)=\frac{\sum_{\tau=1}^n T(x_\tau)}{n}.
\]
These yield
\begin{align}\nonumber
\frac
{ \partial \log p(x^n|\theta)}{ \partial \theta_i}
&=
n(\bar{T}_i-\eta_i),\\ \label{eq:empFisher}
\frac{\partial^2
\log p(x^n|\theta)}{
\partial \theta_i \partial \theta_j }
&=
-n J_{ij}(\theta).
\end{align}
The latter implies that $\log p(x^n|\theta)$
is strictly concave in $\Theta^\circ$,
hence,
the former implies that
$\log p(x^n|\theta)$ takes the unique maximum
at $\eta=\bar{T}$ under the condition
that $\bar{T} \in {\cal H}$.
Let $\hat{\eta}$ $=\hat{\eta}(x^n)$ denote the maximum likelihood estimate
of $\eta$ given $x^n$.
Then, we have
$\hat{\eta}(x^n)=\bar{T}(x^n)$ for 
$x^n$ 
with $\bar{T}(x^n) \in {\cal H}^\circ$.
The identity (\ref{eq:empFisher}) also implies
\[
\hat{J}(\theta,x^n) = J(\theta)
\]
for all $\theta \in \Theta$ and for all $x^n$.
Therefore, Assumptions~\ref{assume:gen:1}, 
\ref{assume:gen:dash:1}, 
\ref{assume:gen:2}, and 
\ref{assume:gen:dash:2}
hold for the natural parameter $\theta$ of
exponential families.

For the expectation parameters
$(\tilde{\eta},\eta) \in {\cal H}^2$, corresponding to
$p(\cdot|\theta)$ and 
$p(\cdot|\tilde{\theta})$,
let $D(\tilde{\eta}||\eta)$
denote 
the Kullback divergence from
$p(\cdot|\tilde{\theta})$ 
to $p(\cdot|\theta)$.
Then, we have
\[
D(\tilde{\eta}||\eta)
=
E_{\tilde{\theta}}\Bigl(\log \frac{p(x|\tilde{\theta})}{p(x|\theta)}\Bigr)
=\tilde{\theta} \cdot \tilde{\eta} -\psi(\tilde{\theta})
-\theta \cdot \tilde{\eta} +\psi(\theta),
\]
We extend the domain  ${\cal H}^2$ to $\Re^d \times {\cal H}$, following
\cite{br86}:
\begin{eqnarray*}
\lefteqn{D(\tilde{\eta}||\eta)}\\
&&=\left\{
\begin{array}{ll}\displaystyle
\lim_{\epsilon \rightarrow +0} 
\inf
\{D(\eta'||\eta): \eta' \in {\cal H}, |\tilde{\eta}-\eta'| < \epsilon  \},\\
 \mbox{\hspace{30mm}for $(\tilde{\eta},\eta) \in \partial {\cal H} \times {\cal H}$,}\\
\infty, \:\:\:\: \:\:\:\: 
\mbox{\hspace{19mm}for $(\tilde{\eta},\eta) \in \bar{\cal H}^c \times {\cal H}$},
\end{array}\right.
\end{eqnarray*}
where $\partial {\cal H}$ is the boundary of $\cal H$
and $\bar{{\cal H}}^c$ is the complement of the closure.

We have the following relation between
differentiation with respect to $\eta$
and
differentiation with respect to $\theta$:
\[
\frac{\partial}{\partial \eta_i}
=
\sum_{j}\frac{\partial \theta_j}{\partial \eta_i}
\frac{\partial }{ \partial \theta_j}
=
\sum_{j}I_{ij}(\eta)
\frac{\partial}{\partial \theta_j}.
\]
This holds for 
$\theta \in \Theta^\circ$ 
(equivalently for $\eta \in {\cal H}$).
Thus
$\theta$ and $\psi(\theta)$
are infinitely differentiable with respect to $\eta$
as well as with respect to $\theta$.
Hence, 
$D(\tilde{\eta}||\eta)$ 
is of class $C^{\infty}$ on ${\cal H}^2$.

The following is a large deviation inequality.
\begin{lemma}\label{ldpmulti}
Suppose that $\eta \in {\cal H}$.
Let $\Lambda$ be a closed half space of $\Re^d$
{\rm (}i.e. $\Lambda=\{x \in \Re^d : x \cdot \xi \geq \gamma   \}$
for any specified $\xi \in \Re^d$ and $\gamma \in \Re${\rm )}.
Then, the following inequality holds
\[
P_\theta ( \bar{T}(x^n) \in \Lambda )
\leq 
\exp(-n \inf_{\tilde{\eta} \in \Lambda }
D(\tilde{\eta}||\eta)).
\]
\end{lemma}
See \cite{csis84,br86} for the proof. 

\subsection{Lower and Upper Bounds for  Exponential Families}
\label{upperboundforexponentialfamily}

By Lemma~\ref{ldpmulti}, we see that Assumptions~\ref{assume:gen:3} 
and \ref{assume:gen:dash:3} hold for regular exponential families.
We continue to assume finiteness of $C_J(K)$
(Assumption~\ref{assume:gen:K;prime}).
For the regular exponential families,
all the remaining assumptions for Lemmas~\ref{gen:lemma:lower} and \ref{lemma:Lower:Jeffreys} 
automatically hold.
Hence, Lemmas~\ref{gen:lemma:lower} and \ref{lemma:Lower:Jeffreys} 
and 
Theorems~\ref{gen:thm:lower} and \ref{st:thm:lower} can be applied
to show the desired lower bound indeed holds for the maximin regret
for regular exponential families.


Next, we consider the upper bound.
For the natural parameter $\theta$ of exponential families,
$\hat{J}(\theta,x^n) = J(\theta)$
holds for any $\theta \in \Theta^\circ$ and for any $x^n$.
Hence, if we use the natural parameter,
we do not need the fiber bundle of local exponential families
to obtain the asymptotic minimax regret.
In fact, we can prove
the following theorem,
where we suppose only Assumptions about $K$.
($K$ is compact, convex, and included in $\Theta^\circ$.)
\begin{theorem}\label{thm:main:multi}
Let $S=\{ p(\cdot|\theta) : \theta \in \Theta\}$
be a $d$-dimensional exponential family.
Let $K$ be a compact and convex subset of $\Theta^\circ$.
Let $m_{K,\epsilon_n,n}$ be the same one
as in Theorem~5.
Then, 
the following holds.
\begin{align}\label{multi:main:(2)}
\varlimsup_{n \rightarrow \infty}
\Bigl(
\sup_{x^n \in {\cal K} }
\log \frac{p(x^n|\hat{\theta})}{m_{K,\epsilon_n,n}(x^n)}
&-\frac{d}{2}\log\frac{n}{2\pi}
\Bigr) \\ \nonumber
& \leq
\log C_J(K).
\end{align}
\end{theorem}
This theorem immediately follows from 
Lemmas~\ref{newlemma12-0} and \ref{newlemma12-1}, 
which hold in much simpler forms than the original ones,
since
any string $x^n$ is an element of the good set $G_{n,\delta}$ by the fact $\hat{J}(\theta,x^n) = J(\theta)$,

\subsection{Lower and Upper Bounds for Curved Exponential Families}

As a more general case than exponential families,
we consider curved exponential families.
It is defined as a surface in the natural parameter
space of an exponential family.
If the surface is a hyperplane,
the surface forms an exponential family again.
Hence, we are interested in cases that
the surface is curved.
Below, we give a formal definition of curved exponential families.
\begin{definition}[Curved Exponential Families]
\label{curve}
Let $S =\{\bar{p}(\cdot|u): u \in {\cal U} \}$
be a $\bar{d}$-dimensional steep exponential family,
where  
$\bar{p}(x|u)=\exp(u\cdot T(x) -\psi(u))$
denotes the density
with respect to a measure $\nu(dx)$
and
$u$ its natural parameter.
Let $\Theta$ be an open subset of $\Re^{d}$ 
($d <  \bar{d}$)
and $\phi=(\phi_1,\ldots,\phi_{\bar{d}})$ be a function from $\Theta$ to ${\cal U}$.
Assume that $\phi$ is three times continuously differentiable
in  $\Theta$
and that the Jacobian of $\phi$ is of rank $d$ for every  $\theta$ in $\Theta$.
Further, we assume that $\phi$ is an injection.
Let 
$p(\cdot|\theta)=
\bar{p}(\cdot|\phii(\theta))$.
Then, the family
$
M = \{ p(\cdot|\theta):  \theta \in \Thetac  \}
$
with densities 
\[
p(x|\theta)=\exp(\phi(\theta) \cdot T(x) -\psi(\phi(\theta)))
\] 
is referred to as a 
$d$-dimensional curved exponential family of densities embedded in 
the exponential family $S$.
\end{definition}

Let $\ep$ denote the function which maps
$u$ to the corresponding expectation parameter of $S$,
that is, $\ep(u) = \int T(x)\bar{p}(x|u)\nu(dx)$.

For the curved exponential family $M$, the score function is
\begin{align}\nonumber
  \frac{\partial \log p(x|\theta)}{\partial \theta_i}
& =
\frac{\partial \phi(\theta)\cdot T(x)}{\partial \theta_i}
-\sum_k
\frac{\partial \psi(\phi(\theta))}{\partial u_k}
\frac{\partial \phi_k(\theta)}{\partial \theta_i}\\ \nonumber
& =
\frac{\partial \phi(\theta)\cdot T(x)}{\partial \theta_i}
-
\eta \cdot
\frac{\partial \phi(\theta)}{\partial \theta_i}\\ \label{eq:curved_score}
& =
\frac{\partial \phi(\theta)}{\partial \theta_i}\cdot (T(x)-\eta).
\end{align}
Then, by the chain rule of differentiation, we have
\begin{align}\nonumber
&  \frac{\partial^2 \log p(x|\theta)}{\partial \theta_j \partial
  \theta_i} \\ \nonumber
& =
\frac{\partial^2 \phi(\theta)}{\partial \theta_j\partial \theta_i}\cdot (T(x)-\eta)
-
\sum_{k,l}
\frac{\partial \phi_l(\theta)}{\partial \theta_i}
 \frac{\partial \eta_l}{\partial u_k}\frac{\partial \phi_k(\theta)}{\partial \theta_j}\\
\label{eq:curved_emp_calc}
& =
\frac{\partial^2 \phi(\theta)}{\partial \theta_j\partial \theta_i}\cdot (T(x)-\eta)
-
\sum_{k,l}
\frac{\partial \phi_l(\theta)}{\partial \theta_i}
\bar{J}_{kl}(u)
\frac{\partial \phi_k(\theta)}{\partial \theta_j},
\end{align}
where
$\bar{J}_{kl}(u)$ is the Fisher information matrix of the 
exponential family $S$.
Taking the expected value, we have
\[
-J_{ij}(\theta) = -\sum_{k,l}
\frac{\partial \phi_l(\theta)}{\partial \theta_i}
\bar{J}_{kl}(u)
\frac{\partial \phi_k(\theta)}{\partial \theta_j}.
\]
By this, we have confirmed that the Fisher information
exists and is finite.
Further, since $\phi$ is twice continuously differentiable
and 
its Jacobian is of rank $d$ for every $\theta \in \Theta$,
the Fisher information
is continuous and is positive definite
for all $\theta \in \Theta$.
(That is, Assumption~\ref{assume:gen:2} holds.)
Further,
(\ref{eq:curved_emp_calc})
yields
\begin{align}\label{empfisher-fisher0}
\hat{J}_{ij}(\theta,x)
=
J_{ij}(\theta)
-
\frac{\partial^2 \phii(\theta)}{\partial \theta_i\partial \theta_j}\cdot
(T(x)-\ep(\phii(\theta))).
\end{align}
This shows how the empirical Fisher information differs from 
the Fisher information in curved exponential families.
Since $T(x)$ and its covariance matrix with respect
to $\bar{p}(\cdot|u)$ for all $u \in \mathcal{U}$
is finite,
Assumptions~\ref{assume:gen:1} and \ref{assume:gen:dash:1} hold.
Assumption~\ref{assume:gen:dash:2} also holds because of (\ref{eq:curved_score}).
Further, note
that the maximum likelihood estimator
for curved exponential families is
consistent uniformly for $\theta \in K$,
when $K$ is compact and interior to $\Theta$.
This can be easily confirmed by Lemma~\ref{ldpmulti} (the large deviation theorem).
Hence, Lemmas~\ref{gen:lemma:lower} and \ref{lemma:Lower:Jeffreys} 
and Theorems~\ref{gen:thm:lower} and \ref{st:thm:lower} work for
curved exponential families
and we have the validity of the desired lower bound.
  
Next, we consider the upper bound.
By (\ref{empfisher-fisher0}), we have
\begin{align}\label{empfisher-fisher_n}
\hat{J}_{ij}(\theta,x^n)
=
J_{ij}(\theta)
-
\frac{\partial^2 \phii(\theta)}{\partial \theta_i\partial \theta_j}\cdot
(\bar{T}-\ep(\phii(\theta))).
\end{align}
Let $H^k$ denote the Hessian of $\phi_k(\theta)$
(the $k$th component of $\phi(\theta)$).
Then we can write
\begin{align}\label{empfisher-fishernn}
\hat{J}(\theta,x^n)
=
J(\theta)
-
\sum_k
(\bar{T}_k-\ep_k(\phii(\theta)))
H^k(\theta).
\end{align}
By this, we have 
\begin{align}\label{VforCurved}
V(x^n|\theta)
=
-
\sum_k
(\bar{T}_k-\ep_k(\phii(\theta)))
J_{\theta}^{-1/2}
H^k(\theta)
J_{\theta}^{-1/2}.
\end{align}
Since the exponential moments of $T(x)$ exist for $u \in \mathcal{U}^\circ$
by definition of the natural parameter space of exponential families,
Assumption~\ref{assume:process:upper:0} holds for a certain $b > 0$.
Further, note that, if $T(x)$ is bounded,
then $\bar{T}(x^n)$ is uniformly bounded for all $n$.
Hence, if $T(x)$ is bounded,
the collection of functions $V(x^n|\theta)$, $x^n \in \mathcal{X}^n$, $n=1,2,\ldots$
is equicontinuous uniformly for $\theta \in K$.
Then, Assumptions~\ref{assume:process:upper:2:for_all:dash}, 
\ref{assume:process:upper:2:for_good}, 
and
\ref{assume:process:upper:3}
hold.
As for the ideal prior, 
Assumption~\ref{assume:st:-upper2} does hold,
since we assume $\phi(u)$ is three times continuously differentiable.
Hence, Assumption~\ref{assume:process:upper:prior} holds.
That is, we can use Theorem~5,
and the general minimax strategy works.

On the contrary, if $T(x)$ is not bounded,
it is not clear whether the assumptions hold or not.
However, to deal with such curved exponential families with unbounded $T(x)$, 
we can establish another minimax strategy.
In this alternative strategy, to deal with the string $x^n$
whose empirical Fisher information differs from the Fisher
information,
we mix in small measure component living in the full 
$\bar{d}$-dimensional exponential family $S$
instead of 
the fiber bundle of local exponential families.

Let $K$ be a compact set included in $\Theta^\circ$
such that $\bar{K}=\bar{K^\circ}$.
Let ${\cal U}_c$ be a compact set included in ${\cal U}^\circ$
such that
${\cal U}_c^\circ \supset \{\phii(\theta) : \theta \in \bar{K} \}$.
Our asymptotic minimax strategy is established
in the following theorem.
Note that we do not assume the boundness of $T(x)$ here.

\begin{theorem}\label{thm:main:curved}
Let $S=\{ p(\cdot|\theta) : \theta \in  \Theta \}$
be a $d$-dimensional curved exponential family of densities
defined in Definition~\ref{curve}
and let 
\begin{align*}
m_n(x^n)
=(1-n^{-r}) m_{K,\epsilon,n}(x^n) 
 + n^{-r} \int \bar{p}(x^n|u)w(u)du,
\end{align*}
where $w(u)$ is the uniform prior over $\mathcal{U}_c$.
Then, the following holds.
\begin{eqnarray}\label{general:main}
\lefteqn{\limsup_{n \rightarrow \infty} 
(\sup_{x^n \in \mathcal{K}}
\log \frac{p(x^n|\hat{\theta})}{m_{n}(x^n)}
-\frac{d}{2}\log\frac{n}{2\pi})} \\ \nonumber
&&\hspace*{30mm} \leq
\log C_J(K).
\end{eqnarray}
\end{theorem}

{\it Proof:}
Define a set of good strings:
\[
G_n= \{ x^n : 
\hat{\theta} \in K
\: \mbox{and} \:
|\hat{\ep} - \ep(\phii(\hat{\theta}))| 
\leq \delta_n 
\}
\]
(recall that $\ep$ denotes the expectation parameter
and that $\hat{\ep}=\bar{T}$),
where we let
$\delta_n = n^{-1/2+\gamma}$ ($0< \gamma <1/2$). 
Define also
\[
G_n^c= \{ x^n : 
\hat{\theta} \in K
\: \mbox{and} \:
|\hat{\ep} - \ep((\phii(\hat{\theta}))| 
> \delta_n 
\}
\]
We can prove the usual asymptotic bound for regret
when $x^n \in G_n$, while
for $x^n \in G_n^c$,
it turns out that the regret of our strategy
becomes negative, that is,
the code length becomes shorter than $-\log p(x^n|\hat{\theta})$.

For $x^n \in G_n$,
from (\ref{empfisher-fishernn}) and the Cauchy-Schwartz inequality,
we have for each $(i,j)$,
\begin{align*}
(\hat{J}_{ij}(\theta,x^n)-J_{ij}(\theta))^2
&\le
|\bar{T}-\ep(\phii(\theta))|^2
\sum_k
(
H_{ij}^k(\theta)
)^2\\
&\le
\delta_n^2
\sum_k
(
H_{ij}^k(\theta)
)^2.
\end{align*}
Since $H^k(\theta)$ is continuous and $K$ is compact,
$\sum_k
(
H_{ij}^k(\theta)
)^2$ is bounded by a constant $C_{H,K}^2$. Accordingly,
for $x^n \in G_n$,
\[
\max_{ij}
\max_{\theta \in K}
|\hat{J}_{ij}(\theta,x^n)-J_{ij}(\theta)|
\le
C_{H,K}\, \delta_n
\]
holds, which implies
\[
\max_{\theta \in K}
||\hat{J}(\theta,x^n)-J(\theta)||
\le
d\, C_{H,K}\delta_n.
\]
Therefore, uniformly over $x^n \in G_n$,
\[
\max_{\theta \in K}
||\hat{J}(\theta,x^n)-J(\theta)||_s
\le
d\, C_{H,K}\delta_n
\]
which means, for all unit vector $z$,
we have
\[
\max_{\theta \in K}
|z^t(\hat{J}(\theta,x^n)-J(\theta))z|
\le
d\, C_{H,K}\delta_n.
\]
Since $J(\theta)$ is continuously differentiable,
for $\epsilon > 0$,
\[
\max_{\theta' \in B_\epsilon(\theta)\cap K}
\max_{\theta \in K}
|z^t(\hat{J}_n(\theta')z-z^tJ(\theta))z|
\le
d\, C_{H,K}\delta_n+
O(\epsilon).
\]
Now since $\delta_n=n^{-1/2 + \gamma}$, the right side is $O(\epsilon)$
for $n \ge (1/\epsilon)^{2/(1-2\gamma)}$.
Therefore, for such $n$, we have the following two inequalities holding uniformly for $x^n \in G_n$
and uniformly for all unit vectors $z$,
\begin{align*}
    \max_{\tilde{\theta} \in B_\epsilon(\hat{\theta})\cap K}
|z^t\hat{J}(\tilde{\theta},x^n)z-z^tJ(\hat{\theta})z|
&\le
O(\epsilon)
\end{align*}
and 
\par\vspace{-0.7cm}
\begin{align*}
|z^t\hat{J}(\hat{\theta},x^n)z-z^tJ(\hat{\theta})z|
&\le
O(\epsilon).
\end{align*}
Since $z^tJ(\hat{\theta}))z$ is lower bounded by a positive constant (even when we take a minimum over all unit vectors $z$),
for sufficiently small $\epsilon$,
the $z^t(\hat{J}(\hat{\theta},x^n)z$ is also lower bounded by a positive constant, uniformly over unit vectors $z$ and $x^n \in 
G_n$.
Consequently, for all $n \ge (1/\epsilon)^{2/(1-2\gamma)}$ 
and all $x^n \in G_n$, we have
\[
\max_{\tilde{\theta} \in B_\epsilon(\hat{\theta})\cap K}
\frac{z^t\hat{J}(\tilde{\theta},x^n)z}{z^t\hat{J}(\hat{\theta},x^n)z}
\le 1+O(\epsilon).
\]
This is the last inequality in Assumption~9 and
is sufficient to obtain the same bound
for good strings as in Theorem~5.
To show that, we have the condition that
$\epsilon$ must converge to $0$ slower than $1/\sqrt{n}$.
Now, we face the new condition $\epsilon \ge \delta_n = n^{-1/2+\gamma}$,
which does not contradict the previous condition.

Next we consider $x^n \in G^c_n$.
Let $N_{\delta}(u)$ denote the $\delta$ neighborhood of $u$
and let ${\cal U}_c'$ a compact set included in ${\cal U}_c^\circ$.
Now we will prove that the following holds for a certain $\kappa > 0$.
\begin{eqnarray}\nonumber
&&\forall n \in \mathbb{N},\:
\forall x^n \in G_n^c,\:
\exists \tilde{u} \in {\cal U}_c',\: \\ \label{key0}
&&\inf_{u \in N_{\kappa \delta_n}(\tilde{u})}
\frac{1}{n}\log\frac
{\bar{p}(x^n|u)}
{p(x^n|\hat{\theta})}
\geq
C_2 \delta_n^2.
\end{eqnarray}
Indeed, by Taylor expansion with respect to $u$ at $\phii(\hat{\theta})$, we have
\begin{align*}
\frac{1}{n}&\log \frac{\bar{p}(x^n|u)}{p(x^n|\hat{\theta})}\\
=&
\frac{1}{n}\log \frac{\bar{p}(x^n|u)}{\bar{p}(x^n|\phii(\hat{\theta}))}
\\
=&
(u-\phii(\hat{\theta}))\cdot \bar{T}- \psi(u)+\psi(\phii(\hat{\theta}))
\\
=&
(\bar{T}-\ep(\phii(\hat{\theta})))\cdot (u-\phii(\hat{\theta}))\\
&-(1/2)(u-\phii(\hat{\theta}))^t\bar{J}(u')(u-\phii(\hat{\theta})),
\end{align*}
where 
$u'=\xi u +(1-\xi)\phii(\hat{\theta})$ 
for some $\xi \in [0,1]$.
Let 
\[
\tilde{u} 
= \phii(\hat{\theta})
+2 \kappa \delta_n    
\frac{\bar{T}-\phii(\hat{\theta}))}{
|\bar{T}  -\phii(\hat{\theta})|}.
\]
Then, for all $u\in N_{\kappa \delta_n}(\tilde{u})$, we have
$\kappa \delta_n \le |u-\phii(\hat{\theta})| \le 3\kappa \delta_n$.
Hence, we have
\begin{eqnarray*}
\inf_{u \in N_{\kappa \delta_n}(\tilde{u})}
\frac{1}{n}\log \frac{\bar{p}(x^n|u)}{p(x^n|\hat{\theta})}
&\geq&
\kappa \delta_n^2
-
9\kappa^2 \lambda^*  \delta_n^2/2\\
&=&
\kappa (\delta_n^2
-
9\kappa \lambda^*  \delta_n^2/2),
\end{eqnarray*}
where $\lambda^*$ denotes the maximum value among
the maximum eigenvalues of $\bar{J}(u)$ for $u \in \mathcal{U}_c$.
Let $\kappa=1/9\lambda^*$, then the above implies (\ref{key0}).
Hence we have
for all $n \in \mathbb{N}$ and for all 
$x^n \in G_n^c$, 
\begin{align*}
&\frac
{n^{-r}\int \bar{p}(x^n|u)w(u)du}
{p(x^n|\hat{\theta})}\\
&\geq
\int_{N_{\kappa \delta_n}(\tilde{u})}
\frac
{n^{-r}\bar{p}(x^n|u)w(u)}
{p(x^n|\hat{\theta})}du
\\
&\geq
C_{3}n^{-r}
\int_{N_{\kappa \delta_n}(\tilde{u})}
\exp(C_2 n \delta_n^2  )du
\\
&\geq
C_{3}n^{-r}
{(\kappa \delta_n)^{\bar{d}}}
\exp(C_2 n \delta_n ^2  )
\\
&\geq
C_{4}
{n^{-r}n^{-\bar{d}/4}}
\exp(C_2 n^{2\gamma} ).
\end{align*}
This means that the regret of $m_n$ is negative
for all large $n$ and can be ignored.
{\it This completes the proof.}

Lastly we give a comment about the case of stochastic processes.
Though Theorem~\ref{thm:main:curved} above is for the i.i.d.\ case only,
it can be extended to the case of Markov sources.
It is known that some parametric model of Markov sources approaches an exponential family
as $n$ goes to infinity,
in the sense that the difference between Fisher and empirical Fisher information matrices 
converges to zero as $n$ goes to infinity.
We call the model of stochastic processes with that condition as an asymptotic exponential family.
In particular, the Markov model defined by a context tree \cite{wst} (a tree model) is interesting.
When the context tree satisfies a certain condition, the model is called a finite state machine X 
(FSMX) model \cite{WRF1995}. It is shown that the tree model is an asymptotic exponential family
if and only if the model is an FSMX model \cite{TN2017}.

As a special case, consider 
the model of $k$th order Markov chains, which is an example of an FSMX model.
For this case, it has been shown that the Jeffreys mixture asymptotically achieves the minimax regret
\cite{TKB2013}.
Note that any tree model can be regarded as a sub family (a surface) of the 
model of a certain order Markov chains. It means that
a tree model can be regarded as a curved exponential family embedded in an
asymptotic exponential family. It suggests that we can establish 
a similar theorem as Theorem~\ref{thm:main:curved} concerning tree models.
In fact, such a theorem is in \cite{TB2014b}.

\subsection{Lower and Upper Bounds for Models with Hidden Variables}


The contents of this section were presented in part
in 2014 IEEE International Symposium on Information Theory \cite{TB2014a}.

First we introduce the notion of models with hidden variables below.
\begin{definition}[Model with Hidden Variables]
Let 
$(\mathcal{X},\mathcal{B}_x,\nu_x)$
and
$(\mathcal{Y},\mathcal{B}_y,\nu_y)$
be measurable spaces
with reference measures 
$\nu_x$
and $\nu_y$, assumed to be sigma finite.
Let $q(y|\theta)$ be a density of a
$d$-dimensional exponential family over $\mathcal{Y}$ with natural parameter $\theta$.
Define a class $S$ of probability density functions over $\mathcal{X}$ 
with respect to $\nu_x$
by
\begin{align}\label{mwh}
S = 
\Bigl\{ 
p(x|\theta) = \int \! \! \kappa(x|y)q(y|\theta)\nu_y(dy) \; 
\colon \;
\theta \in \Theta 
\Bigr\},
\end{align}
where $\kappa(x|y)$ is a fixed conditional probability density function of 
$x$ given $y$. 
Then, $S$ is called a model with a hidden variable $y$.
\end{definition}

For the above definition, we will assume that the Fisher information is positive definite over $\Theta^\circ$.
(Assumption~2.)
Whether this holds or not depends on the property of $\kappa(x|y)$.
Later we will see that if $x$ and $y$ are not independent random variables
with respect to 
the joint density $\kappa(y|x)q(y|\theta)$, then $J(\theta)$ is positive definite.
This is reasonable because we can draw information about $\theta$ from $x$,
if $x$ and $y$ are not independent.

If $q(y|\theta)$ is the multinomial model over
the finite set $\mathcal{Y}=\{0,1,\ldots,d  \}$, then
$p(x|\theta)$ is in the following form:
\begin{align}\label{eq:mixturefamily}
p(x|\theta) = \sum_{y=0}^d \theta_y \kappa(x|y) 
.    
\end{align}
Here, $\theta$ 
denotes a $d$-dimensional vector
$(\theta_1,\ldots,\theta_d) \in \mathbb{R}^d$ satisfying
$0 \le \sum_{i=1}^d \theta_i \le 1$, $\theta_i \ge 0$ for all $i$.
and $\theta_0$ denotes $1- \sum_{i=1}^d \theta_i$.
This model is called a mixture family in information geometry \cite{amari90,an2000}.
The mixture family is the dual object of the exponential family
in information geometry \cite{amari90,an2000}.
Note that this mixture family is a simpler object
than the more general class of mixture models such as Gaussian mixtures, binomial mixtures, etc.,
which are used in practical applications. 

Typically, models with a hidden variable are examples of non exponential families.
But note that
if it is the mixture family
and 
if
$\al$ is a finite subset
of an Euclidean space and each $\kappa(\cdot|y)$ is a point mass at an $x=x_y$ in $\al$,
then it is also an exponential family.
That is the case of the multinomial Bernoulli model.


In this section we first argue the general case of
models with hidden variables, assuming $T(y)$ is bounded,
for which Theorem~\ref{thm:main:process} gives the minimax solution
for the set of strings $\mathcal{K} = \mathcal{X}^n(K)$
for a compact $K$ interior to $\Theta$.

After that, 
for the mixture family case,
we show a minimax strategy
for the problem
without restriction to the set of strings $\mathcal{X}^n$.
The shown strategy is different from the one used in
Theorem~\ref{thm:main:process}.

\subsubsection{Lower Bound for Models with Hidden Variables}

Here, we consider the lower bound.
We examine Assumptions for Theorems~1 and 2.
Here, we assume $\theta$ in $q(y|\theta)$ is the natural parameter
of exponential family $q(y|\theta)$:
\[
q(y|\theta) = \exp(\theta \cdot T(y)- \psi(\theta)).
\]
Below, let $\eta$ denote the expectation parameter:
$\eta = \int T(y)q(y|\theta)\nu_y(dy)$
and let $G(\theta)$ denote the Fisher information
matrix for $q(y|\theta)$ with respect to $\theta$.

We will show some nice properties of 
the score function, the Fisher information, and the empirical
Fisher information.
Let
${\rm Cov}_\theta [\sqrt{n}\bar{T}(Y^n)|x^n]$ 
denote the covariance matrix of $\sqrt{n}\bar{T}(y^n)$ with respect to the conditional
probability distribution $p(y^n|x^n,\theta)$ with a fixed $x^n$, where
$\bar{T}=\sum_{t=1}^n T(y_t) /n$.
Further, define
\[
{\rm Cov}_\theta [T(Y)|X]
=\int
{\rm Cov}_\theta [\bar{T}(Y)|x]p(x|\theta)\nu(dx).
\]
Then, we have the following lemma.
\begin{lemma}\label{lemma_for_mdh}
For the model with hidden variable (\ref{mwh}),
the following equalities hold.
    \begin{align}\label{eq:hiddenscore}
\frac{1}{n}  \frac{\partial \log p(x^n|\theta)}{\partial \theta_i}
&= \tilde{t}_i - \eta_i,\\ \label{eq:empiricalFisherofmwh}
\hat{J}(\theta,x^n)
& =
G(\theta) - \, {\rm Cov}_\theta [\sqrt{n}\bar{T}(Y^n)|x^n],\\ \label{eq:Fisherofmwh}
J(\theta) &= G(\theta) - {\rm Cov}_\theta[T(Y)|X]
\end{align}
and
\[
\frac{\partial \hat{J}_{ij}(\theta,x^n)}{\partial \theta_k}
=
\frac{\partial G_{ij}(\theta)}{\partial \theta_k} - \sqrt{n} \tilde{w}_{ijk}(\theta,x^n),
\]
where
\begin{align*}
    \tilde{t}_i&=\tilde{t}_i(\theta,x^n)
 = \int \bar{T}(y^n)p(y^n|x^n,\theta)\nu_y(dy^n)
\end{align*}
and
\begin{align*}
&\tilde{w}_{ijk} \\
=& \tilde{w}_{ijk}(\theta,x^n) \\
=& n^{3/2}
\int (\bar{T}_i-\tilde{t}_i)(\bar{T}_j-\tilde{t}_j) 
(\bar{T}_k-\tilde{t}_k)
p(y^n|x^n,\theta)\nu_y(dy^n).
\end{align*}
\end{lemma}

{\it Proof:}
Note that
\begin{align}\label{eq:hiddenkey}
\frac{\partial p(x^n|\theta)}{\partial \theta_i}
=
\int \kappa(x^n|y^n)n(\bar{T}_i-\eta_i)q(y^n|\theta)\nu(dy^n).
\end{align}
Hence,
we have
\begin{align}
  \frac{\partial \log p(x^n|\theta)}{\partial \theta_i}
& = \frac{\int \kappa(x^n|y^n)n(\bar{T}_i-\eta_i)q(y^n|\theta)\nu(dy^n)}{p(x^n|\theta)} \\ \nonumber
& =
\int n(\bar{T}_i-\eta_i)p(y^n|x^n,\theta)\nu(dy^n) \\ \nonumber 
& = n(\tilde{t}_i - \eta_i),
\end{align}
where $p(y^n|x^n,\theta) = \kappa(x^n|y^n)q(y^n|\theta)/p(x^n|\theta)$.
This yields (\ref{eq:hiddenscore}).
To further differentiate, we need $(\partial/\partial \theta_i ) p(y^n|x^n,\theta)$.
By the chain rule of differentiation 
\begin{align*}
 & \frac{\partial p(y^n|x^n,\theta)}{\partial \theta_i}\\
= &
\frac{\kappa(x^n|y^n)n(\bar{T}_i-\eta_i)q(y^n|\theta)}{p(x^n|\theta)}\\
&-
\frac{\kappa(x^n|y^n)q(y^n|\theta)
}{(p(x^n|\theta))^2}
\frac{\partial p(x^n|\theta)}{\partial \theta_i}\\
= &
n(\bar{T}_i-\eta_i)p(y^n|x^n,\theta)\\
&-
\frac{\kappa(x^n|y^n)q(y^n|\theta)}
{p(x^n|\theta)}
\frac{\partial \log p(x^n|\theta)}{\partial \theta_i}\\
= &
n(\bar{T}_i-\eta_i)p(y^n|x^n,\theta)\\
&-
p(y^n|x^n,\theta)\int n(\bar{T}_i-\eta_i)p(y^n|x^n,\theta)\nu(dy^n)\\
= &
n(\bar{T}_i-\eta_i)p(y^n|x^n,\theta)-
n(\tilde{t}_i-\eta_i)p(y^n|x^n,\theta)\\
= & n(\bar{T}_i-\tilde{t}_i)p(y^n|x^n,\theta).
\end{align*}
Using this formula, we have
\begin{align*}
 & \frac{\partial^2 \log p(x^n|\theta)}{\partial \theta_j \partial \theta_i}\\
 = &
\int \frac{\partial n(\bar{T}_i-\eta_i)}{\partial \theta_j}
p(y^n|x^n,\theta)\nu(dy^n)\\
& + 
\int n(\bar{T}_i-\eta_i)\frac{\partial p(y^n|x^n,\theta)}{\partial \theta_j}\nu(dy^n)\\
= &
\int (-n G_{ij}(\theta))
p(y^n|x^n,\theta)\nu(dy^n)\\
& + 
\int n(\bar{T}_i-\eta_i)n(\bar{T}_j-\tilde{t}_j) p(y^n|x^n,\theta)\nu(dy^n)\\
= &
-n G_{ij}(\theta)
\\
& + 
\int n(\bar{T}_i-\eta_i)n(\bar{T}_j-\tilde{t}_j) p(y^n|x^n,\theta)\nu(dy^n)\\
= &
-n G_{ij}(\theta)
\\
& + 
\int n(\bar{T}_i-\tilde{t}_i)n(\bar{T}_j-\tilde{t}_j) p(y^n|x^n,\theta)\nu(dy^n)\\
& + 
\int n(\tilde{t}_i-\eta_i)n(\bar{T}_j-\tilde{t}_j) p(y^n|x^n,\theta)\nu(dy^n)\\
= &
-n G_{ij}(\theta)
\\
& + 
\int n(\bar{T}_i-\tilde{t}_i)n(\bar{T}_j-\tilde{t}_j) p(y^n|x^n,\theta)\nu(dy^n)\\
= &
-n G_{ij}(\theta) + n \tilde{v}_{ij},
\end{align*}
where $\tilde{v}_{ij} = \tilde{v}_{ij}(\theta,x^n)$
denotes 
\[
\int \sqrt{n}(\bar{T}_i-\tilde{t}_i)\sqrt{n}(\bar{T}_j - \tilde{t}_j)p(y^n|x^n,\theta)\nu_y(dy^n).
\]
That is, we have
\begin{align}\label{eq:hiddensecond}
\frac{\partial^2 \log p(x^n|\theta)}{\partial \theta_j \partial \theta_i}
= 
-n G_{ij}(\theta)
 + n\tilde{v}_{ij},
\end{align}
Since $\tilde{v}_{ij}$ is $ij$-entry of ${\rm Cov}_\theta [\sqrt{n}\bar{T}(Y^n)|x^n]$,
this is (\ref{eq:empiricalFisherofmwh}),
which directly yields (\ref{eq:Fisherofmwh}).

To differentiate (\ref{eq:hiddensecond})
again, note the following
\begin{align*}
    \frac{\partial \tilde{t}_i}{\partial \theta_j}
    &=
    \frac{\partial }{\partial \theta_j}
    \int \bar{T}_i p(y^n|x^n,\theta)\nu(dy^n)\\
    &=
    \int
    \bar{T}_i
    n(\bar{T}_j-\tilde{t}_j)p(y^n|x^n,\theta)\nu(dy^n)\\
    &= n
    \int
    (\bar{T}_i-\tilde{t}_i)
    (\bar{T}_j-\tilde{t}_j)p(y^n|x^n,\theta)\nu(dy^n)
    \\
    & =  \tilde{v}_{ij}.
\end{align*}
Then,
\begin{align*}
 & \frac{\partial^3 \log p(x^n|\theta)}{\partial \theta_k \partial \theta_j \partial \theta_i}\\
= &
-n \frac{\partial G_{ij}(\theta)}{\partial \theta_k}
\\
& + 
\int \frac{\partial n(\bar{T}_i-\tilde{t}_i)}{\partial \theta_k}
n(\bar{T}_j-\tilde{t}_j) p(y^n|x^n,\theta)\nu(dy^n)\\
& + 
\int n(\bar{T}_i-\tilde{t}_i) 
\frac{\partial n(\bar{T}_j-\tilde{t}_j)}{\partial \theta_k}
p(y^n|x^n,\theta)\nu(dy^n)\\
& + 
\int n(\bar{T}_i-\tilde{t}_i)n(\bar{T}_j-\tilde{t}_j) \frac{\partial 
p(y^n|x^n,\theta)}{\partial \theta_k}\nu(dy^n).
\end{align*}
Since
$\partial \bar{T}/\partial \theta_k = 0$ holds,
and since
$\partial \tilde{t}_i/\partial \theta_k = \tilde{v}_{ik}$
dose not depend on $y^n$, the second and third terms 
in the last expression
in the integral yield $0$.
Hence,
\begin{align*}
 & \frac{\partial^3 \log p(x^n|\theta)}{\partial \theta_k \partial \theta_j \partial \theta_i}\\
= &
-n \frac{\partial G_{ij}(\theta)}{\partial \theta_k}
\\
& + 
\int n(\bar{T}_i-\tilde{t}_i)n(\bar{T}_j-\tilde{t}_j) \frac{\partial 
p(y^n|x^n,\theta)}{\partial \theta_k}\nu(dy^n).
\end{align*}
The second term is
\[
\int n(\bar{T}_i-\tilde{t}_i)n(\bar{T}_j-\tilde{t}_j) 
n(\bar{T}_k-\tilde{t}_k)p(y^n|x^n,\theta)
\nu(dy^n),
\]
which equals $n^{3/2} \tilde{w}_{ijk}$.
Then, we have
\begin{align}\label{eq:hiddenthird}
 \frac{\partial^3 \log p(x^n|\theta)}{\partial \theta_k \partial \theta_j \partial \theta_i}
= 
-n \frac{\partial G_{ij}(\theta)}{\partial \theta_k} + n^{3/2} \tilde{w}_{ijk}.
\end{align}
{\it This completes the proof of the lemma.}

Recall that $G(\theta)$ equals the covariance of $T(y)$
with respect to $q(y|\theta)$
and note that the expectation of ${\rm Cov}_\theta[T(y)|x]$
with respect to $p(x|\theta)$, which we denote by ${\rm Cov}_\theta[T(y)|X]$,
is not more than ${\rm Cov}_\theta[T(y)] = G(\theta)$ in the sense of positive definiteness.
It confirms the fact that $J(\theta)$
is positive semi-definite.
Here, the equality holds only if $x$ and $T(y)$ are independent of each other.
To be more precisely, if and only if $x$ and $z^tT(y)$
(the linear combination of  each entry of $T(y)$) is independent, 
$z^tG(\theta)z = z^t {\rm Cov}_\theta[T(y)|X]z$ holds.
Therefore, we assume that $x$ and $z^tT(y)$ with any non-zero $z$ is not independent 
for all $\theta \in \Theta^\circ$.

Since we assume that $T(y)$ is bounded, 
\[
E_\theta[
\max_{\theta' \in B_\epsilon(\theta)}|\hat{J}_{ij}(\theta,x)|]
\]
and
\[
E_\theta[
\max_{\theta' \in B_\epsilon(\theta)}(\hat{J}_{ij}(\theta,x)|)^2]
\]
are finite.
This means that Assumptions~1 and 1' hold.
By (\ref{eq:hiddenscore}),
\[
\sup_{n \in \mathbb{N}}
\sup_{x^n \in \al^n}
\sup_{\theta \in K}\Bigl|
\frac{1}{n}\frac{\partial \log p(x^n|\theta)}{\partial \theta_i}
\Bigr|
\]
is finite for any compact set $K \subset \Theta^\circ$. That is, 
the log likelihood function devided by $n$ is 
equicontinuous for all $\theta \in K$, $n$, and $x^n$. 
Hence, noting positive definiteness of $J(\theta)$, we can prove the maximum likelihood
estimator is uniformly consistent for all $\theta \in K$
with the tail probability of $o(1/\log n)$. 
In conclusion, all the assumptions for Theorems~1 and 2 are satisfied
and we have demanded lower bound results.

\subsubsection{Upper Bounds for Models with Hidden Variables}
Note that any order's moments of $\bar{T}(Y^n)$ exist
at $p(y^n|x^n,\theta)$.
(This is true even if $T(x)$ is not bounded.)
Hence, 
by (\ref{eq:empiricalFisherofmwh})
of Lemma~\ref{lemma_for_mdh},
Assumption~7 holds for any $b >0$.
Further,
we can show that
$|\hat{J}(x^n,\theta)|$
and
$|{\partial \hat{J}(\theta,x^n)/\partial \theta_k}|$ is uniformly
bounded for all $x^n$ such that $\hat{\theta} \in K$,
which means all the assumptions for the upper bound results hold.
In fact, by (\ref{eq:hiddenthird}),
we have
\[
\frac{\partial \hat{J}_{ij}(\theta,x^n)}{\partial \theta_k}
=
\frac{\partial G_{ij}(\theta)}{\partial \theta_k} - n^2 \tilde{w}_{ijk}(\theta,x^n).
\]
Since the posterior density $p(y^n|x^n,\theta)$
is i.i.d.,
$n^2\tilde{w}(\theta,x^n)$ equals
the third order moment of $T(y)$,
which does not depend on $n$.
Recalling we assume $T(y)$ is bounded,
$|\partial \hat{J}_{ij}(\theta,x^n)/\partial \theta_k|$
is bounded for $\theta \in K$.

Hence, we see that our general minimax strategy
works for models with hidden variables
assuming that $T(y)$ is bounded and $z^tT(y)$ for any nonzero $z$
is not independent of $x$ for all $\theta \in \Theta$.



\subsubsection{Extension to the Whole Strings Set Case for
Mixture Families}

Next, we consider the mixture family case
defined as (\ref{eq:mixturefamily}).
Here, we let $p_y$ denote $\kappa(x|y)$.

Recall the assumption that
$z^tT(y)$ and $x$ are not independent 
to each other for all $z \in \mathbb{R}^{d}\setminus \{ 0\}$
and for all $\theta$.
This condition is equivalent to the positive definiteness
of the Fisher information matrix at each $\theta \in \Theta^\circ$.
Note that
\begin{align}\nonumber
\hat{J}_{ij}(\theta,x)
&=
  -\frac{\partial^2 \log p(x|\theta) }{\partial \theta_i \partial
  \theta_j}\\ \label{mixture_hessian}
&=
\frac{(p_i(x)-p_0(x))(p_j(x)-p_0(x))}{(p(x|\theta))^2}.
\end{align}
Hence, for $z \in \Re^d$,
\begin{align}\nonumber
z^t\hat{J}(\theta,x)z 
&= \frac{\sum_{ij} z_iz_j (p_i(x)-p_0(x))(p_j(x)-p_0(x)) }{p(x|\theta)^2}\\ \label{mixture_quadra}
&= \frac{(\sum_{i} z_i(p_i(x)-p_0(x)))^2 }{p(x|\theta)^2} \geq 0
\end{align}
holds for all $\theta \in \Theta^\circ$,
that is, $\hat{J}(\theta,x^n)$
is semi-positive definite for all $\theta \in \Theta^\circ$.
In fact, we can show it is positive definite
as follows.

When $x^n$ is drawn according to $p(x|\theta)$,
\begin{align*}
\hat{J}_{ij}(\theta,x^n)
=
\frac{1}{n}
\sum_{t=1}^n 
\frac{p_i(x_t)-p_0(x_t)}{p(x_t|\theta)}
\frac{p_j(x_t)-p_0(x_t)}{p(x_t|\theta)}
\end{align*}
converges to the positive definite matrix
$J(\theta)$ in probability.
Now, assume that $\hat{J}(\theta,x^n)$
is positive definite for a certain $n$.
This assumption implies that the rank of the 
$d \times n$
matrix whose $(i,t)$-entry is
\[
\frac{p_i(x_t)-p_0(x_t)}{p(x_t|\theta)}
\]
equals $d$. That is equivalent to the proposition that the rank 
of the matrix whose $(i,j)$-entry is
\[
\frac{p_i(x_t)-p_0(x_t)}{p(x_t|\theta')}
\]
equals $d$ for any $\theta'$.
This implies that $\hat{J}(\theta',x^n)$ is positive definite
for all $\theta'$ and the log likelihood function is strictly concave.
Since the rank of this matrix is non-decreasing in $n$,
the proposition holds for all $n' \ge n$.
Hence, we can conclude that the MLE is consistent.
Therefore, we can apply Theorem~1 for $K = \Theta$ case. 

Unfortunately, 
we can construct an example for which
the convergence of $\hat{J}(\theta,x^n)$
is not uniform for $\theta \in \Theta^\circ$.
Hence, 
we did not determine 
whether
Assumption~3' holds or not for
the mixture family for the whole parameter space.
Hence, it is open whether
Theorem~\ref{gen:thm:lower:new} is applicable for this case
or not.

Now we consider the upper bound for the case that ${\mathcal K}= \al^n$.
Define a subset $\Theta_\tau$ of $\Theta$ as
\begin{eqnarray*}
\Theta_{\tau} = \{ \theta \in \Theta : \, \theta_i \geq \tau, \, 
i=0,1,\ldots, d\}.
\end{eqnarray*}
Since $\Theta_\tau$ for $\tau > 0$
is a compact subset of $\Theta^\circ$,
it may be possible
to design a minimax strategy for
the model $\{ p(x|\theta) : \theta \in \Theta_\tau  \}$
utilizing the result for the general case.
In fact, we can do it as we show later.

Then we argue the situation that $\tau$ converges to zero as $n$
goes to infinity. 
Under that situation we utilize Lemmas~2 and 3.
Then we have to control
the behavior of $\kappa_J(K)$ and $\lambda^*_K$
(the maximum of the largest eigenvalue of $J(\theta)$ among $\theta \in K$), 
since $K$ changes as $n$ increases.

To treat the strings with the maximum likelihood estimate being near
the boundary, we employ the technique introduced 
in \cite{xb96b},
which utilizes the Dirichlet($\alpha$) prior 
$
w_{(\alpha)}(\theta) \propto
\prod_{i=0}^d \theta_i^{-(1-\alpha)}
$
with $\alpha < 1/2$.
Note that this prior with $\alpha=1/2$
has the same form as the Jeffreys prior for the multinomial
model. 
When $\alpha < 1/2$
it has higher density than the Jeffreys prior as $\theta$ approaches the
boundary of $\Theta$.

Our asymptotic minimax strategy is the mixture
\begin{eqnarray}\label{solution}
  m_n(x^n) &=&
(1-2n^{-r})
m_{\Theta}(x^n)\\ \nonumber
&+&n^{-r}
\int \! \! p_e (x^n|\theta,\beta)\bar{w}(\theta,\beta)d\theta d\beta \\  \nonumber
&+&n^{-r}
\int \! \! p(x^n|\theta)w_{(\alpha)}(\theta)d\theta,
\end{eqnarray}
where $m_\Theta$ is the Jeffreys mixture over $\Theta$.
Here, the first and second terms are for the strings with the MLE
being away from the boundary, while the third term works when the MLE
approaches the boundary.

We can show the following theorem.
\begin{theorem}\label{mixturetheorem}
  The strategy $m_n$ defined as (\ref{solution}) 
  with $r < (1/2 -\alpha)(1-p)$
asymptotically achieves the
minimax regret for the mixture family, i.e.\
\[
\sup_{x^n \in \al^n}\log\frac{p(x^n|\hat{\theta})}{m_n(x^n)}
\leq \frac{d}{2}\log\frac{n}{2\pi}
+\log C_J(\Theta) +o(1).
\]
\end{theorem}



To prove Theorem~\ref{mixturetheorem}, we use the following useful inequalities
for the model with hidden variables. 
Here $A \leq B$ for two matrices $A$ and $B$ means that
$B-A$ is positive semidefinite.
\setcounter{forlemmamonotone}{\thelemma}
\begin{lemma}\label{ineq_for_mhv}
\input{lemma_ineq_monotone}
\end{lemma}

The proof is in Appdendix~\ref{proof_of_lemma_ineq_for_mhv}.

Note that,
for
(\ref{mixture_hessian}) we have
\begin{eqnarray*} 
\Bigl| \frac{p_i(x)-p_0(x)}{p(x|\theta)}\Bigr|
\leq
\frac{p_i(x)+p_0(x)}{\sum_{i=0}^d \theta_i p_i(x)}
\leq \frac{1}{\theta_i}+\frac{1}{\theta_0}.
\end{eqnarray*}
Hence $\hat{J}_{ij}(\theta,x^n)$ is 
not more than $2 \max_{\theta \in K}\max_i\theta_i^{-1}$,
which is finite.
Further, $\partial V(x^n|\theta)/\partial \theta_i$ is similarly
bounded, so $V(x^n|\theta)$ is 
equicontinuous for all $x^n$ : $\hat{\theta}
\in K$.
This implies the general mixture strategy works for mixture families
with a compact $K$ interior to $\Theta$.

Next, we give a discussion on the case with the set of strings
$\{ x^n : \hat{\theta}(x^n) \in {\Theta}^\circ \}$.
We are to examin Assumptions~\ref{assume:process:upper:2:for_all}
and \ref{assume:process:upper:2:for_good}
for $\theta \in \Theta_\tau$.

Because of (\ref{mixture_quadra}),
 we have
\begin{align}\label{semi_positive}
z^t\hat{J}(\theta,x^n)z\geq  0
\end{align} 
for all $z \in \Re^d$ and for all $\theta \in \Theta$.


Since
\[
\frac{\partial \hat{J}_{ij}(\theta,x)}{\partial \theta_k}
=
\frac{-2(p_k(x)-p_0(x))}{(p(x|\theta))}\hat{J}_{ij}(\theta,x),
\]
we have
\begin{eqnarray}\label{mfsection:derivative2} 
\frac{\partial z^t \hat{J}(\theta,x)z}{\partial \theta_k}
=
\frac{-2(p_k(x)-p_0(x))}{(p(x|\theta))}z^t\hat{J}(\theta)z,
\end{eqnarray}
which yields
\begin{align}\label{mfsection:derivative}
& \frac{\partial z^t \hat{J}(\theta,x^n)z}{\partial \theta_k}\\ \nonumber
= &
\frac{1}{n}
\sum_t \frac{-2(p_k(x_t)-p_0(x_t))}{(p(x_t|\theta))}z^t\hat{J}(\theta,x_t)z.
\end{align}
Hence we have for $\theta \in \Theta_\tau$,
\begin{eqnarray*}
 \frac{\partial z^t \hat{J}(\theta,x^n)z}{\partial \theta_k}
\leq \frac{1}{n}\sum_t \frac{2z^t\hat{J}(\theta,x_t)z}{\tau}
=
 \frac{2z^t\hat{J}(\theta,x^n)z}{\tau}
\end{eqnarray*}
By this inequality and the fact $z^t \hat{J}(\tilde{\theta},x^n)z \ge 0$,
we have the following inequalities.
\begin{align}\label{mixture:jhatbound}
e^{-2\sqrt{d}|\tilde{\theta}-\hat{\theta}|/\tau} z^t \hat{J}(\hat{\theta},x^n)z
& \le 
z^t \hat{J}(\tilde{\theta},x^n)z \\ \nonumber
\le & \,
e^{2\sqrt{d}|\tilde{\theta}-\hat{\theta}|/\tau}
z^t \hat{J}(\hat{\theta},x^n)z.
\end{align}
By this, we can see that $\bar{h}_K$ for $K = \Theta_\tau$ 
is not more than $e^{2\sqrt{d}}$,
as long as $\epsilon \le \tau$.
Further,  if $|\tilde{\theta}-\hat{\theta}| \leq \tau/2\sqrt{d}$,
\[
z^t \hat{J}(\tilde{\theta},x^n)z \le
z^t \hat{J}(\hat{\theta},x^n)z \Bigl( 1+\frac{e \tau}{2\sqrt{d}}\Bigr)
\]
holds for all $z \in \Re^d \setminus \{ 0 \}$.
Note that
\[
\max_{\tilde{\theta} \in B_\epsilon(\hat{\theta})}|\tilde{\theta}-\hat{\theta}|
= \frac{\epsilon}{ \lambda_{\rm min}},
\]
where $\lambda_{\rm min}$ is the minimum value among $\theta \in \Theta_\tau$ of the smallest eigenvalue of $J(\theta)$.
Hereafter, we assume 
$\epsilon \le \tau \lambda_{\rm min}/ 2\sqrt{d}$.

When $\epsilon \leq \tau \lambda_{\rm min}/2\sqrt{d}$ is satisfied,
\[
z^t \hat{J}(\tilde{\theta},x^n)z \le
z^t \hat{J}(\hat{\theta},x^n)z
\Bigl(
1+\frac{e \epsilon}{\lambda_{\rm min}}
\Bigr)
\]
holds for all $\tilde{\theta} \in B_\epsilon(\hat{\theta})$.
This implies Assumption~\ref{assume:process:upper:2:for_good} holds with
$\kappa_J = e/\lambda_{\rm min}$ for 
$\epsilon \leq \tau \lambda_{\rm min}/2\sqrt{d}$.
Note that the argument here works even if $\tau$ and $\epsilon$ depend on $n$.
%
%

Next we examine Assumption~\ref{assume:process:upper:3}.
Concerning this matter, we have the following lemma.
\begin{lemma}\label{lemma:mixture_assume10}
Assume that $||V(x^n|\hat{\theta})||_s \ge \delta$,
$\delta \ge 8 \sqrt{d}\epsilon/\tau$,
$4\sqrt{d}\epsilon/\tau \le 1/2$,
and $\delta \le 1/2$.
Then, we have for $\tilde{\theta} \in B_\epsilon(\hat{\theta})$
    \[
    ||T(x^n|\tilde{\theta})||_s
\geq ||T(x^n|\hat{\theta})||_s/4.
    \]
\end{lemma}

{\it Proof:}
Note that there exists a unit vector $\bar{z} \in \Re^d$ such that
$
|\bar{z}^t V(x^n|\hat{\theta})\bar{z}| = ||V(x^n|\hat{\theta})||_s
$.
Here we have two cases;
i) $\bar{z}^t V(x^n|\hat{\theta})\bar{z} = ||V(x^n|\hat{\theta})||_s$
and
ii) $-\bar{z}^t V(x^n|\hat{\theta})\bar{z} = ||V(x^n|\hat{\theta})||_s$.

First consider the case i), for which we have
\begin{align*}
||V(x^n|\hat{\theta})||_s
&=
\bar{z}^t J(\hat{\theta})^{-1/2} \hat{J}(\hat{\theta})
J(\hat{\theta})^{-1/2}\bar{z} -1\\
&=
\frac{\bar{z}^t J(\hat{\theta})^{-1/2} \hat{J}(\hat{\theta})
J(\hat{\theta})^{-1/2}\bar{z}}{\bar{z}^t\bar{z}} -1\\
&=
 \frac{\tilde{z}^t 
   \hat{J}(\hat{\theta})\tilde{z}}{\tilde{z}^t J(\hat{\theta})\tilde{z}}-1,
\end{align*}
where $\tilde{z}$ denotes $J(\hat{\theta})^{-1/2}\bar{z}$.
Hence,
\begin{align}\label{eq:ztilde}
 \frac{\tilde{z}^t  \hat{J}(\hat{\theta})\tilde{z}}{\tilde{z}^t J(\hat{\theta})\tilde{z}}
=
1 + ||V(x^n|\hat{\theta})||_s
\end{align}
holds.
For the numerator in the left side, 
from (\ref{mixture:jhatbound}) we have 
\[
\tilde{z}^t  \hat{J}(\tilde{\theta})\tilde{z}
\geq
e^{-2\sqrt{d}|\tilde{\theta}-\hat{\theta}|/\tau}
\tilde{z}^t  \hat{J}(\hat{\theta})\tilde{z}.
\] 
As for the denominator, similarly as (\ref{mixture:jhatbound})
we can show by (\ref{mfsection:derivative2}) 
\begin{eqnarray}\label{mixture:jbound}
e^{-2\sqrt{d}|\tilde{\theta}-\hat{\theta}|/\tau}
\leq
\frac{ \tilde{z}^t J(\tilde{\theta})\tilde{z}}{ \tilde{z}^t  J(\hat{\theta})\tilde{z} } \leq
e^{2\sqrt{d}|\tilde{\theta}-\hat{\theta}|/\tau}
\end{eqnarray}
for $\tilde{\theta} \in \Theta_\tau$.
Then, we have for all $\tilde{\theta} \in B_\epsilon(\hat{\theta}) \cap \Theta_\tau$,
\begin{align*}
 \frac{\tilde{z}^t  \hat{J}(\tilde{\theta})\tilde{z}}{\tilde{z}^t J(\tilde{\theta})\tilde{z}}
&\ge (1 + ||V(x^n|\hat{\theta})||_s )e^{-4\sqrt{d}|\tilde{\theta}-\hat{\theta}|/\tau}\\
&\ge (1 + ||V(x^n|\hat{\theta})||_s )e^{-\delta/2}\\
& \ge (1 + ||V(x^n|\hat{\theta})||_s )(1-\delta/2)\\
& = 1 +||V(x^n|\hat{\theta})||_s - \delta/2 -\delta ||V(x^n|\hat{\theta})||_s/2\\
& \ge 1 +||V(x^n|\hat{\theta})||_s/2 -\delta ||V(x^n|\hat{\theta})||_s/2 \\
& = 1 +(1-\delta)||V(x^n|\hat{\theta})||_s/2\\
& = 1 +||V(x^n|\hat{\theta})||_s/4.
\end{align*}
Letting $\xi = J(\tilde{\theta})^{1/2}\tilde{z}/|J(\tilde{\theta})^{1/2}\tilde{z}|$,
that is, $\tilde{z} = |J(\tilde{\theta})^{1/2}\tilde{z}|J(\tilde{\theta})^{-1/2}\xi$,
we have
\begin{align*}
\frac{\tilde{z}^t \hat{J}(\tilde{\theta})\tilde{z}}
{\tilde{z}^t J(\tilde{\theta})\tilde{z}} 
& =
\frac{\xi^t J(\tilde{\theta})^{-1/2} \hat{J}(\tilde{\theta})J(\tilde{\theta})^{-1/2}\xi}
{\xi^t \xi}\\
& = \xi^t J(\tilde{\theta})^{-1/2} \hat{J}(\tilde{\theta})J(\tilde{\theta})^{-1/2}\xi\\
&\le 1+ ||V(x^n|\tilde{\theta})||_s
\end{align*}
Hence,
we have
$
||T(x^n|\tilde{\theta})||_s
\geq ||T(x^n|\hat{\theta})||_s/4
$.

For the case ii), we can show the same conclusion.
{\it This completes the proof.}

{\it Remark:}
This means that Assumption~10 holds with $\zeta = 1/4$.

To see Assumption~\ref{assume:process:upper:2:for_all} 
holds
is easy because of (\ref{mixture:jhatbound}).

Finally, we evaluate $a =
\min \{1/ 2 \lambda^*, 1  \}$ in Lemma~\ref{newlemma3}.
Since $|\hat{J}_{ij}(\theta)|$ is bounded by $4/\tau^2$ for $\theta
\in \Theta_\tau$ and since the smallest eigenvalue of $J(\theta)$ is
lower bounded by a certain positive constant,
$\lambda^*$
is of order $\tau^{-4}$.
Hence, $a$ is lower bounded by $\tau^4$ times a
certain constant.

Now, we will give the proof of Theorem~\ref{mixturetheorem}.

{\it Proof of Theorem~\ref{mixturetheorem}:}
We divide the strings to the following three categories:
\begin{align*}
    A_n &= \{x^n : \hat{\theta} \in \Theta_\tau \text{ and } ||V(x^n|\hat{\theta})||_s \le \delta \},\\
        B_n &= \{x^n : \hat{\theta} \in \Theta_\tau \text{ and } ||V(x^n|\hat{\theta})||_s > \delta \},\\
        C_n &= \{ x^n : \hat{\theta} \not\in \Theta_\tau \}.
\end{align*}
Set
$ \tau = n^{-(1-p)}$,
$ \delta = n^{-1/2+\gamma}$, and
$ \epsilon = n^{-1/2 + \iota}$,
assuming 
$1/2 + \iota < p < 1$, 
$0< \gamma < 1/2$,
and
$1-p < \gamma/2$, that is,
$2(1-p) < \gamma < 1/2$.
Note that $\epsilon/\tau = n^{-1/2 + \iota +1 -p}= n^{-(p-1/2-\iota)} \rightarrow 0$
as $n$ goes to infinity under this setting.
Let $\tilde{\epsilon}=\epsilon/\tau$
and 
assume $n$ is large enough
so that $\tilde{\epsilon}=\epsilon /\tau \leq \lambda_{\rm min}/2\sqrt{d}$ holds.
Let $C_1 = 2\sqrt{d}/\lambda_{\rm min}$.
Then, $\tilde{\epsilon} \le 1/C_1$.

When $x^n$ belongs to $A_n$, 
the Laplace approximation for the integral with respect to $\theta$
is correct enough.
In fact by (\ref{mixture:jhatbound}), 
\begin{align*}
    z^t \hat{J}(\tilde{\theta},x^n)z 
&\le 
e^{2\sqrt{d}|\tilde{\theta}-\hat{\theta}|/\tau}
z^t \hat{J}(\hat{\theta},x^n)z\\
&\le 
e^{2\sqrt{d}\epsilon/\lambda_{\rm min}\tau}
z^t \hat{J}(\hat{\theta},x^n)z\\
&=
e^{C_1 \tilde{\epsilon}}
z^t \hat{J}(\hat{\theta},x^n)z
\end{align*}
holds for all $\tilde{\theta} \in B_\epsilon(\hat{\theta})$.
Hence for $x^n \in A_n$,  similarly as the proof of Lemma~\ref{newlemma1},
we have
\begin{align*}
\log \frac{p(x^n|\hat{\theta})}{m_{\Theta}(x^n)}
&\leq
\frac{d}{2}\log\frac{n}{2\pi} 
+\log
\frac{|J(\hat{\theta})|^{1/2}}{w_J(\hat{\theta})
\Phi(U_{\Theta,\epsilon,n}(\hat{\theta}))}\\
&+\log\frac{e^{dC_1 \tilde{\epsilon}/2}(1+\delta)^{d/2}}{1-\eta_{\epsilon}}.
\end{align*}
Here, since $B_{\epsilon}(\hat{\theta}) \subset \Theta$ holds when $n$ is large,
$\Phi(U_{\Theta,\epsilon,n}(\hat{\theta}))$ converges to $1$ as
$n$ goes to infinity, the above means
\begin{align*}
\log \frac{p(x^n|\hat{\theta})}{(1-2n^{-r})m_{\Theta}(x^n)}
&\leq
\frac{d}{2}\log\frac{n}{2\pi} 
+\log C_J(\Theta)
+ o(1).
\end{align*}

Next, we consider the case that $x^n \in B_n$.
By Lemma~\ref{lemma:mixture_assume10}, 
Assumption~\ref{assume:process:upper:3} 
holds with $\zeta = 1/4$.
Then, similarly as the proof of Lemma~\ref{newlemma3},
we can show (\ref{eq:newlemma3})
with 
$\bar{m}(x^n) = \int \! \! p_e (x^n|\theta,\beta)\bar{w}(\theta,\beta)d\theta d\beta$,
$\zeta = 1/4$
and $\bar{h}_K = \bar{h}_{\Theta} < \infty$.
Hence, we can show the same inequality as (\ref{eq:upper_for_not_good})
for $x^n \in B_n$.
In this case, with a certain $C>0$, we have 
\[
a n^{2\gamma} \ge C \tau^4 n^{2\gamma} = C n^{2\gamma -4(1-p)},
\]
which is polynomially large, since we set $1-p < \gamma/2$.
This makes the regret of $m_n$ go to negative infinity as $n$ goes to infinity.

Finally, we consider the case that $x^n \in C_n$, in which
there exists $y$ such that
$\hat{\theta}_y \le \tau = n^{-(1-p)}$.
Then, 
we can use the help from the third term in (\ref{solution})
as follows.
Note that the following holds from (\ref{forBer}).
\[
\frac{\int \! \! p(x^n|\theta)w_{(\alpha)}(\theta)d\theta}
{p(x^n|\hat{\theta})}
\geq
\frac{
{\int\prod_{i=0}^d\theta_i^{n\hat{\theta}_i}w_{(\alpha)}(\theta)d\theta}}
{\prod_{i=0}^d {\hat{\theta}_i^{n\hat{\theta}_i}}}.
\]
Then, by Lemma~4 of \cite{xb96b},
the right side is not less than
\[
\frac{1}{R n^{{d/2} -( {1/2}-\alpha )(1-p) }
}
\]
for $x^n \in C_n$,
where $R$ is a constant depending only on $d$.
Let $r$ be smaller than
$(1/2-\alpha )(1-p)$,
then for large $n$
the third term 
of the right side of (\ref{solution}) is larger than
$
{(2\pi)^{d/2}/(C_J(\Theta) n^{d/2})}.
$
This provides an upper bound on regret of $m_n$
smaller than the maximin value for all $x^n \in C_n$.
{\it This completes the proof.}
%
%

\subsection{Mixture with Fixed Weights}

We consider the following model.
\[
p(x|\theta) = 0.5 g_0(x) + 0.5 g_1(x|\theta),
\]
where $q(x|\theta)$ is an exponential family with the natural parameter $\theta$:
 $g_1(x|\theta) 
 = \exp(\theta \cdot T(x) - \psi(\theta))$,
 and $g_0(x)$ is a certain fixed probability density function.
This is another example of non-exponential families.
Let $\eta$ denote the expectation parameter for $g_1(x|\theta)$.
For this case, we assume $T(x)$ is bounded again. 

We consider the lower and upper bounds for this model with
the compact parameter space $K$ interior to $\Theta$. 
Then, it is easy to see the assumptions for lower bound
holds and we can use Theorems~\ref{gen:thm:lower}
and
\ref{gen:thm:lower:new}.

We have
\begin{align*}
    \frac{\partial \log p(x|\theta)}{\partial \theta_i}
    &=\frac{0.5 (T_i - \eta_i)g_1(x|\theta)}{p(x|\theta)}\\
    &= (T_i - \eta_i)r(x|\theta),
\end{align*}
where we have defined
\[
r(x|\theta)=\frac{0.5 g_1(x|\theta)}{p(x|\theta)}.
\]
Note that $0 \le r(x|\theta) \le 1$ 
and
\[
\int r(x|\theta)p(x|\theta)\nu(dx) = 0.5
\]
hold. 
The score function for $x^n$ is
\begin{align*}
    \frac{\partial \log p(x^n|\theta)}{\partial \theta_i}
    = \sum_{t=1}^n(T_i(x_t) - \eta_i)r(x_t|\theta),
\end{align*}

We have
\begin{align*}
\frac{\partial r(x|\theta)}{\partial \theta_i}
= &
\frac{0.5 (T_i-\eta_i)g_1(x|\theta)}{p(x|\theta)} \\
& -
\frac{0.5 g_1(x|\theta)0.5(T_i(x)-\eta_i)g_1(x|\theta)}{(p(x|\theta))^2} \\
 = & (T_i-\eta_i)(r(x|\theta) - r(x|\theta)^2)\\
 = &(T_i-\eta_i)r(x|\theta) (1- r(x|\theta))\\
  = &(T_i-\eta_i)q(x|\theta).
\end{align*}
where $q(x|\theta)=r(x|\theta)(1-r(x|\theta))$.
Hence,
\begin{align*}
    \frac{\partial^2 \log p(x|\theta)}{\partial \theta_j \partial \theta_i}
    = & -G_{ij}(\theta)r(x|\theta) \\
    &+ (T_i-\eta_i)(T_j-\eta_j)q(x|\theta),
\end{align*}
by which, we have
\begin{align}\label{eq:empricalFisher:mfw}
     \hat{J}(\theta,x^n)
    = & \, G_{ij}(\theta)\bar{r} \\
    - & \frac{1}{n}\sum_{t=1}^n (T_i(x_t)-\eta_i)(T_j(x_t)-\eta_j)q(x_t|\theta),
\end{align}
where $\bar{r}=(1/n)\sum_{t=1}^n r(x_t|\theta)$.
We also have
\[
J_{ij}(\theta) = \frac{G_{ij}(\theta)}{2} - \int (T_i-\eta_i)(T_j-\eta_j)q(x|\theta)p(x|\theta)\nu(dx).
\]
Hence, $\hat{J}(\theta,x^n) - J(\theta)$ is bounded for any $\theta \in \Theta^\circ$,
which means that Assumption~7 holds for any $b > 0$.
Further, 
since $
\int (T_i-\eta_i)(T_j-\eta_j)p(x|\theta)\nu(dx)
=G_{ij}(\theta)$ and $0 \le q(x|\theta) \le 1/4$,
$J(\theta)$ is positive definite for all $\theta$.

Now we have
\begin{align*}
    \frac{\partial q(x|\theta)}{\partial \theta_i}
    &= (T_i - \eta_i)q(x|\theta)(1-r(x|\theta))\\
    & -(T_i - \eta_i)r(x|\theta)q(x|\theta)\\
    &= (T_i - \eta_i)q(x|\theta)(1-2r(x|\theta)).
\end{align*}
Using the above, we have
\begin{align*}
   & \frac{\partial^3 \log p(x|\theta)}{\partial \theta_k \partial \theta_j \partial \theta_i}\\
    = & -\frac{\partial G_{ij}(\theta)}{\partial \theta_k}r(x|\theta)
    -G_{ij}(\theta)(T_k - \eta_k)q(x|\theta)\\
    & -G_{ik}(\theta)(T_j-\eta_j)q(x|\theta)\\
    & -  (T_i-\eta_i)G_{jk}(\theta)q(x|\theta)\\
    &+ (T_i-\eta_i)(T_j-\eta_j)(T_k - \eta_k)q(x|\theta)(1-2r(x|\theta)).
\end{align*}
Hence, noting $T(x)$ is assumed to be bounded, 
$\hat{J}(\theta,x^n)$ is equicontinuous in $\theta \in \Theta$
for all $n$ and for all $x^n$.
Therefore, we can apply Theorem~5 to show the asymptotic minimax strategy.

\subsection{Contaminated Gaussian Location Families}

In this subsection, we treat a simple case of contamination Gaussian location families,
which is a basic model for robust estimation.

Let us define the model formally below.
Let
\begin{align}\label{eq:contami_model}
    p(x|\theta) = (1-\nu)g_0(x|\theta) + \nu g_1(x|\theta),
\end{align}
where
\begin{align*}
g_0(x|\theta) &= N(\theta,I),\\
g_1(x|\theta) &= N(\theta,s^2I)
\end{align*}
are Gaussian densities over $\mathbb{R}^d$,
$I$ the unit matrix of order $d$,
$\nu$ a small positive number,
and 
$s$ a number larger than $1$.
Here, $(1-\nu)g_0(x|\theta)$ is the main target for estimation and
$\nu g_1(x|\theta)$ represents a noise distribution.
This $p(x|\theta)$ is an example of non-exponential families,
In \cite{MT2020}, MDL estimation based on 
two-stage codes for this model is discussed. 

Note that our setting that $\nu$ and the forms of $g_0$ and $g_1$
are known is the simplest one in robust statistics.
See \cite{Huber2004} for more general settings.

Similarly as the previous subsection,
we consider the lower and upper bounds for this model with
the compact parameter space $K$ interior to $\Theta$. 
Then, the assumptions for lower bound seems eqsy to confirm,
we give the discussion about the upper bound.

First, we evaluate the score function and the empirical Fisher information.
Let $\partial_i$ denote $\partial / \partial \theta_i$
and $r = r(x|\theta)$ $= (1-\nu) g_0(x|\theta)/p(x|\theta)$.
We have
\begin{align*}
    \partial_i \log p(x|\theta) &= \frac{\partial_i (1-\nu)g_0(x|\theta) +\partial_i \nu g_1(x|\theta)}{p(x|\theta)}\\
    & = - r (\theta_i -x_i) -\frac{(1-r)(\theta_i -x_i)}{s^2} \\
    & = -(\theta_i - x_i) w,   
\end{align*}
where 
\[
w= w(x|\theta)
= r(x|\theta) +\frac{1-r(x|\theta)}{s^2}
=\Bigl( 1-\frac{1}{s^2}\Bigr)r(x|\theta)+\frac{1}{s^2}.
\]
Hence,
\begin{align*}
    \partial_j \partial_i \log p(x|\theta) 
    & = -I_{ji} w - (\theta_i - x_i)\partial_j w \\
    & = -I_{ij} w - (\theta_i - x_i)\Bigl(1-\frac{1}{s^2}\Bigr)\partial_j r.
\end{align*}
For the last term, we have
\begin{align}\nonumber
    \partial_j r & = -\frac{\partial_j p(x|\theta)}{p(x|\theta)^2}(1-\nu)g_0(x|\theta)
    +  \frac{\partial_j (1-\nu)g_0(x|\theta)}{p(x|\theta)}\\ \nonumber
    & = -r(x|\theta) \partial_j \log p(x|\theta) - r(x|\theta) (\theta_j - x_j)\\ \nonumber
    & = r(x|\theta)( (\theta_j - x_j)w -(\theta_j -x_j)   )\\ \nonumber
    & = r(x|\theta)(\theta_j -x_j)(w-1)\\ \nonumber
    & = r(x|\theta)(\theta_j -x_j)(1-r(x|\theta))\Bigl( \frac{1}{s^2} -1 \Bigr)\\ \label{derofr}
    & = (\theta_j -x_j)q \Bigl( \frac{1}{s^2} -1 \Bigr),
\end{align}
where $q=q(x|\theta)=r(x|\theta)(1-r(x|\theta))$.
Hence, we have
\begin{align*}
    & \partial_j \partial_i \log p(x|\theta) \\
    &= -wI_{ij}  + (\theta_i - x_i)(\theta_j - x_j)\Bigl(1-\frac{1}{s^2}\Bigr)^2q,
\end{align*}
which yields
\begin{align}\label{eq:empiricalFisher:contami:d}
    &\hat{J}_{ij}(\theta,x)\\ \nonumber
    &= wI_{ij} - \Bigl(1-\frac{1}{s^2}\Bigr)^2  (\theta_i - x_i)(\theta_j - x_j)q.
\end{align}
By this, we will evaluate the Fisher information.
First note that $J(\theta)$ dose not depend on $\theta$.
Since $E r = 1$, we have $E w = 1$. Hence, our task is to evaluate 
the expectation of the second term.
Since
\[
r(x|\theta)(1-r(x|\theta))p(x|\theta) \le \min \{ (1-\nu)g_0(x|\theta), \nu g_1(x|\theta)  \}
\]
holds, we have 
\begin{align*}
    \int \! \! (\theta_i - x_i)(\theta_j - x_j)q \, p(x|\theta)dx 
     \le \min \{ 1-\nu, \nu s^2  \}I_{ij}.
\end{align*}
Hence, the following holds.
\[
I \ge
J(\theta) \ge 
\max\{ \nu, 1-\nu s^2 \} I.
\]
The third part is positive definite and equals $I$ if and only if $\nu =0$.
Note that $I$ equals the Fisher information of $g_0(x|\theta)=N(x|\theta,I)$.

The empirical Fisher information $\hat{J}_{1}(\theta) = \hat{J}(\theta,x)$ has an interesting property.
To see it, consider the $d=1$ case.
Assume $x=0$ without loss of generality.
Then, we have
\begin{align}\label{eq:empricalFisher:contami:1}
    \hat{J}_1(\theta)
    = r+\frac{1-r}{s^2} - \Bigl(1-\frac{1}{s^2}\Bigr)^2 \theta^2 q.
\end{align}

Assume
$s$ is very large and $\nu$ is very small.
Then, if $|\theta| < 1$ and $r(0|\theta) \fallingdotseq 1$, then 
$\hat{J}_{1}(\theta)$ nearly equals the empirical Fisher information of $N(0|\theta,1)$, which is $1$.
On the other hand, let us consider the case that $r(0|\theta) \fallingdotseq 0$,
which is realized when $\theta^2$ is large.
In particular, assume that $\theta^2$ is much larger than $\log (s/\nu)$,
then 
\begin{align*}
   \theta^2r(0|\theta) &\le \frac{\theta^2 (1-\nu)g_0(0|\theta)}{\nu g_1(0|\theta)}\\
   &\le \frac{s\theta^2}{\nu} \exp\Bigl(-\frac{\theta^2}{2}+\frac{\theta^2}{2s^2} \Bigr)  \fallingdotseq 0. 
\end{align*}
Hence, $\hat{J}(\theta,0) \fallingdotseq 1/s^2$, when $\theta^2$ is much larger than $\log(s/\nu)$.

The above is reasonable observations and not surprising.
More interestingly, when $\theta$ takes an intermediate value, $\hat{J}_1(\theta)$ can be a large negative value.
Let us see it below.

Let $c$ be a solution to the equation $r(0|\theta)= 1/2$,
which is
\[
\frac{1-\nu}{\sqrt{2\pi}}\exp\Bigl(-\frac{c^2}{2}\Bigr) = \frac{\nu}{\sqrt{2\pi s^2}}\exp\Bigl( -\frac{c^2}{2s^2}\Bigr).
\]
This is modified to
\[
s^2 =\frac{\nu^2}{(1-\nu)^2} \frac{}{}\exp \Bigl( c^2 -\frac{c^2}{s^2}\Bigr).
\]
Hence, we have
\[
c^2 =\Bigl(1-\frac{1}{s^2}\Bigr)^{-1}\Bigl(\log s^2 +\log \frac{(1-\nu)^2}{\nu^2}\Bigr),
\]
which implies $c^2 \ge \log s^2$, because we assume $\nu$ is small.
Therefore, the following holds.
\[
\hat{J}_1(\pm c) \le \frac{1}{2}\Bigl( 1 +\frac{1}{s^2} \Bigr)
-\frac{\log s^2}{4}\Bigl(1-\frac{1}{s^2} \Bigr)^2.
\]
The right side is negative when $\log s^2 \ge 4$.
Further, its absolute value can be arbitrarily large
by letting $s$ be large.

Now, we can show that the MLE is not unique for the 
string $x^2=x_1x_2$ with $x_1 =c$ and $x_2=-c$.
When $s^2$ is large, $\hat{J}(0,x^2)$ is negative,
which means that $\theta = 0$ is not the maximizer of
the log likelihood, because $-\hat{J}(0) > 0$
is the second derivative of the log likelihood at $\theta = 0$.
Hence, the maximum is taken at a certain $\tilde{\theta} \neq 0$.
Then by symmetry, $-\tilde{\theta}$ also achieves the maximum.


Next, we consider $\hat{J}(\theta,x^n)$. From (\ref{eq:empiricalFisher:contami:d})
we can obtain the following representation.
\begin{align}\label{eq:empiricalFisher:contami:d:xn}
    \hat{J}(\theta,x^n)
    = \Bigl( \Bigl( 1-\frac{1}{s^2}\Bigr)\bar{r}+\frac{1}{s^2} \Bigr) I 
    - \Bigl(1-\frac{1}{s^2}\Bigr)^2 \hat{\Sigma},
\end{align}
where we have defined
\begin{align*}
    \bar{r} &= \frac{1}{n}\sum_{t=1}^n r(x_t|\theta),\\
\tilde{\Sigma} & = \frac{1}{n} \sum_{t=1}^n q(x_t|\theta)(x_t -\theta)(x_t - \theta)^t.
\end{align*}
Now, we concern whether (\ref{eq:empiricalFisher:contami:d:xn}) is equicontinuous
or not for $x^n \in {\mathcal K}$.
By
(\ref{derofr}), the derivative of the first term with respect to $\theta_i$
is proportional to $\theta_i - x_i$, which is unbounded.
It may seem to cause a problem, but it is not,
since $\theta_i - x_i$ is with a factor $q(x|\theta)$
which is of order $\exp(- C |\theta -x|^2)$.
As a result, the derivative of the first term is bounded.
This holds for all the terms which contain the factor $\theta_i - x_i$.
Hence, the derivative of the empirical Fisher information is bounded,
which implies it is equicontinuous for all $x^n$ and for all $n$
as functions of $\theta$.

\appendices


\section{On Assumptions for lower bounds of Theorem 1}

Suppose that Assumption~\ref{assume:gen:1} holds. This means that $E_\theta
\sup_{\theta' \in B_{r}(\theta)} |\hat{J}_{ij}(\theta',x)|$ is finite for $r$ not more than $r(\theta)$, for all $i,j$ in $\{1,\ldots,d\}$.
Now $J(\theta)^{-1/2}\hat{J}_{1}(\theta',x)J(\theta)^{-1/2}$ is a linear combination of the $\hat{J}_{ij}(\theta',x)$.
Consequently, 
\begin{align} \label{eq:supremum_sequence}
E_\theta \sup_{\theta' \in B_{\delta}(\theta) } 
\pm ((J(\theta)^{-1/2}\hat{J}(\theta',x)J(\theta)^{-1/2})_{ij} -I_{ij})
\end{align}
is finite for $\delta$ not more than $r(\theta)$.
Also by Assumption~\ref{assume:gen:1}, $\hat{J}(\theta',x)$ is continuous in $\theta'$ for each $x$. Accordingly, as $\delta$ decreases to $0$, the supremum in the above expression convergences monotonically to $\pm ((J(\theta)^{-1/2}\hat{J}(\theta,x)J(\theta)^{-1/2})_{ij} -I_{ij})$ which has zero expectation. 
Then by the monotone convergence theorem, the expected supremum in (\ref{eq:supremum_sequence}) converges to $0$ as $\delta \rightarrow 0$ for each $\theta$ in $\Theta^\circ$.

The plus-minus $\pm$ means that the indicated property holds with each sign choice.  It is the flipped sign case $\sup_{\theta' \in B_{\delta}(\theta) } 
(I_{ij} - ((J(\theta)^{-1/2}\hat{J}(\theta',x)J(\theta)^{-1/2})_{ij})$ that especially matters for our development.  Nevertheless, it is appropriate to see that it can be handled with either sign. One might be tempted to take an absolute value, 
but $|(J(\theta)^{-1/2}\hat{J}(\theta,x)J(\theta)^{-1/2})_{ij} -I_{ij}|$ does not have zero expectation.

On $K$, the assumption that $C_J (K)$ is finite means that the Jeffreys measure is a finite measure on $K$.  Accordingly, by Egorov's Theorem \cite{Bartle66,Bartle80}, the convergent sequences (\ref{eq:Jsecond}), (\ref{eq:Jfirst}), and (\ref{eq:supremum_sequence}) are uniformly convergent to $0$ as $\delta \rightarrow 0$, except for $\theta$ in sets of arbitrarily small measure.  In particular, for any $\epsilon >0$, there is a $\delta >0$ and a good set $G \subset K$, such that for all $\theta$ in $G$ the quantities in (\ref{eq:Jsecond}) and (\ref{eq:Jfirst}) are less than $\epsilon$, the expected suprema in (\ref{eq:supremum_sequence}) are less than $\epsilon/(2d)$ and the Jeffreys measure of the complement of $G$, which is $\int_{K \smallsetminus G}  |J(\theta)|^{1/2} d\theta$, is less than $\epsilon\, C_J (K)$.

Next consider $(J(\theta)^{-1/2} \hat{J}(\theta',x^n) J(\theta)^{-1/2})_{ij}$ for a sample of size $n$. By the definition of the empirical Fisher information, this is 
$$\frac{1}{n} \sum_{t=1}^n  \bigl(J(\theta)^{-1/2} \hat{J}(\theta',x_t) J(\theta)^{-1/2}\bigr)_{ij}.$$  
The law of large numbers informs us that it is near its expectation with high probability. We use the expected supremum property to get 
uniformity over $B_\delta (\theta)$. Indeed, the supremum of an average is less than the average of the supremum. In particular, $\sup_{\theta' \in B_\delta(\theta)} \pm (I_{ij} - (J(\theta)^{-1/2} \hat{J}(\theta',x^n) J(\theta)^{-1/2})_{ij} $ is not more than
$$\frac{1}{n} \sum_{t=1}^n \sup_{\theta' \in B_\delta(\theta)} \pm \bigr(I_{ij} - (J(\theta)^{-1/2} \hat{J}(\theta',x_t) J(\theta)^{-1/2})_{ij}\bigr).$$ 
When the $x_t$ are independent from the distribution $P_\theta$,
this is a sample average having expectation bounded by $\epsilon/(2d)$.  Accordingly, by the law of large numbers, for each $i,j$, and for each sign choice, it is within $\epsilon/d$ of $0$ except in an event of ($P_\theta$) probability tending to $0$ as $n\rightarrow \infty$. Accordingly, taking the exception event as the finite union of these tail events for $1\le i,j\le d$ and for each sign, we have that 
$\max_{i,j} \sup_{\theta' \in B_\delta(\theta)} | (I_{ij} - (J(\theta)^{-1/2} \hat{J}(\theta',x^n) J(\theta)^{-1/2})_{ij} |$ is not more than $\epsilon/d$ except in an event having a ($P_\theta$) probability, say $\delta_n$, which tends to zero as $n \rightarrow \infty$.

Now the spectral norm of a $d$ by $d$ matrix is within a factor $d$ of the maximum absolute value of its entries.  Accordingly,
\begin{align} \label{eq: supremumofaverageJ}
\sup_{|\zeta|=1} \sup_{\theta' \in B_\delta(\theta)} \zeta^t \bigl(I- (J(\theta)^{-1/2} \hat{J}(\theta',x^n) J(\theta)^{-1/2})\bigr)\zeta
\end{align}
is less than $\epsilon$, except in the event of probability $\delta_n$.  
Any unit vector $\zeta$ may be obtained as $J(\theta)^{1/2} z /|\!|J(\theta)^{1/2} z|\!|$ for some non-zero vector $z$. Accordingly, switching from $\zeta$ to $z$, this gives
$$\sup_{\theta' \in B_\delta(\theta)} \inf_{z\ne 0} \frac{z^t \hat{J}(\theta',x^n)z}{z^t J(\theta)z} \ge 1-\epsilon$$ except in the event of vanishing $P_\theta$ probability, asymptotically.

\section{On Assumptions for lower bounds of Theorem 2}

The analysis in the proof of Theorem 2 is similar. The extra assumptions of continuity of the expected supremum function and continuity of the Fisher information are used, along with Dini's Theorem, to deduce that the convergence to $0$ of the expected supremum as well as the convergences to $0$ of the expressions in (\ref{eq:Jsecond}) and (\ref{eq:Jfirst}), are in fact uniform convergences within the whole presummed compact $K$ (with no need for an Egorov style exception set $K\smallsetminus G$). In particular there is a $\delta>0$, sufficiently small, such that the expected value of the supremum in expression (\ref{eq: supremumofaverageJ}) is less than $\epsilon/2$, uniformly over $\theta$ in $K$.

Another distinction is the use of the assumption of the continuous and finite expected square, to permit an appeal to Chebyshev's inequality. The variance of the  supremum in expression (\ref{eq: supremumofaverageJ}) is not more than $v_\theta/n$, where $v_\theta$ is the variance of the supremum when $n=1$.  From Assumption $1^\prime$ this variance is finite and continuous so it has a finite bound $v_K$ on the compact domain $K$.  Accordingly, by Chebyshev's inequality, the probability that expression (\ref{eq: supremumofaverageJ}) exceeds $\epsilon$ is not more than $4 v_K /(n\epsilon^2)$, uniformly over $\theta$ in $K$. This converges to zero faster than the required $1/\log n$.  [A simlar sufficiently fast tail probability conclusion can be arranged if using a $1+\alpha$ moment of the empirical Fisher information for any $\alpha > 0$].

\section{Asymptotic Normality and Uniform Tightness of the Distribution of the MLE}


First we restate and then prove Lemma~2 from Section~IV, which is for general i.i.d.\ families.

\setcounter{forlemma}{\thelemma}
\setcounter{lemma}{\theforlemmaasymptnormality}
\begin{lemma}
    \input{lemma_asympt_normality}
\end{lemma}
\setcounter{lemma}{\theforlemma}


{\it Proof:}
Assume that $x^n$ is drawn from $p(\cdot|\theta)$. 
Let $l_{x^n}(\theta)$ be the log likelihood function $\log p(x^n|\theta)$.
Then, the first order Taylor expansion of the score
function $\nabla l_{x^n}(\theta)$ at $\hat{\theta}$
shows
\begin{align*}
  \nabla l_{x^n}(\theta)
=   \nabla \nabla^t l_{x^n}(\tilde{\theta})(\theta - \hat{\theta}),
\end{align*}
where 
$\nabla f(\theta)$ and
$\nabla \nabla^t f(\theta)$ denote
the gradient and the Hessian of
 $f$, respectively,
and
$\tilde{\theta}$ is a point on the line between $\theta$ and $\hat{\theta}$. 
From the definition of the empirical Fisher information, this equation
is 
\begin{align}
  \nabla l_{x^n}(\theta)
=  n \hat{J}(\tilde{\theta},x^n)(\hat{\theta} - \theta),
\end{align}
which is equivalent to
\begin{align*}
 & J(\theta)^{-1/2}\frac{1}{\sqrt{n}}  \nabla l_{x^n}(\theta) \quad \\
& \quad =  [ J(\theta)^{-1/2} \hat{J}(\tilde{\theta},x^n)J(\theta)^{-1/2} ]
[\sqrt{n}J(\theta)^{1/2}   (\hat{\theta} - \theta)].
\end{align*}
Let's call the random vector on the left side $Z_n$.
It is
asymptotic normal by the central limit theorem (for sums of i.i.d.\ random
vectors with finite covariance) and it has
covariance $EZ_n Z_n^t=J(\theta)^{-1/2}  
I(\theta)
J(\theta)^{-1/2}
$.

By Assumption~$3^\prime$ 
the estimator
$\hat{\theta}$  is in a $\delta$ neighborhood of $\theta$
with probability greater than $1-o(1/\log n)$ uniformly for $\theta
\in K$. 
Then by (\ref{seriesBdash}), which is a consequence of Assumption~1',
$J(\theta)^{-1/2} \hat{J}(\tilde{\theta},x^n)J(\theta)^{-1/2}$
is  close to $I$, indeed not less than $(1-\epsilon)I$,
for all $\theta \in K$ with probability at least $1-o(1/\log n)$.
(The inequality of matrices is in the sense that the difference is a non-negative definite matrix.)
Accordingly, the random vector of interest defined by 
$\xi_n = \xi_n(\theta)=\sqrt{n}J(\theta)^{1/2} (\hat{\theta} - \theta)$
satisfies $\xi_n = A_nZ_n$  where $A_n \le (1-\epsilon)^{-1}I$. It
follows that $E \xi_n \xi_n^t\le (1-\epsilon)^{-2}EZ_nZ_n^t$
and that $E ||\xi_n||^2 \le (1-\epsilon)^{-2}E||Z_n||^2$ which is
$(1-\epsilon)^{-2}
{\rm trace }(J(\theta)^{-1}I(\theta))$, not more than
$(1-\epsilon)^{-2}c$.
Thus by Chebyshev's inequality
\begin{align*}
& \max_{\theta \in K}
P_\theta \bigl ( \sqrt{n}||J(\theta)^{1/2}(\hat{\theta}-\theta)|| >
b \sqrt{\log n}
\bigr)\\
& \quad \le   \frac{(1-\epsilon)^{-2}c}{b^2 \log n}  +o\Bigl(\frac{1}{\log n} \Bigr).
\end{align*} 
This means that $\limsup_n (\log n)\max_{\theta \in K}P_\theta(||\xi_n(\theta)|| \ge
b\sqrt{\log n}) \le {(1-\epsilon)^{-2}c/b^2 }$.
Since this bound holds for all positive $\epsilon$, it follows that
this limsup is less than $c/b^2$, which is the desired result.
{\em This completes the proof of
Lemma 2} 

\par\vspace{0.3cm} \noindent
{\it Remark on Uniform Tightness:}
Consider the family of approximately standardized random variables 
$\xi_n(\theta) = \sqrt{n}J(\theta)^{1/2} (\hat{\theta} - \theta)$.
Convergence in distribution provides a form of stochastic tightness, that is, $P_\theta(||\xi_n(\theta)|| > a)$ 
tends to zero as $a \rightarrow \infty$, uniformly in $n$.
Here we used slightly stronger assumptions to get that
the convergence is also uniformly valid for $\theta \in K$ 
and that moderately sized tails with $a_n = b \sqrt{\log n}$
have tail probability bounded by a small multiple of $1/\log n$.

\section{Consistency of the MLE}

The present setting is for a model in which the $x_1,\ldots,x_n$ given $\theta$ are independent and identically distributed.

Here we state conditions that are sufficient for the consistency of the MLE for each $\theta$. The subsequent section addresses uniform consistency for $\theta$ in compacta.

For consistency, we use Assumptions~\ref{assume:gen:3.1}-\ref{assume:gen:6}
described below.  
The first assumption concerns continuity.

\begin{assumption}\label{assume:gen:3.1} \emph{Continuity.}
For almost every $x \in \al$ and for all $\theta \in \Theta$,
$p(x|\theta)$ is upper half continuous
at $\theta$:
\[
\lim_{\delta \rightarrow 0}
\sup_{\theta': |\theta-\theta'| < \delta}
p(x|\theta') = p(x|\theta).
\]
Moreover, the relative entropy $D(\theta||\theta')$ is a continuous function of $\theta$ and $\theta'$ in $\Theta$.
\end{assumption}

\begin{assumption} \label{assume:gen:5}
For each $\theta$ and $\theta'$ in the interior of $\Theta$, there
is a $\delta > 0$
such that the following holds
\[
E_\theta
\Bigl(
\sup _{\theta'' \in \Theta : |\theta''-\theta'| < \delta}
\log \frac{p(x|\theta'')}{p(x|\theta)} 
\Bigr) < \infty.
\]
\end{assumption}

The above two assumptions are sufficient for consistency when the parameter space $\Theta$ is compact.  Accordingly, for bounded parameter spaces, it is natural to include in the parameter space any boundary points that do correspond to probability densities.

For non-compact parameter spaces, the following additional assumption is used to handle those cases in which the densities lose mass as the parameter approaches boundary points not in $\Theta$.  For $B\ge 1$, we let $\Theta_B$ be an increasing sequence of compact subsets whose union is $\Theta$. 

For one example, consider the family of exponential densities $p(x|\theta)= \theta e^{- \theta x}$ for $x>0$.  It has the natural parameter space $\Theta =\{\theta: \theta >0\}$ and the sets $$\Theta_B =\{\theta: 1/B \le \theta \le B\}$$ are increasing compact sets whose union is $\Theta$. Moreover, as $\theta' \rightarrow 0$ or as $\theta' \rightarrow \infty$ it has $\lim p(x|\theta') =0$ for each $x>0$.  

When the parameter space is $\Theta=R^d$ we may take 
$$\Theta_B = \{\theta: |\theta|\le B\}.$$

\begin{assumption}\label{assume:gen:6}
Let an increasing sequence of compact subsets $\Theta_B$ be given
whose union is $\Theta$.
For each $\theta \in \Theta^\circ$,
there is a $B$ sufficiently large
such that
\[
E_\theta
\Bigl(
\sup_{\theta' \in \Theta \setminus \Theta_B}
\log \frac{p(x|\theta')}{p(x|\theta)} 
\Bigr) < \infty .
\]
Moreover, there is a value $a > 0$, such that, for any sequence $\theta_B$, $B\ge 1$, with $\theta_B \in \Theta \setminus \Theta_B$, in the limit the likelihood ratio satisfies 
$\limsup_B p(x|\theta_B)/p(x|\theta) \le \exp\{-a\}$, for each $x \in \al$.
\end{assumption}

We call this latter property the loss of mass for any sequences diverging from the set. Often, as in the example above, the indicated limit property holds trivially, with full loss of mass $\lim p(x|\theta_B)=0$, if there be sequences $\theta_B$ heading out of compacta in $\Theta$.

For compact $\Theta$, taking $\Theta_B=\Theta$, the set $\Theta \setminus \Theta_B$ is empty and Assumption \ref{assume:gen:6} is regarded as holding vacuously, with the supremum of an empty set taken to be $- \infty$.

Under 
\ref{assume:gen:3.1}, \ref{assume:gen:5}, \ref{assume:gen:6},
the maximum likelihood estimate is
a consistent estimator of $\theta$.
We prove the following.
\begin{lemma}\label{lemma:ce}
  For a family of densities
$S=\{p(\cdot|\theta) : \theta \in \Theta  \} $,
suppose Assumptions \ref{assume:gen:3.1}, \ref{assume:gen:5}, and \ref{assume:gen:6} hold.
Then, for each $\theta$ in $\Theta^\circ$, we have convergence to zero in probability of the relative entropy between the densities at the maximum likelihood estimate $\hat \theta (x^n)$ and at $\theta$. That is,
$$\lim_{n \rightarrow \infty} D(\theta||\hat \theta) = 0, \quad \hbox{in} \:\: P_\theta \:\: \hbox{probability.}$$
Moreover, if $D(\theta||\theta')$ is continuous in $\theta'$, and if $P_{\theta'} = P_\theta$ only at $\theta'=\theta$, then also $\hat \theta (x^n)$ converges to $\theta$ in probability. That is, for every $\epsilon > 0$.
\[
P_\theta (|\hat{\theta}(x^n)-\theta| > \epsilon  ) 
=o(1).
\]
\end{lemma}

 {\it Proof:}
For positive $\xi$,
let $U$ be a subset of $\Theta$ defined as $ U \deff \{ \theta' \in \Theta : D(\theta||\theta')\ge \xi \}$. By the continuity of $D(\theta||\theta')$ as a function of $\theta'$, when $\Theta$ is compact, so also is $U$. When $\Theta$ is not compact, in place of $U$ we use $U \cap \Theta_B$
for which will discuss the size of $B$ later below.
By Assumption~\ref{assume:gen:3.1}, we have, for each $\theta'$ in $U$,
\[
\lim_{\delta \rightarrow 0}
\sup_{\theta'' : |\theta''-\theta'|< \delta}
\log \frac{p(x|\theta'')}{p(x|\theta)}
=
\log \frac{p(x|\theta')}{p(x|\theta)}.
\]
Therefore, we have
\[
\lim_{\delta \rightarrow 0}
E_\theta 
\Bigl(
\sup_{\theta'' : |\theta''-\theta'|< \delta}
\log \frac{p(x|\theta'')}{p(x|\theta)}
 \Bigr)
=\,-\,D(\theta||\theta'),
\]
by the monotone convergence theorem, since 
$
\sup_{\theta'' : |\theta''-\theta'|< \delta}
\log (p(x|\theta'')  / p(x|\theta))
$ decreases as $\delta$ decreases for each $x \in \al$
and since it has finite expectation by Assumption~\ref{assume:gen:5}.

For each $\xi > 0$ and each $\theta' \in U$, let $\delta(\theta')=\delta(\theta',\xi)>0$ 
be so small that
\[
E_\theta 
\Bigl(
\sup_{\theta'' : |\theta''-\theta'|< \delta(\theta')}
\log \frac{p(x|\theta'')}{p(x|\theta)}
\Bigr) 
\leq -D(\theta||\theta')+\xi/2 
\]
holds.
Then, since $D(\theta||\theta') \ge \xi$ in $U$, we have
\[
E_\theta 
\Bigl(
\sup_{\theta'' : |\theta''-\theta'|< \delta(\theta')}
\log \frac{p(x|\theta'')}{p(x|\theta)}
\Bigr) 
\leq -\xi/2. 
\] 
Now consider $\frac{1}{n}
\sup_{\theta'' : |\theta''-\theta'|< \delta(\theta')}
\log \frac{p(x^n|\theta'')}{p(x^n|\theta)}$ for samples of size $n$. Since we are presently treating models in which the $x_1$ through $x_n$ are independent. It is
$$\sup_{\theta'' : |\theta''-\theta'|< \delta(\theta')}
\frac{1}{n} \sum_{i=1}^n \log \frac{p(x_i|\theta'')}{p(x_i|\theta)}.$$
Then since the supremum of the average is less than the average of the supremum, we can apply the law of large numbers and deduce that it has a limit in ($P_\theta$) probability not greater than $-\xi/2$, for each $\theta'$ in $U$.  We will need to arrange an analogous property holding uniformly over $\theta'$ in $U$, and we will come back to that matter momentarily.

For non-compact domains $\Theta$, consider
the sequence $\sup_{\theta' \in \Theta \setminus \Theta_B} \log \frac{p(x|\theta')}{p(x|\theta)}$. It is decreasing in $B$ and it has limit less than or equal to $-a$ as $B\rightarrow \infty$, by Assumption \ref{assume:gen:3.1}.  
Then with $0 \le \xi \le a$, by the monotone convergence theorem, we can arrange that $B$ is sufficiently large that
\[
E_\theta
\Bigl(
\sup _{\theta' \in \Theta \setminus \Theta_B }
\log \frac{p(x|\theta')}{p(x|\theta)} 
\Bigr) < -\xi/2.
\]
Accordingly, again using that a supremum of an average is less than the average of a supremum,
we can deduce that $\frac{1}{n} \sup_{\theta' \in \Theta\setminus \Theta_B}
\log \frac{p(x^n|\theta'')}{p(x^n|\theta)}$ has a limsup in probability not greater than $-\xi/2$.

To obtain that $D(\theta||\hat \theta)$ is less than $\xi$ it is enough to confirm that the likelihood at $\theta$ is higher than the supremum of the likelihoods for all $\theta'$ with $D(\theta||\theta')\ge \xi$. Indeed, this gives an instance of a parameter value in the Kullback ball (namely at the center point $\theta$) that has higher likelihood than at all points outside the Kullback ball. Accordingly the maximum must be in that ball.  So if suffices for Kullback consistency to show that with high probability 
$$\sup_{\theta' : D(\theta||\theta') \ge \xi} \log \frac{p(x^n|\theta')}{p(x^n|\theta)} < 0.$$
In the compact $\Theta$ case, this supremum is over all $\theta'$ in $U$. Now the compact $U$ is covered by the (infinite) union of all the variable radius balls $\{\theta'':|\theta''-\theta'|\le \delta(\theta')\}$ for $\theta'$ in $U$.  For a compact set every cover has a finite subcover.
Thus, there exist a finite number of points
$\theta_1, \theta_2,...,\theta_N$ such that
the family of sets
${\cal U} = \{ U_i \} $ ($i=1,2,...,N $) 
covers $\bar{U}$, where $U_i \deff \{ \theta : |\theta -\theta_i| < \delta(\theta_i)\}$.  

Similarly, in the non-compact $\Theta$ case, we obtain such a cover of the compact $U \cap \Theta_B$ 
and append the set $U_0= \{\theta'\in \Theta \setminus \Theta_B : D(\theta||\theta') \ge \xi\}$  to obtain thereby $U_0 \cup U_1 \cup \ldots \cup U_N$ as a cover of $\{\theta' \in \Theta : D(\theta||\theta') \ge \xi\}$.

Now, we have
\begin{eqnarray*}
&&\sup_{\theta' : D(\theta||\theta) > \xi} \left(
\frac{1}{n}
\log \frac{p(x^n|\theta')}{p(x^n|\theta)} \right) \\
&\leq&
\max_{0\le k \le N} \left(
\frac{1}{n}
\sup_{\theta' \in U_k} 
\log \frac{p(x^n|\theta')}{p(x^n|\theta)} \right) \\
&\leq&
\max_{0\le k \le N} \left( \frac{1}{n} \sum_{i=1}^n \sup_{\theta' \in U_k} 
\log \frac{p(x_j|\theta')}{p(x_j|\theta)} \right). \\
\end{eqnarray*}
Now as we showed each of these finitely many averages 
$\frac{1}{n} \sum_{i=1}^n \sup_{\theta' \in U_k} 
\log \frac{p(x_j|\theta')}{p(x_j|\theta)}$ 
is less than $0$ except in an event $E_{k,n}$ of probability 
$P_\theta (E_{k,n})$ which tends to zero as $n\rightarrow \infty$.  
Accordingly, the finite union of these exception sets $E_n = \cup_{k=0}^N U_{k,n}$ 
has probability $P_\theta (E_n)$ tending to zero as well.  Outside $E_n$, we have
$\sup_{\theta' : D(\theta||\theta) > \xi} 
\log \frac{p(x^n|\theta')}{p(x^n|\theta)} < 0$. 

So for each $\xi > 0$, the maximum likelihood estimate
is in the Kullback ball of size $\xi$ except in an event of probability tending to zero 
as $n\rightarrow \infty$.  This means that $D(\theta||\hat \theta)$ converges to $0$ in probability.

Now for any $\epsilon >0$, with $D(\theta||\theta')$ continuous in $\theta'$ and zero only at $\theta'=\theta$, it follows that in any compact $K \subseteq \Theta$ (such as $K=\Theta_B$)
$$\inf_{\theta' \in K: |\theta'-\theta| \ge \epsilon} D(\theta||\theta') > 0$$
Moreover, for non-compact $\Theta$ the condition on the limit of $p(x|\theta')/p(x|\theta)$ as $\theta'$ diverges from compacta, prevents $D(\theta||\theta')$ from tending to zero for such sequences.  Accordingly, the convergence of $D(\theta||\hat \theta)$ to zero (in probability) implies $|\hat \theta(x^n) - \theta| \rightarrow 0$ (in probability).
{\it This completes the proof of the Lemma.}

\par\vspace{0.3cm}\noindent
{\emph Remarks}:  The first consistency proof of this type is in Wald \cite{wald1949}.
Our conditions are similar, but slightly weaker in that we use suprema of log density ratios (demonstrated close to relative entropies). In contrast Wald use suprema of log densities together with finite entropy conditions.  Secondly, here we organized the proof to exhibit the primacy of the conclusion of information consistency ($D(\theta||\hat \theta)$ tending to zero in probability) with the parameter consistency conclusion $\hat \theta \rightarrow \theta$ of Wald as a consequence. 

Extension to parameter spaces that are separable metric spaces is straightforward, provided the finite balls of radius $B$ are compact (so that covers have finite subcovers). Extension to almost sure convergence is also straightforward.
Also, the conclusion extends to the case that the governing distribution $P$ is not in the family $\{P_\theta:\theta \in \Theta\}$, but is an information limit of a sequence of members of the family (that is, there exists $P_{\theta_k}$ with $D(P||P_{\theta_k})$ tending to $0$ as $k$ grows).

\section{Consistency Conditions and the Shtarkov Value}

Recall that the minimax regret value is defined by $r_n(K)=\min_{q}  \sup_{x^n \in {\cal{X}}^n} \sup_{\theta \in K} \log p(x^n|\theta)/q(x^n)$, where the minimum is over $q$ which are non-negative and integrate to not more than $1$ with respect to $\nu$. Per the theory of Shtarkov \cite{shtarkov}, the exact minimax value is $r_n(K)= \log c_{n,K}$ achieved by the normalized maximum likelihood distribution $q(x^n)= \sup_{\theta \in K} p(x^n|\theta)/c_{n,K}$ where $c_{n,K} = \int \sup_{\theta \in K} p(x^n|\theta) \eta(d\theta)$ is the Shtarkov constant.  As explained in the introduction, it is the properties of $r_n(K)$ that are of interest to us here.  In particular, the use of Bayes mixture approximations to the minimax distribution arises from aim of approximate implementation as well as the aim of determining approximations of the Shtarkov value suitable for use in Rissanen's stochastic complexity formulation of the minimum description length (MDL) principle.

In this appendix we show a relationship between the consistency conditions and finiteness of the Shtarkov constant.

Let's focus attention on the case that the parameter space is taken to be a set $K$ (possibly a subset of the whole parameter space) for which $c_{n,K}$ is finite.

For independent random variables, if the total sample size $N=nm$ is a multiple of a block size $m$, one has the option to organize the sequence $x_1,x_2,\ldots,x_N$ as the sequence of independent blocks $\underline x^m_1, \underline x^m_2,\ldots \underline x^m_n$, each of size $m$, where 
$$\underline x^m_i = (x_{(i-1)m+1},x_{(i-1)m+2},\ldots,x_{im}).$$

Accordingly, for $m\ge 1$, the Assumptions \ref{assume:gen:5} and \ref{assume:gen:6} generalize as follows. 

{\emph{Assumption}} \ref{assume:gen:5}[\emph{m}]:
For the specified blocksize $m$ and for each $\theta$ and $\theta'$ in the interior of $K$, there
is a $\delta > 0$
such that the following holds
\[
E_\theta
\Bigl(
\sup _{\theta'' \in \Theta : |\theta''-\theta'| < \delta}
\log \frac{p(x^m|\theta'')}{p(x^m|\theta)} 
\Bigr) < \infty.
\]

{\emph{Assumption}} \ref{assume:gen:6}[\emph{m}]:
Let an increasing sequence of compact subsets $K_B$ be given
whose union is $K$. For the specified $m$ and 
for each $\theta \in K^\circ$,
there is a $B$ sufficiently large
such that
\[
E_\theta
\Bigl(
\sup_{\theta' \in K \setminus K_B}
\log \frac{p(x^m|\theta')}{p(x^m|\theta)} 
\Bigr) < \infty .
\]
Moreover, there is the loss of mass property for sequences diverging from $K$. That is, there is a value $a > 0$, such that, for any sequence $\theta_B$, $B\ge 1$, with $\theta_B \in K \setminus K_B$, in the limit, the likelihood ratio satisfies 
$\limsup_B p(x^m|\theta_B)/p(x^m|\theta) < \exp\{-a\}$, for each $x^m \in \al$.

With these assumptions we have the following consistency conclusion.

\begin{lemma}\label{lemma:ce:m}
For a family of densities
$S=\{p(\cdot|\theta) : \theta \in \Theta  \} $, and a blocksize $m$,
suppose Assumptions \ref{assume:gen:3.1}, \ref{assume:gen:5}[\emph{m}] and \ref{assume:gen:6}[\emph{m}]. Consider the sequence of maximum likelihood estimates $\hat \theta (x^N)$ with sample sizes $N=nm$ a multiple of $m$.
Then, for each $\theta$ in $\Theta^\circ$, we have convergence to zero in probability of the relative entropy between the densities at the maximum likelihood estimate $\hat \theta$ and at $\theta$. That is,
$$\lim_{n \rightarrow \infty} D(\theta||\hat \theta) = 0, \quad \hbox{in} \:\: P_\theta \:\: \hbox{probability.}$$
Moreover, if $D(\theta||\theta')$ is continuous in $\theta'$, and if $P_{\theta'} = P_\theta$ only at $\theta'=\theta$, then also $\hat \theta (x^n)$ converges to $\theta$ in probability. That is, for every $\epsilon > 0$.
\[
P_\theta (|\hat{\theta}(x^n)-\theta| > \epsilon  ) 
=o(1).
\]
\end{lemma}

{\it Proof:} Taking the sequence of sample sizes $nm$ which are a multiple of $m$, the proof is essentially the same as in the $m=1$ case. 

\begin{lemma}
If the Shtarkov value $c_{1,K}$ at $m=1$ is finite, then it implies the satisfaction of assumptions \ref{assume:gen:5} and \ref{assume:gen:6}, and hence (together with the continuity assumption \ref{assume:gen:3.1} and the loss of mass assumption if $K$ is not compact), a finite Shtarkov value implies the full consistency of the MLE, per the conclusions of Lemma \ref{lemma:ce}. Likewise, for $m\ge 1$, finiteness of the Shtarkov value $c_{m,K}$ implies the satisfaction of assumptions \ref{assume:gen:5}[\emph{m}] and \ref{assume:gen:6}[\emph{m}] and, hence (along with the continuity assumption \ref{assume:gen:3.1} and the loss of mass assumption if $K$ is not compact), it implies the consistency of the MLE for the sequence of sample sizes that are multiples of $m$, per the conclusions of Lemma \ref{lemma:ce:m}.
\end{lemma}

{\it Proof:}  The expected suprema in \ref{assume:gen:5}[\emph{m}] and \ref{assume:gen:6}[\emph{m}] are
\[
E_\theta
\Bigl(
\log \frac{\sup_{\theta'' \in K \cap B_{\theta',\delta}} p(x^m|\theta'')}{p(x^m|\theta)} 
\Bigr)
\]
and
\[
E_\theta
\Bigl(
\log \frac{\sup_{\theta' \in K \setminus K_B} p(x^m|\theta')}{p(x^m|\theta)} 
\Bigr)
\]
which by Jensen's inequality have the upper bounds
\[
\log \int \sup_{\theta'' \in K \cap B_{\theta',\delta}} p(x^m|\theta'') \nu(dx^m)
\]
and
\[
\log \int {\sup_{\theta' \in K \setminus K_B} p(x^m|\theta')} \nu(dx^m),
\]
respectively.  Thus (local) domination of the joint densities is sufficient for 
(local) domination of the log likelihood ratios. 

Next, we can further upperbound these by a global dominating quantity (when finite)
\[
\log \int {\sup_{\theta' \in K} p(x^m|\theta')} \nu(dx^m)
\]
which is $\log c_{m,K}$, the Shtarkov minimax regret.  The indicated consistency 
properties then follow when it is finite.

\section{Uniform Consistency of the MLE}

\renewcommand{\theassumption}{\arabic{assumption}$^\prime$}
\setcounter{assumption}{10}

We turn our attention to assumptions for uniform consistency, on compact subsets of $\Theta$, as used in the development of Theorem 2.

Recall that we assumed continuity of the mean of the log density ratio $D=D(\theta||\theta')$ in assumption \ref{assume:gen:3.1}.
Here we will also assume continuity of its variance $$V(\theta||\theta') = E_\theta [(\log p(x|\theta)/p(x|\theta') -D)^2].$$

\begin{assumption} \label{assume:gen:3.1:prime} $V(\theta||\theta')$ is a continuous function of $\theta'$ and $\theta$ in $\Theta$.
\end{assumption} 

\begin{assumption} \label{assume:gen:5:prime}
Within each compact subset of $\Theta$, there exists a $\bar \delta > 0$ such that for
$0 < \delta \le \bar \delta$ the following expressions are finite and continuous as a function of $\theta$ and $\theta'$
\[
E_\theta
\Bigl[
\sup _{\theta'' \in \Theta : |\theta''-\theta'| < \delta}
\log \frac{p(x|\theta'')}{p(x|\theta)} 
\Bigr] 
\]
and
\[
E_\theta \Bigl[
\Bigl(
\sup _{\theta'' \in \Theta : |\theta''-\theta'| < \delta}
\log \frac{p(x|\theta'')}{p(x|\theta)} 
\Bigr)^2 \Bigr].
\]
\end{assumption}

The above assumption \ref{assume:gen:5:prime}, along with \ref{assume:gen:3.1} and \ref{assume:gen:3.1:prime},  is sufficient to show uniform consistency for $\theta$ within any compact subset when the parameter space on which the likelihood is maximized is also compact.  

There can also be interest to know whether uniform consistency within compact subsets $K$ can hold when the parameter space $\Theta$ is not compact.  For that purpose we have the following assumption.

\begin{assumption}\label{assume:gen:6:prime}
Let $K$ be a compact subset of  $\Theta$.
There exists a certain positive number $B$ such that
\[
\sup_{ \theta \in K}
E_\theta
\Bigl[ \Bigl(
\sup _{\theta' \in \Theta \setminus \Theta_B}
\log \frac{p(x|\theta')}{p(x|\theta)} 
\Bigr)^2 \Bigr] < \infty
\]
and
\[
\sup_{ \theta \in K}
E_\theta
\sup _{\theta'  \in \Theta \setminus \Theta_B}
\log \frac{p(x|\theta')}{p(x|\theta)} 
< 0
\]
hold.
\end{assumption}

\renewcommand{\theassumption}{\arabic{assumption}}


We prove the following uniform consistency of maximum likelihood estimators.

\begin{lemma}\label{lemma:uce}
Consider a family of densities
$S=\{p(\cdot|\theta) : \theta \in \Theta  \} $. Consider any global maximizer $\hat \theta (x^n)$ of the likelihood over $\theta$ in $\Theta$. When $\Theta$ is compact,
suppose Assumptions 
\ref{assume:gen:3.1}, \ref{assume:gen:3.1:prime} and \ref{assume:gen:5:prime}. 
Then for each $\xi>0$
\[
\sup_{\theta \in \Theta}
P_\theta ( D(\theta||\hat{\theta}) > \xi  ) 
=O(1/n).
\]
and likewise, for each $\epsilon > 0$,
\[
\sup_{\theta \in \Theta}
P_\theta (|\hat{\theta}(x^n)-\theta| > \epsilon  ) 
=O(1/n).
\]
If $\Theta$ is not compact, suppose Assumptions \ref{assume:gen:3.1}, \ref{assume:gen:3.1:prime}, \ref{assume:gen:5:prime} and \ref{assume:gen:6:prime}.
Then, for any compact subset $K$, we have $$\sup_{\theta \in K} P_\theta ( D(\theta||\hat{\theta}) > \xi  ) =O(1/n)$$ and
\[
\sup_{\theta \in K}
P_\theta (|\hat{\theta}(x^n)-\theta| > \epsilon  ) 
=O(1/n).
\]
\end{lemma}

{\it Proof:}
In the compact $\Theta$ case, let $\bar \delta >0$ be a choice such that the square of the expected supremum in Assumption \ref{assume:gen:5:prime} is finite and and continuous in $\theta'$ and $\theta$. We could proceed as in the proof of the previous lemma, arranging, for each $\xi > 0$, positive $\delta(\theta,\theta') \le \bar \delta$ sufficiently small to achieve the negative expected supremum as explained there.  But we can arrange matters a little better, because we have convergence as $\delta$ goes to $0$ of functions, here assumed to be continuous on the compact set, which are converging to the continuous function $-D(\theta||\theta')$.  So by Dini's theorem it is a uniform convergence. Consequently, we can deduce that for each $\xi > 0$, there is a common choice of positive $\delta < \bar \delta$ with which the expected supremum is less than $-D(\theta||\theta')+\xi/2$ (and moreover, the variance of the expected supremum is arbitrarily close to its continuous limit $V(\theta||\theta')$).  From the common choice $\delta$ of the radii of the balls, when we arrange the finite cover $U_1,\ldots, U_N$, if desired, we have control on the number of them.
Then, as in the proof of the previous lemma,
\begin{eqnarray*}
&&\sup_{\theta' : D(\theta||\theta) > \xi} \left(
\frac{1}{n}
\log \frac{p(x^n|\theta')}{p(x^n|\theta)} \right) \\
&\leq&
\max_{1\le k \le N} \left( \frac{1}{n} \sum_{i=1}^n \sup_{\theta' \in U_k} 
\log \frac{p(x_j|\theta')}{p(x_j|\theta)} \right), \\
\end{eqnarray*}
where each of these suprema have expectation less than $-\xi/2$.  Then by Chebyshev's inequality each of these averages is less than $0$ except in an event of probability not more than $4v_\delta /(n \xi^2)$, where $v_\delta$ is the maximum over $\theta'$ and $\theta$ of the variance of the expected suprema (this maximum being finite since it is a maximum over a compact set of a function assumed continuous).  So we have, crudely, the union bound on the exception event $4N v_\delta /(n \xi^2)$ of order $1/n$.  This proves the conclusion in the compact $\Theta$ case.  

Maximization of the likelihood over non-compact $\Theta$ is handled with an analogous incorporation of the Assumption \ref{assume:gen:6:prime}.  This completes the proof.

\par\vspace{0.3cm} \noindent
{\emph {Remarks}:}
Note that in the non-compact $\Theta$ setting, with Assumption \ref{assume:gen:6:prime}, the domain for maximization of the likelihood is allowed to be larger than the set $K$ of $\theta$ values for which we bound $P_{\theta} \{|\hat \theta (x^n) -\theta|| \ge \epsilon \}$.  That Assumption may be dropped if the likelihood is maximized only within the compact $K$. 

\section{On Assumptions in the Non-i.i.d. Setting}

Stochastic processes for the joint densities $p(x^n|\theta)$, $n\ge 1$, provide considerable generality, beyond the i.i.d. model setting, in which the regret could potentially be analyzed. Two natural settings include Markov models with time homogeneous transitions and stationary ergodic processes.  The core ingredients are the conditional densities $p(x_{t+1}|x_t,\ldots x_1,\theta)$.

Such conditional densities provide the heart of what is needed for arithmetic coding and predictive gambling procedures. Indeed, to facilitate implementation, the rational for Bayes procedures that approximate the mimimax regret is that these yield predictive densities $q(x_{t+1}|x_t,\ldots,x_1)$ which can be computed (e.g. via posterior Monte Carlo) as the posterior means of the $p(x_{t+1}|x_t,\ldots x_1,\theta)$ using the posterior $w(\theta|x_t,\ldots,x_1)$ built from the (suitably modified) Jeffreys prior.

For stationary processes, there is, for each $k\ge 1$, a function $p(s_{k+1}|s_{k},\ldots,s_{1},\theta)$ on ${\cal{X}}^{k+1}$ such that the conditional probability density for $X_{t+1}$, evaluated at $X_{t+1}=s_k$, given $X_{t}=s_{k},\ldots, X_{t+1-k}=s_{1}$ and given $\theta$ is the same for all $t\ge k$, equal to this $p(s_{k+1}|s_{k},\ldots,s_{1},\theta)$.  

These stationary conditionals exist also for Markov processes that do not necessarily start in the invariant distribution. More generally, for asymptotically mean stationary processes (in the sense of 
Gray and Kieffer \cite{GrayKieffer}), 
associated to the process with densities $p(x^n|\theta)$ there is a stationary process with densities ${\bar p}(x^n|\theta)$ with corresponding stationary conditionals.

For simplicity, here we present results for irreducible Markov processes (of arbitrary order $k$). And leave to briefer mention settings in which for large $k$ these approximate more general asymptotically mean stationary processes which have a stationary ergodic ${\bar p}(x^n|\theta), n\ge 1$.  

We consider the following setting for the Markov model of order $k$.  The model of joint densities takes the form $$p(x^n|\theta)=p_{init}(x^k) \prod_{t=k+1}^n p(x_{t}|x_{t-1},\ldots x_{t-k},\theta),$$ where the conditional (transition) density function is time-homogeneous as mentioned above.  It depends on the parameter $\theta$. It is assumed that there is an invariant probability distribution $P^*_{X^k|\theta}$, which, in general, would depend on the parameter. It may have a density $p^*(x^k|\theta)$. It will assumed that the Markov model is irreducible, so the invariant distribution is unique.

As for the initial density $p_{init}(x^k)$, with additional complication, we could consider the case that the initial is the invariant density $p^*(x^k|\theta)$.  But for simplicity here, let's simply take the case that $p_{init}(x^k)$ is fixed (not depending on the parameter) and known.

Now the empirical Fisher information matrix takes the form 
$$\hat{J}_{ij}(\theta,x^n) = - \frac{1}{n} \sum_{t=k+1}^n \frac{\partial^2}{\partial \theta^i \partial \theta^j} \log p(x_{t}|x_{t-1},\ldots x_{t-k},\theta),$$
where it will be assumed that $\log p(s_{k+1}|s^k,\theta)$ is twice continuously differentiable for every $s^{k+1}$ in ${\cal{X}}^{k+1}$.  The second derivatives in the expression above will also be denoted $\partial_i \partial_j \log p(x_{t}|x_{t-1},\ldots x_{t-k},\theta)$.

In investigating the expectation of the empirical Fisher information, let $J_{ij}(\theta|s^k)$ denote the conditional expectation of minus the second derivatives of the log of transition density, given an arbitrarily specified vector of previous values $s^k$.  Then the Fisher information matrix $J_{ij,n}(\theta) = E_\theta [\hat{J}_{ij}(\theta,x^n)]$ takes the form 
$$
J_{ij,n}(\theta)= E_{{\bar P}_{n,\theta}} [J_{ij}(\theta|\cdot )] =\int J_{ij}(\theta|s^k) {\bar p}_n (s^k|\theta) \nu (s^k)$$
where ${\bar P}_{n,\theta} = (1/n) \sum_{t=k+1}^n P_{X_{t-k}^{t-1}|\theta}$ is the average across $t$ of the distributions for $X_{t-k}^{t-1}=(x_{t-k},\ldots,x_{t-1})$ for $t\ge k+1$ having joint density
$${\bar p}_n (s^k|\theta) = \frac{1}{n} \sum_{t=k+1}^n p_{X_{t-k}^{t-1}}(s^k|\theta).
$$
Continuity of $J_{ij,n}(\theta)$ as a function of $\theta$ is inherited from the continuity of $J_{ij}(\theta|s^k)$, and of the distributions $P_{X_{t-k}^{t-1}|\theta}$. These distributions $P_{X_{t-k}^{t-1}|\theta}$ with density functions $p_{X_{t-k}^{t-1}}(s^k|\theta)$ are induced by the transitions starting from the specified initial distribution $P_{init}$ for the first $k$.  

In general the distribution ${\bar P}_{n,\theta}$ converges (weakly) to the invariant $P^*_\theta$ distribution on ${\cal{X}}^k$. 

Let's assume that $J_{ij}(\theta|s^k)$ is a bounded function of $s^k$ in ${\cal{X}}^k$ for each $\theta$ (which allows that the expectation with respect to the any initial distribution is finite).  With $J_{ij}(\theta|s^k)$ is bounded and continuous for $s^k$ in ${\cal{X}}^k$ for each $\theta$, from the convergence of ${\bar P}_{n,\theta}$ to the invariant $P^*_\theta$, it follows that the Fisher information $J_{ij,n}(\theta)$ converges to the limit $J_{ij}(\theta)$ as $n\rightarrow \infty$, where the limiting Fisher information matrix
$$J_{ij}(\theta)= E_{P^*_{\theta}} [J_{ij}(\theta|\cdot )]$$
is the expectation using the invariant distribution.

Moreover, by the ergodic theorem (for asymptotically mean stationary processes \cite{GrayKieffer}), if $\partial_i \partial_j \log p(x_{t}|x_{t-1},\ldots x_{t-k},\theta)$ has finite expected absolute value, with respect to the stationary $P_\theta^*$, it follows that the empirical Fisher information $\hat{J}_{ij}(\theta,x^n)$ converges to $J_{ij}(\theta)$ in $P_\theta$ probability. 
The following assumption allows the extension of that conclusion via local domination.

\begin{assumption}\label{assume:gen:1:st}
The stochastic process is $k$th order Markov with a time homogeneous one-step-ahead transition density that is twice continuously differentiable for $\theta$ in $\Theta^\circ$ and has a unique invariant distribution. The conditional Fisher information $J_{ij}(\theta|\cdot )$ is assumed bounded as a function ${\cal{X}}^k$ for each $\theta$ in $\Theta$, and continuous as a function of 
$\theta$ for each $s^k$.
Moreover, for every $\theta \in \Theta^\circ$ there is an $r=r(\theta)$
such that, for every $i$, $j$,
\begin{align}\label{eq:assume:gen:1:st}
E^*_{\theta} \Biggl[
\sup_{\theta' \in B_{r}(\theta) } 
|\partial_i \partial_j \log p(x_{k+1}|x_k,\ldots x_1,\theta')|\Biggr]
\end{align}
is finite, where the expectation is taken with respect to the stationary distribution. The Fisher information $J_n(\theta)$ and its limit $J(\theta)$ are assumed to be continuous as a function of $\theta$ in $\Theta$. Moreover, $J(\theta)$ is strictly positive definite.
\end{assumption}

Accordingly, as in Appendix A, given any $\epsilon$, there exists $\delta(\theta)$ sufficiently small that the expected supremum in Assumption \ref{assume:gen:1:st} is less than any prescribed positive value, and likewise for the expected supremum when the matrix of second derivatives of $\log p(x_{k+1}|x_k,\ldots x_1,\theta')$ is hit on the left and right by $J(\theta)^{-1/2}$. Moreover, using that a supremum of averages is less than the average of suprema, and appealing to the ergodic Theorem, we find again that, for each $i,j$, and for each sign choice,
\[
\sup_{\theta' \in B_\delta(\theta)} \pm (I_{ij} - (J(\theta)^{-1/2} \hat{J}(\theta',x^n) J(\theta)^{-1/2})_{ij})
\]
is within $\epsilon/d$ of $0$ except in an event of ($P_\theta$) probability tending to $0$ as $n\rightarrow \infty$.

In compact sets, as we have previously indicated (via Dini's Theorem), convergence of continuous functions to a continuous limit is uniformly convergent.  Accordingly, the convergence of $J_n(\theta)$ to $J(\theta)$ is a uniform convergence in $K$.  Accordingly, the collection of $J_n(\theta)$ for large $n$ is equicontinuous, uniformly in $K$, and they share a positive lower bound on their minimum eigenvalue.  Moreover $\int_K |J_n(\theta)|^{1/2} d\theta$ converges to $\int_K |J(\theta)|^{1/2}$.

For consistency, one proceeds similarly, using the relative entropy rate. The relative entropy rate $D(\theta||\theta')$ defined by $ \lim_n (1/n) D(P_{X^n|\theta}||P_{X^n|\theta'}$, in the Markov case, is equal to 
$$D(\theta||\theta') = E_{P^*_{X^k|\theta}} E_{X_{k+1}|X^k,\theta} \Bigl[ \log \frac{p(X_{k+1}|X^k,\theta)}{p(X_{k+1}|X^k,\theta')}\Bigr].$$

\begin{assumption} \label{assume:consistency:st}
For the Markov model, the relative entropy rate $D(\theta||\theta')$ is assumed to be continuous as a function of $\theta'$ and $\theta$, and to satisfy the identifiability property that it is zero only at $\theta'=\theta$.  Moreover, local domination of log-likelihood ratios is assumed for each $\theta$ and $\theta'$ in a compact $K$.  Namely for each $\theta$ and $\theta'$ there is an $r=r(\theta,\theta')$ such that 
$$
E^*_\theta \left[\sup_{\theta'' \in K\cap B_r(\theta')} \log \frac{p(X_{k+1}|X^k,\theta'')}{p(X_{k+1}|X^k,\theta)} \right]
$$
is finite and continuous as a function of $\theta$ and $\theta'$ in $K$.
\end{assumption}

These Assumptions \ref{assume:gen:1:st} and \ref{assume:consistency:st} for Markov models are sufficient for the conditions of 
Theorem \ref{st:thm:lower} for the lower bound conclusion for the regret for stochastic processes, for compact parameter sets. 

Further details are omitted since they closely parallel the previous analysis from the i.i.d. setting.

One may wonder about further generalization beyond the Markov model setting. For consistency, the core matter is the convergence of (1/n) log density ratios to a relative entropy (an asymptotic equipartion property (AEP)). One setting beyond Markov, is when  target stationary ergodic processes $P^*$ are approximated by possibly variable order Markov models $P_{\theta}$.  For this setting one can appeal to the moderately general AEP of 
\cite{Barron85}, 
\cite{gray}, and 
\cite{CoverAlgoet}. 
Presummably, shifting back to a time $0$ reference, the domination condition would be of the form of an assumption of finiteness of
\[
E^*_\theta 
\Bigl[
\sup_{\theta'' \in K\cap B_r(\theta')} 
\log \frac{p(X_0|\theta'', X_{-1},X_{-2},\ldots,X_{-k})}{p^*(X_{0}|X_{-1},X_{-2},\ldots)} 
\Bigr].
\]
where perhaps the Markov order $k=k(\theta')$ would depend on $\theta$.  To deal more generally for families of non-Markov stationary ergodic processes, there were early attempts in 
\cite{Perez}, 
and counterexamples for pairs of stationary (but not ergodic) processes in 
\cite{Kieffer}.  
Nevertheless the question of formulation of suitably general conditions for an AEP for pairs of stationary ergodic processes $P_\theta$ and $P_\theta'$ is largely open.  Until such is obtained, the matter of consistency of maximum likelihood in non-Markov models requires that it be addressed on a case-by-case basis.

\section{Lower Bound on Gaussian Measure (Proposition~\ref{factor})}
\label{proofoffactorporoposition}

\setcounter{proposition}{0}

Recall that $\Phi$ denotes the measure of
the $d$-dimensional standard normal distribution
and $N_r(0)=\{z : |z| \leq r \}$.
The following holds.
\begin{proposition}\label{factor:appendix}
\input{proposition1}
\end{proposition}

{\it Proof:}
Note that
\[
\Phi(N_{\sqrt{n}\epsilon}(0))
=
\Pr\Bigl\{ \sum_{i=1}^d X_i^2 \le n\epsilon^2  \Bigr\},
\]
where each $X_i$ is an independent standard normal random variable.
The density function of normal distribution 
with mean $0$ and variance $\sigma^2$ is
\begin{align*}
  \frac{1}{(2\pi \sigma^2)^{1/2}}\exp\Bigl( \frac{-x^2}{2\sigma^2} \Bigr)
&=
\exp\Bigl( \frac{-x^2}{2\sigma^2} -\frac{1}{2}\log (2\pi \sigma^2) \Bigr)\\
&=
\exp(u x^2 -\psi(u)),
\end{align*}
where
$u = -1/(2\sigma^2)$ and 
$\psi(u) = (1/2)\log(2\pi \sigma^2)
=(1/2)\log(-\pi/u)
$.
Let $q(x|u)=\exp(u x^2 -\psi(u))$,
which forms an exponential family, for which 
the natural parameter is $u=-1/(2\sigma^2)$.
Let $v=E_u(x^2)=\sigma^2$.
Then $v$ is the expectation parameter coresponding to $u$.
We have
\[
q(x^d|u)=\exp(d(u ||x||^2/d -\psi(u))),
\]
where $x^d=(x_1,\dots,x_d)$
and
$||x||^2$ denotes the square of $x^d$'s Euclidian norm.
Note that the maximum likelihood estimate
of $v=\sigma^2$ given $x^d$ is 
$\hat{\sigma^2}=||x||^2/d$.
Then by the large deviation inequality
(Lemma~\ref{ldpmulti}),
we have
\[
\Pr\Bigl\{ \sum_{i=1}^d X_i^2 \ge n\epsilon^2  \Bigr\}
\le \exp( -d D(n\epsilon^2/d||1) ),
\]
where $D(v||v')$ denotes
the Kullback-Leibler divergence from
$q(x|u)$ to $q(x|u')$ ($v'$ is the expectation parameter corresponding
to $u'$).
Here, we have
\begin{align*}
D(v||v')
&=(u -u')v -\psi(u)+\psi(u')  \\
&
=-\frac{1}{2} +\frac{1}{2}\frac{v}{v'} +\frac{1}{2}\log\frac{v'}{v}\\
&
=\frac{1}{2}
\Bigl(-1 +\frac{v}{v'} -\log\frac{v}{v'}\Bigr).
\end{align*}
Hence we have
\begin{align*}
d D(n\epsilon^2/d||1)
&=
\frac{d}{2}
\Bigl(-1 +\frac{n\epsilon^2}{d} -\log\frac{n\epsilon^2}{d} \Bigr)\\
&=
-\frac{d}{2}
+\frac{n\epsilon^2}{2}
\bigl(1 -\frac{d}{n\epsilon^2}\log\frac{n\epsilon^2}{d} \bigr),
\end{align*}
which is not less than
\[
-\frac{d}{2}
+
\frac{n\epsilon^2}{2}
\bigl(1 -\frac{\log 2}{2} \bigr)
\geq
-\frac{d}{2}
+\frac{n\epsilon^2}{4},
\]
when $n\epsilon^2/d \geq 2$.
{\it This completes the proof.}

\section{Normalization Constant of The Ideal Priors}
\label{appendix:normalizaionconstant}

\setcounter{forlemma}{\thelemma}
\setcounter{lemma}{\theforlemmafornormalization}
\begin{lemma}
    \input{lemma_for_normalization}
\end{lemma}
\setcounter{lemma}{\theforlemma}

{\it Proof:}
Assume $\theta \in K_{\epsilon\alpha}$.
Then, by $\epsilon^2\alpha^2 \le \ubar{\lambda}$,
$N_{\sqrt{n}\epsilon}(0) \subset
(\sqrt{n}/\alpha )(J_{n,\theta})^{1/2}(K-\theta)$ is satisfied,
which means
\begin{align*}
\Phi(U_{K,\epsilon,n}^{(\alpha)}(\theta))
&=
\Phi(N_{\sqrt{n}\epsilon}(0))
\end{align*}
holds for $\theta \in K_{\epsilon \alpha}$.
Hence by Proposition~\ref{factor}
and the assumption $n\epsilon^2/d \ge 2$, the following holds.
\begin{align*}
\inf_{\theta \in K_{\epsilon\alpha}}
\Phi(U^{(\alpha)}_{K,\epsilon,n}(\theta))
\geq 1-e^{-n\epsilon^2/4+d/2}.
\end{align*}

Using this inequality, we have
\begin{align*}
&  \int_K \frac{|J_{n,\theta}|^{1/2}}{\Phi(U^{(\alpha)}_{K,\epsilon,n}(\theta))}d\theta\\
&\le
  \int_{K_{\epsilon/\alpha}} \frac{|J_{n,\theta}|^{1/2}}{1-e^{-n\epsilon^2/4+d/2}}d\theta
  +
  \int_{K\setminus K_{\epsilon \alpha}} \frac{|J_{n,\theta}|^{1/2}}{\ratio_n^{(\alpha)}(\epsilon)}d\theta\\
&\le
  \frac{C_{J,n}(K)}{1-e^{-n\epsilon^2/4+d/2}}
  +
  \frac{ C_{J,n}(K \setminus K_{\epsilon \alpha }) }{\ratio_n^{(\alpha)}(\epsilon)}.
\end{align*}
{\it This completes the proof of the Lemma.}

\section{Lower Bounds on $\rho^{(\alpha)}_n(\epsilon)$}

For a subset $A$ of $\mathbb{R}^d$,
let ${\rm diam}(A)$ and ${\rm vol}(A)$ denote the diameter of $A$ and  the volume of $A$,
respectively.

\setcounter{forlemma}{\thelemma}
\setcounter{lemma}{\theforlemmaratio}
\begin{lemma}
\input{lemma_for_ratio}
\end{lemma}
\setcounter{lemma}{\theforlemma}



{\it Proof:}
Let $L = J_{n,\theta}^{1/2}K$.
Then, we have
\begin{align*}
\ratio_n^{(\alpha)}(\epsilon,\theta)
&=
\Phi(U^{(\alpha)}_{K,\epsilon,n}(\theta)) 
\\
 &=
 \Phi(N_{\sqrt{n}\epsilon}(0) \cap (\sqrt{n}/\alpha) (L - J_{n,\theta}^{1/2}\theta) )
\\
 &=
 \Phi\Bigl( (\sqrt{n}/\alpha)  \bigl( N_{\epsilon \alpha}(0) \cap (L - J_{n,\theta}^{1/2}\theta)\bigr) \Bigr).
\end{align*}

Define for $\eta \in L$ and for $\epsilon > 0$,
\begin{align*}
A(\eta) & = \{ u \in \Re^d : |u|=1, \exists \kappa > 0,\; \eta + \kappa u  \in L^\circ \},\\
C(\eta,\epsilon) & = \{ u \in \Re^d : |u|=1,\;  \eta + \epsilon u  \in L^\circ \},
\end{align*}
and
\[
\tilde{C}(\eta,\epsilon) = \{ v \in \Re^d : |v|=\epsilon,\;  \eta + v  \in L^\circ \}.
\]
By definition, we have
\[
A(\eta) \subset \bigcup_{\epsilon > 0}C(\eta,\epsilon).
\]
By the convexity of $L$, 
\[
A(\eta) \supset C(\eta,\epsilon_1) \supset C(\eta,\epsilon_2)
\] holds for $0< \epsilon_1 \le \epsilon_2$.
Hence, $A(\eta) = \bigcup_{\epsilon > 0}C(\eta,\epsilon)$ holds.

Note that $\tilde{C}(\eta,\epsilon)$ is an open subset of the surface of $N_\epsilon(0)$,
since it is the intersection of $L^\circ - \eta$ (an open set) and the surface of $N_\epsilon(0)$.
Hence it is measurable in the surface of $N_\epsilon(0)$, which means $C(\eta,\epsilon)$ is also measurable.

Let $r(\eta,\epsilon)$ denote the ratio of the $d-1$ dimensional volume of $C(\eta,\epsilon)$
to that of the surface of $N_1(0)$ ($d\pi^{d/2}/\Gamma(d/2 +1)$).
Since $C(\eta,\epsilon)$ does not decrease as $\epsilon$ decreases,
$r(\eta,\epsilon)$ does not decrease as $\epsilon$ decreases.

For $\eta \in L$, define
\[
{\rm cone}(\eta,\kappa) = \{ \eta + \xi u :  u \in C(\eta,\kappa), \xi \ge 0\}.
\]
If $C(\eta,\kappa)$ is connected,
this is an unbounded cone with the vertex $\eta$. If not, it is a union of unbounded cones.
Note that 
\begin{align}\label{eq:for_rho_lowerbound}
N_\epsilon(\eta) \cap L \supset N_\kappa(\eta) \cap {\rm cone}(\eta,\kappa)
\end{align}
holds for all $\kappa \in (0,\epsilon]$.
This is shown as follows.
Let $\eta + v$ be an element of ${\rm cone}(\eta,\kappa)$ with $|v| \le \kappa$.
Then, $v/|v| \in C(\eta,\kappa)$ holds by the definition of $C(\eta,\kappa)$.
This means that $\eta + \kappa v /|v| \in L^\circ$. 
Since $\kappa /|v| \ge 1$,
$\eta + v$ is on the line segment between $\eta \in L$ and $\eta +\kappa v /|v| \in L^\circ$.
Hence $\eta +v \in L^\circ$ holds. 
Since $|v| \le \kappa \le \epsilon$, (\ref{eq:for_rho_lowerbound}) holds.

Because of (\ref{eq:for_rho_lowerbound}),
for any nonnegative integrand,
the integral over
$(\sqrt{n}/\alpha)(N_{\epsilon \alpha}(0) \cap (L-\eta))$ is lower bounded by
that over $(\sqrt{n}/\alpha)(N_{\epsilon \alpha}(0) \cap ({\rm cone}(\eta,\epsilon \alpha) -\eta))$.
Then due to the symmetry of $\Phi$, we have for each $\theta \in K$,
\[
\ratio^{(\alpha)}_n(\epsilon,\theta )
\ge
r(\eta,\epsilon \alpha)
\Phi(N_{\sqrt{n}\epsilon}(0)),
\]
where
$\eta = J_{n, \theta}^{1/2} \theta$.

Finally, we will show that $\inf_{\eta \in L}r(\eta,\epsilon \alpha)$ is lower bounded by
the positive constant determined by ${\rm vol}(L)$ and ${\rm diam}(L)$.
Let $\epsilon_0$ so small that
${\rm vol}(N_{\epsilon_0}(0)) \le {\rm vol}(L)/2$
and assume that $\epsilon \alpha \le \epsilon_0$.
Let $\Delta$ denote ${\rm diam}(L)$.
Recall that $C(\eta,\epsilon \alpha) \supset C(\eta,\epsilon_0) \supset C(\eta,\kappa)$ holds
for all $\kappa \ge \epsilon_0$ and for all $\epsilon$ such that $\epsilon \alpha \le \epsilon_0$.
Since
\[
\frac{\eta' - \eta}{|\eta' -\eta|} 
\in C(\eta,\kappa)
\]
holds for all $\eta' \in L^\circ$ with
$|\eta' - \eta| = \kappa$,
the set $L^\circ \setminus N_{\epsilon_0}(\eta)$
is included in ${\rm cone}(\eta,\epsilon_0)$.
Note that $L$ is included in $N_{\Delta}(\eta)$.
Then, we see that
$L^\circ \setminus N_{\epsilon_0}(\eta)$ is included in
${\rm cone}(\eta,\epsilon_0) \cap N_{\Delta}(\eta)$.
Comparing the volumes of both sets, we have
\[
{\rm vol}(L) - {\rm vol}(N_{\epsilon_0}(\eta))
\le
V_d \Delta^d r(\eta,\epsilon_0).
\]
Since the left side is not less than
${\rm vol}(L) - {\rm vol}(L)/2 = {\rm vol}(L)/2$, we have
\[
r(\eta,\epsilon \alpha) \ge 
\frac{{\rm vol}(L)}{ 2({\rm diam}(L))^d V_d }
\]
holds for all $\epsilon$ such that
$\epsilon \alpha \in (0,\epsilon_0)$.
{\it The proof is completed.}

\section{Continuity of $\Phi(U_{K,\epsilon,n}^{(\alpha)})$}
\label{appendix_continuity_of_factor}

\setcounter{forlemma}{\thelemma}
\setcounter{lemma}{\theforlemmaideal}
\begin{lemma}
\input{lemma_for_ideal_prior}
\end{lemma}
\setcounter{lemma}{\theforlemma}

{\it Proof:}
First assume that $J_n(\theta)$ does not depend on $\theta$.
Let $J_n$ denote the value.
Note that
\begin{align*}
&U^{(\alpha)}_{K, \epsilon,n}(\theta')
\setminus
 U^{(\alpha)}_{K, \epsilon,n}(\theta) \\
&= 
N_{\sqrt{n}\epsilon}(0) \cap (\sqrt{n}/\alpha)
 J_n^{1/2} ( (K-\theta') \setminus (K-\theta))\\
 &= 
N_{\sqrt{n}\epsilon}(0) \cap (\sqrt{n}/\alpha)
  ((L-\eta') \setminus (L-\eta)),
\end{align*}
holds,
where $L = J_n^{1/2}K$, $\eta = J_n^{1/2}\theta$, and $\eta' = J_n^{1/2}\theta'$.
Then, we have
\[
\Phi(U^{(\alpha)}_{K, \epsilon,n}(\theta')
\setminus
 U^{(\alpha)}_{K, \epsilon,n}(\theta)) \le
 \Phi( (\sqrt{n}/\alpha)
  ((L-\eta') \setminus (L-\eta))).
\]
Recall that $\Phi$ is the $d$ variate
standard Gaussian measure.
Note that $L-\eta'$ is given by displacing $L-\eta$ by $\eta-\eta'$.
Hence, a point $\tilde{\eta}$ in $L - \eta$
has a representation as a scalar amount
in the direction  $\eta' -\eta$ together with $d-1$ amounts
in directions orthogonal to $\eta' - \eta$.
The displacement from $L - \eta$ to $L - \eta'$
only affects that scalar translation by the amount $|\eta' - \eta|$.
Hence, we have
\begin{align*}
\Phi & \Bigl( (\sqrt{n}/\alpha)
  \bigl((L-\eta') \setminus (L-\eta)\bigr) \Bigr) \\
 & \le 
  \Pr\{  |y| \le (\sqrt{n}/\alpha)|\eta' -\eta|/2 \}\\
  & \le  (\sqrt{n}/\alpha)|\eta' -\eta|,
\end{align*}
where $y$ denotes the (univariate) standard Gaussian variable.
Since $|\theta' - \theta| \le r \le \epsilon$ 
and since the eigenvalues of $J_n(\theta)$ are bounded by $\lambda_n$,
\[
\Phi  \Bigl( (\sqrt{n}/\alpha)
  \bigl((L-\eta') \setminus (L-\eta)\bigr) \Bigr) 
  \le \frac{\sqrt{n\lambda_n}r}{\alpha}.
\]
Recalling that 
$\Phi(U^{(\alpha)}_{K,\epsilon,n}(\theta)) \ge \ratio_n(\epsilon,\theta)$,
we have the claim of Lemma.

Next, consider the general $J_n(\theta)$ case.
We have
\begin{align*}
J_n(\theta')^{1/2}&(K-\theta') \setminus J_n(\theta)^{1/2}(K-\theta)\\
\subset &
J_n(\theta)^{1/2}(K-\theta') \setminus J_n(\theta)^{1/2}(K-\theta)\\
& \cup
J_n(\theta')^{1/2}(K-\theta') \setminus J_n(\theta)^{1/2}(K-\theta').
\end{align*}
The quantity produced by the first component of the union
in the last expression
is bounded as in the same way as the constant $J_n$ case.
As for the second component,
we have
\begin{align*}
&    J_n(\theta')^{1/2}(K-\theta') \setminus J_n(\theta)^{1/2}(K-\theta')\\
&    =
J_n(\theta)^{1/2}\Bigl( \bigl( J_n(\theta)^{-1/2}J_n(\theta')^{1/2}(K-\theta') \bigr) \setminus (K-\theta') \Bigr) \\
&    \subset
J_n(\theta)^{1/2}\Bigr( \bigl( (1+ g_{\theta,r} )(K-\theta') \bigr) \setminus (K-\theta')\Bigr)    \\
& =
 (1+ g_{\theta,r} ) J_n(\theta)^{1/2} (K-\theta')  \setminus J_n(\theta)^{1/2}(K-\theta') .
\end{align*}
Let $L' = J_n(\theta)(K-\theta')$, then we have
\begin{align*}
  &  J_n(\theta')^{1/2}(K-\theta') \setminus J_n(\theta)^{1/2}(K-\theta') \\
&\subset 
(1+ g_{\theta,r} )L'  \setminus L'.
\end{align*}

Hereafter, assume $g_{\theta,r} > 1$.
Otherwise, the analysis below is not necessary, because 
$ (1+ g_{\theta,r} )L' \setminus L'$ is the empty set in that case.

Note that $L'$, which is compact, includes the origin.
Assume now that the origin is included in ${L'}^\circ$, that is, $\theta' \in K^\circ$.
Since $L'$ is convex,
the set 
$ (1+ g_{\theta,r} )L' \setminus L'$ is a shell outside $L'$
as follows.
\begin{align}\nonumber
& (1+ g_{\theta,r} )L'  \setminus L' \\ \label{eq:defshell}
= & \{   y u : u \in \partial L', 1 < y \le  1 + g_{\theta,r}   \}\\ \nonumber
= & \{   y_{u} u/|u| : u \in \partial L', |u| < y_{u} \le |u|( 1 + g_{\theta,r})   \},
\end{align}
where $\partial L' = L' - {L'}^\circ$.
Note that $\{ u/|u| : u \in \partial L' \}$ is the unit sphere,
when the origin is in the interior of $L'$.
The equality (\ref{eq:defshell}) itself
holds even if the origin is not included in the interior of $L'$.
Hence, we do not need the assumption $\theta' \in K^\circ$
for the argument below.

Now, we can evaluate 
\begin{align*}
 &   \Phi\Bigl( (\sqrt{n}/\alpha)
\bigl(
J_n(\theta')^{1/2}(K-\theta') \setminus J_n(\theta)^{1/2}(K-\theta')\bigr)\Bigr) \\
\le &
\Phi\Bigl( (\sqrt{n}/\alpha) \bigl( (1+ g_{\theta,r} )L'  \setminus L' \bigr) \Bigr)
\end{align*}
Let $\xi$ be a $d$-variate standard Gaussian variable.
Then, 
we transform it to $(v,y) =(\xi/|\xi|,|\xi|)$.
Here, $v$ and $y$ are independent of each other,
the marginal density of $v$ is uniform on the surface of the unit sphere,
and that of $y$ is 
\[
C_d \,
y^{d-1}\exp\Bigl( - \frac{y^2}{2} \Bigr)
\le
C_d (d-1)^{(d-1)/2}e^{-(d-1)/2},
\]
where 
$C_d = 2^{(1-d)/2}/\Gamma(d/2)$ is the normalization constant
and
the equality of the above holds if and only if $y=\sqrt{d-1}$.
Since $|u|$ ($u \in \partial L'$) is upper bounded by ${\rm diam}(L')$, we have
the following, where $y_0 =\sqrt{n} |u|/\alpha$ and $y_1 = \sqrt{n}|u|( 1 + g_{\theta,r})/\alpha$.
\begin{align*}
    & \Phi\Bigl( (\sqrt{n}/\alpha)\bigl( (1+ g_{\theta,r} )L'  \setminus L' \bigr) \Bigr)\\
    & \le 
    \max_{u \in \partial L'}\Pr\{ y_0 < y \le y_1 \}\\
    &\le 
    \max_{u \in \partial L'}C_d \int_{y_0}^{y_1} y^{d-1}\exp\Bigl( - \frac{y^2}{2} \Bigr) dy
    \\
    &\le 
    \max_{u \in \partial L'}C_d \int_{y_0}^{y_1} (d-1)^{(d-1)/2}e^{-(d-1)/2} dy\\
    &\le C_d \, {\rm diam}(L') \sqrt{n}g_{\theta,r}/\alpha\\
    &= C_d {\rm diam}(K)\sqrt{n\lambda_n}g_{\theta,r}/\alpha.
\end{align*}
{\it This completes the proof.}

\section{Some Inequality for Models with Hidden Variables (Lemma~\ref{ineq_for_mhv})}
\label{proof_of_lemma_ineq_for_mhv}

Recall that we defined the model with hidden variable in (\ref{mwh}) as
\[
p(x|\theta) = \int \! \! \kappa(x|y)q(y|\theta)\nu_y(dy), 
\]
where $q(y|\theta)$ is the density of an exponential family.
We have the following, where $G(\theta)$ is the Fisher information of $q(y|\theta)$.
\setcounter{forlemma}{\thelemma}
\setcounter{lemma}{\theforlemmamonotone}
\begin{lemma}\label{ineq_for_mhv:appendix}
Given a data string $x^n$, let $\hat{\theta}$ 
denote the MLE for 
a model with hidden variables 
$p(x^n|\theta)$ defined by (\ref{mwh}).
Then, the following holds for all $x^n \in \mathcal{X}^n$.
\begin{align}\label{divergence:appendix}
\forall \theta \in \Theta, \: \:
\frac{1}{n}
\log
\frac{p(x^n|\hat{\theta})}{p(x^n|\theta)}
&\leq D(q(\cdot|\hat{\theta})||q(\cdot|\theta)),\\ \label{empirical:appendix}
\forall \theta \in \Theta, \:\:
\hat{J}(\theta,x^n)
&\leq
G(\theta).
\end{align}
In particular, when $q(y|\theta)$ is the multinomial model,
the following holds
\begin{eqnarray}\label{forBer:appendix}
\frac{p(x^n|\hat{\theta})}{p(x^n|\theta)}
\leq \exp(n D(q(\cdot|\hat{\theta})||q(\cdot|\theta)))
=\prod_{y \in \mathcal{Y}} \frac{\hat{\eta}_y^{n\hat{\eta}_y}}{\eta_y^{n\hat{\eta}_y}},
\end{eqnarray}
where $\eta_y = q(y|\theta)$ and $\hat{\eta}_y = q(y|\hat{\theta})$.
\end{lemma}
\setcounter{lemma}{\theforlemma}

{\it Proof:}
Note that
\begin{align*}
q(x^n|\theta)
&=\prod_{t=1}^n \int \kappa(x_t|y_t)p(y_t|\theta)\nu_y(dy_t) \\
&=
\int \prod_{t} \kappa(x_t|y_t)p(y_t|\theta)\nu_y(dy^n)\\
&=
\int \kappa(x^n|y^n)p(y^n|\theta)\nu_y(dy^n).
\end{align*}
We have
\begin{align*}
    \frac{q(x^n|\theta)}{q(x^n|\theta')}
&=
\frac{\int \kappa(x^n|y^n)p(y^n|\theta)\nu_y(dy^n)}
{\int \kappa(x^n|y^n)p(y^n|\theta')\nu_y(dy^n)} \\
&=
\int
\frac{ p(y^n|\theta)}{p(y^n|\theta')}
\frac{\kappa(x^n|y^n)p(y^n|\theta')\nu_y(dy^n)}{\int \kappa(x^n|z^n)p(z^n|\theta')\nu_y(dz^n)}.
\end{align*}
Define $q(y^n|x^n,\theta')$ by
\[
q(y^n|x^n,\theta')
=
\frac{\kappa(x^n|y^n)p(y^n|\theta')}{\int \kappa(x^n|z^n)p(z^n|\theta')\nu_y(dz^n)},
\]
which is the posterior distribution of $y^n$ given $x^n$
provided $x^n$ is drawn from $q(x^n|\theta')$.

Using it, we can write
\[
\frac{q(x^n|\theta)}{q(x^n|\theta')}
=
\int
q(y^n|x^n,\theta')
\frac{ p(y^n|\theta)}{p(y^n|\theta')}\nu_y(dy^n).
\]
Then by Jensen's inequality, we have
\begin{align}\label{jensen}
&\frac{1}{n}
\log \frac{q(x^n|\theta)}{q(x^n|\theta')}\\ \nonumber
&\geq
\frac{1}{n}
\int
q(y^n|x^n,\theta')
\log \frac{ p(y^n|\theta)}{p(y^n|\theta')}\nu_y(dy^n).
\end{align}
Let $f(\theta,\theta')$ denote the left side, and $g(\theta,\theta')$ the right side.
Then, we have
\begin{eqnarray}\label{jensen2}
\forall \theta, \theta' \in \Theta,\:\:
f(\theta,\theta') - g(\theta,\theta') \geq 0,
\end{eqnarray}
where equality holds when $\theta=\theta'$.
Hence, Hessian of the left side is semi positive-definite.
That is,
the matrix whose $ij$ entry is
\begin{eqnarray}\label{matrix}
\frac{\partial^2 \log   f(\theta,\theta')}{\partial \theta_i\partial \theta_j}
-\frac{\partial^2 \log
  g(\theta,\theta')}{\partial \theta_i\partial \theta_j}
\end{eqnarray} 
is semi positive definite.
Note that 
\begin{eqnarray}\label{exp}
g(\theta,\theta') = \theta \bar{\eta}^t -\psi(\theta) -(\theta' \bar{\eta}^t -\psi(\theta')),
\end{eqnarray}
where 
\[
\bar{\eta} = \frac{1}{n}\int q(y^n|x^n,\theta') \sum_{t=1}^ny_t \nu_y(dy^n).
\]
From (\ref{exp}), we have
\[
-\frac{\partial^2 \log
  g(\theta,\theta')}{\partial \theta_i\partial \theta_j}
=
\frac{\partial^2 \psi(\theta)}{\partial \theta_i\partial \theta_j}
=
G_{ij}(\theta).
\]
Hence, semi positive-definiteness of (\ref{matrix}) implies
\[
\forall \theta \in \Theta,\:\:
\hat{J}(\theta,x^n) \leq G(\theta),
\]

Plugging in $\hat{\theta}$ to $\theta'$ in (\ref{jensen2})
and noting $f(\theta,\hat{\theta}) \leq 0$, we have
\begin{eqnarray}\label{jensen3}
\forall \theta  \in \Theta,\:\:
0 \geq f(\theta,\hat{\theta}) \geq g(\theta,\hat{\theta}),
\end{eqnarray}
where both inequality hold as equality, when $\theta=\hat{\theta}$.
That is,
\[
g(\theta,\hat{\theta}) \leq g(\hat{\theta},\hat{\theta}) = 0.
\]
Together with (\ref{exp})
the following holds
\begin{align}\label{exp2}
&g(\theta,\hat{\theta}) = \theta \bar{\eta}^t -\psi(\theta) 
-(\hat{\theta} \bar{\eta}^t -\psi(\hat{\theta})) \\
&\leq
\hat{\theta} \bar{\eta}^t -\psi(\hat{\theta}) 
-(\hat{\theta} \bar{\eta}^t -\psi(\hat{\theta})) = 0,
\end{align}
which implies $\bar{\eta}=\hat{\eta}$.
Here $\hat{\eta}$ denotes the coo responding value of expectation
parameter $\eta$ to $\hat{\theta}$.

Note that
\[
g(\theta,\hat{\theta}) = -D(p(\cdot|\hat{\theta})||p(\cdot|\theta)),
\]
where $D(p(\cdot|\hat{\theta})||p(\cdot|\theta))$ is the Kullback divergence
from $p(y|\hat{\theta})$ to $p(y|\theta)$
defined as
\[
D(p(\cdot|\hat{\theta})||p(\cdot|\theta))
= \int p(y|\hat{\theta})\log\frac{p(y|\hat{\theta})}{p(y|\theta)}\nu_y(dy).
\]
Hence from (\ref{jensen}), we have
\begin{eqnarray}\label{jensen100}
\frac{1}{n}
\log \frac{q(x^n|\hat{\theta})}{q(x^n|\theta)}
\leq
D(\hat{\theta}||\theta).
\end{eqnarray}
{\it This completes the proof.}

\section*{Acknowledgement}
The authors thank Hiroshi Nagaoka, who
gave valuable advises for the part of the models with hidden variables.

%








\end{document}

%% file: lemma_asympt_normality.tex
Under Assumptions~\ref{assume:gen:dash:1}, \ref{assume:gen:dash:2}, \ref{assume:gen:dash:3}, 
$\sqrt{n}J(\theta)^{1/2}(\hat{\theta}-\theta)$
converges in distribution to a mean zero normal random vector
and there is a constant $c$ such that
the following inequality holds 
for any positive $b$,
\begin{align*}
& \max_{\theta \in K}
P_\theta \bigl ( \sqrt{n}||J(\theta)^{1/2}(\hat{\theta}-\theta)|| >
b \sqrt{\log n}
\bigr)\\
& \quad \le   \frac{c}{b^2 \log n}  +o\Bigl(\frac{1}{\log n} \Bigr).
\end{align*} 
The constant $c$ equals
$\max_{\theta \in K}{\rm trace}(J(\theta)^{-1}I(\theta))$ which is the
dimension $d$, the trace of an identity matrix, when $I(\theta)=J(\theta)$.

%% file: proposition1.tex
For all $n > 0$ and $\epsilon > 0$,
\[
\Phi(N_{\sqrt{n}\epsilon}(0))
\geq 
1-\exp\Bigl(
-\frac{n\epsilon^2}{2}
\bigl(1 -\frac{d}{n\epsilon^2}\log\frac{n\epsilon^2}{d} \bigr)
+\frac{d}{2}
\Bigr),
\]
which is larger than
\[
1-\exp\Bigl( -\frac{n\epsilon^2}{4}+\frac{d}{2}\Bigr),
\]
when $n\epsilon^2/d \ge 2$.

%% file: lemma_for_normalization.tex
Let $K$ be an arbitrary compact set in $\Theta^\circ$.
Suppose Assumption~\ref{assume:st:-upper2} holds
and let $\ubar{\lambda}$
denote a lower bound on the smallest eigenvalue of $J_{n,\theta}$
among $n \in \mathbb{N}$ and $\theta \in K$.
Let
\[
K_\epsilon= \{ \theta : B_\epsilon(\theta) \subset K  \},
\]
Then for $\epsilon$ such that 
$n\epsilon^2/d \ge 2$ and $\epsilon^2\alpha^2 \le \ubar{\lambda}$, we have
\[
C_{J,n}(K)
\le
C_{K,\epsilon,n}^{(\alpha)}
\le
  \frac{C_{J,n}(K)}{1-e^{-n\epsilon^2/4+d/2}}
  +
  \frac{ C_{J,n}(K\setminus K_{\epsilon \alpha}) }{\ratio_n^{(\alpha)}(\epsilon)},
\]
where
\[
C_{J,n}(A) = \int_A |J_{n,\theta}|^{1/2}d\theta.
\]

%% file: lemma_for_ratio.tex
    Assume that $K$ is compact, convex, and the closure of an open set in $\Re^d$.
    Then, 
    for all $\epsilon$ such that ${\rm vol}(N_\epsilon(0)) \le {\rm vol}(J^{1/2}_{n,\theta}K)/2$,
    the following holds.
    \[
    \ratio^{(\alpha)}_n(\epsilon,\theta)
\ge
 \frac{{\rm vol}(J^{1/2}_{n,\theta}K)}{2  ({\rm diam}(J^{1/2}_{n,\theta}K))^d V_d }
\Phi(N_{\sqrt{n}\epsilon}(0)).
    \]

%% file: lemma_for_ideal_prior.tex
Let $\lambda_n$ denote the maximum of the largest eigenvalue of $J_n(\theta)$  among $\theta \in K$.
For a certain $\epsilon > 0$,
    for all $r \le \epsilon$, 
    for all $\theta' \in B_r(\theta) \cap K$, and
    for all $\theta \in K$,
\begin{align*}
&\frac{
\Phi(U^{(\alpha)}_{K,\epsilon,n}(\theta'))}
{\Phi(U^{(\alpha)}_{K,\epsilon,n}(\theta))} \\
\le & 1 + \frac{\sqrt{n\lambda_n}r}{\ratio_n(\epsilon,\theta) \alpha}
+\frac{C_d  \, {\rm diam}(K)\sqrt{n\lambda_n} \max\{ g_{\theta,r},0\}}{\ratio_n(\epsilon,\theta) \alpha}
\end{align*}  
 holds,
 where $C_d = 2^{1-d/2}\Gamma(d/2)$ and
 \[
 g_{\theta,r} =\max_{\theta' \in B_r(\theta)\cap K}||J_n(\theta)^{-1/2}J_n(\theta')^{1/2}||_s  - 1.
 \]

%% file: lemma_ineq_monotone.tex
Given a data string $x^n$, let $\hat{\theta}$ 
denote the MLE for 
a model with hidden variables 
$p(x^n|\theta)$ defined by (\ref{mwh}).
Then, the following holds for all $x^n \in \mathcal{X}^n$.
\begin{align}\label{divergence}
\forall \theta \in \Theta, \: \:
\frac{1}{n}
\log
\frac{p(x^n|\hat{\theta})}{p(x^n|\theta)}
&\leq D(q(\cdot|\hat{\theta})||q(\cdot|\theta)),\\ \label{empirical}
\forall \theta \in \Theta, \:\:
\hat{J}(\theta,x^n)
&\leq
G(\theta).
\end{align}
In particular, when $q(y|\theta)$ is the multinomial model,
the following holds
\begin{eqnarray}\label{forBer}
\frac{p(x^n|\hat{\theta})}{p(x^n|\theta)}
\leq \exp(n D(q(\cdot|\hat{\theta})||q(\cdot|\theta)))
=\prod_{y \in \mathcal{Y}} \frac{\hat{\eta}_y^{n\hat{\eta}_y}}{\eta_y^{n\hat{\eta}_y}},
\end{eqnarray}
where $\eta_y = q(y|\theta)$ and $\hat{\eta}_y = q(y|\hat{\theta})$.